\pdfoutput=1
\documentclass[unpublished,letterpaper, nopdfoutputerror]{quantumarticle}

\usepackage[T1]{fontenc}
\usepackage[utf8]{inputenc}
\usepackage{graphicx}
\usepackage{multirow}
\usepackage{amsmath}
\usepackage{mathtools}
\usepackage{amssymb}
\usepackage{amsfonts}
\usepackage{bm}
\usepackage{setspace}
\usepackage{import}
\usepackage{tikz}
\usetikzlibrary{quantikz2}
\usetikzlibrary{positioning}
\usepackage{nicefrac}
\usepackage{xspace}
\usepackage{array}
\usepackage{dcolumn}
\usepackage{booktabs}
\usepackage{siunitx}
\usepackage[table,dvipsnames]{xcolor}
\usepackage{tabularx}
\usepackage{enumitem}
\usepackage{adjustbox}
\usepackage[normalem]{ulem}
\usepackage{multirow}
\usepackage{xurl}

\definecolor{titlecolor}{gray}{0.85}
\newcolumntype{M}[1]{>{\centering\arraybackslash}m{#1}}
\newcommand{\RowHT}{0.55cm}   
\newcommand{\ImgHT}{0.7cm}  
\newcommand{\RowStrut}{\rule{0pt}{\RowHT}}
\newcommand{\numcell}[1]{\adjustbox{valign=c}{#1}}
\newcommand{\imgcell}[1]{\adjustbox{valign=c}{\includegraphics[height=\ImgHT]{#1}}}
\newcommand{\opI}{\mathmakebox[1.05em][c]{I}}
\newcommand{\opX}{\mathmakebox[1.05em][c]{X}}
\newcommand{\opZ}{\mathmakebox[1.05em][c]{Z}}

\usepackage{hyperref}
\hypersetup{colorlinks=true, linkcolor=blue, citecolor=blue, urlcolor=blue}

\makeatletter
\patchcmd{\@maketitle}
  {\noindent{\huge\hyphenpenalty=50000 \@printtitle\par}}
  {\noindent{\huge\raggedright\hyphenpenalty=50000 \@printtitle\par}}
  {}{}
\makeatother

\title{High-Rate and Resource-Efficient All-Photonic Quantum Repeater Architectures with 9 km Repeater Spacing}

\author{Ryosuke Shiina}
\affiliation{Department of Physics, University of Massachusetts Amherst, Amherst, MA 01003, USA}
\affiliation{Manning College of Information and Computer Sciences, University of Massachusetts Amherst, Amherst, MA 01003, USA}
\email{rshiina@umass.edu}

\author{Kenneth Goodenough}
\affiliation{Manning College of Information and Computer Sciences, University of Massachusetts Amherst, Amherst, MA 01003, USA}
\affiliation{Naturwissenschaftlich-Technische Fakultät, Universität Siegen, 57068 Siegen, Germany}

\author{Nathan Arnold}
\affiliation{Photon Queue Inc., Champaign, IL 61820, USA}

\author{Filip Rozp\k{e}dek}
\affiliation{Department of Physics, University of Massachusetts Amherst, Amherst, MA 01003, USA}
\affiliation{Manning College of Information and Computer Sciences, University of Massachusetts Amherst, Amherst, MA 01003, USA}

\begin{document}

\begin{abstract}
Quantum communication between two distant parties will serve as a cornerstone of the future quantum internet. However, many existing quantum communication schemes face a critical bottleneck: the need to generate a sufficient number of entangled Bell pairs over long distances. Although photons are ideal carriers of quantum information, overcoming photon loss and the exponential attenuation of signals remains a major challenge. We propose an all-photonic quantum repeater architecture that enables quantum communication over a distance of 1,000 km with an equidistant repeater spacing of 9 km. This repeater spacing is enabled by elementary entangled Bell pairs protected through the concatenation of continuous-variable and discrete-variable quantum error correction codes---namely, the bosonic Gottesman--Kitaev--Preskill (GKP) code and the $[[7,1,3]]$ Steane code---whose combination yields a synergistic improvement in robustness against photon loss. This architecture incorporates a new ranking criterion and a multi-reflection mirror-based optical cavity as a free-space photonic memory module, which we model in terms of its length and mirror-reflection efficiency. Additionally, we propose two heuristic construction methods for the elementary entangled Bell pairs. One method introduces up to two-qubit correlated errors within each logical qubit but requires a large number of GKP qubits, while the other allows up to three-qubit correlated errors within each logical qubit but requires fewer GKP qubits. To more accurately capture realistic physical conditions during photonic resource preparation, we include switching-induced imperfections in our simulations, in addition to other standard optical imperfections. In the presence of these imperfections, our realization requires only a few thousand GKP qubits per repeater station per protocol run---a resource requirement significantly smaller than the corresponding resource requirements of prior third-generation all-photonic repeater proposals.

\end{abstract}

\maketitle

\section{Introduction}
\label{sec:introduction}
The future quantum internet will critically depend on quantum communication protocols~\cite{azuma2023quantum} such as quantum teleportation and quantum key distribution~\cite{bennett2014quantum}. To enable such a network, quantum repeaters must play a central role~\cite{munro2015inside}, as they circumvent the need for signal amplification---which is forbidden by the no-cloning theorem---and mitigate photon loss during transmission between adjacent repeaters.

Recently, it has been found that the costly quantum memories once considered essential at each repeater can be eliminated by preparing graph states in advance~\cite{azuma2015all}. This all-photonic quantum repeater approach avoids the need for quantum memories such as 
atomic ensembles~\cite{davidson2023single}, 
neutral atoms~\cite{zhou2024long},
rare-earth-doped crystals~\cite{liu2022demand},
nitrogen-vacancy (NV) centers~\cite{fuchs2011quantum},
silicon-vacancy (SiV) centers~\cite{knaut2024entanglement},
trapped ions~\cite{drmota2023robust}, 
and superconducting cavity-based memories~\cite{milul2023superconducting}---technologies that are typically challenging to operate and scale while maintaining high precision. Instead, this all-photonic repeater scheme effectively uses optical fibers themselves as a form of quantum memory, allowing many photonic qubits to be stored without requiring a separate quantum memory for each qubit. Furthermore, by performing all entanglement-swapping operations simultaneously, the overall communication rate can approach the local repetition rate determined by the slowest among the single-photon sources, photon detectors, optical switches, and active feedforward, effectively removing the bottleneck associated with heralding delays.

Subsequently, an all-photonic quantum repeater design combining Gottesman--Kitaev--Preskill encoded qubits (GKP qubits)---replacing single photons---with tree encoding has emerged~\cite{fukui2021all}. In other words, GKP qubits constitute the repeater graph states (RGSs) in this scheme. Here, RGSs refer to highly entangled photonic states that serve as resources for all-photonic quantum repeaters, enabling multiplexed entanglement-generation attempts. Although schemes based solely on bare GKP qubits have also been considered~\cite{fukui2021all}, they typically require very dense repeater spacing (approximately $1.5~\mathrm{km}$) to operate effectively. In the GKP error correction code~\cite{gottesman2001encoding}, quantum information is encoded into a bosonic harmonic oscillator mode, and the corresponding continuous-variable states can be effectively represented in phase space. GKP qubits are well known for their robustness against photon loss and their ability to correct shift errors induced by such loss. Moreover, GKP qubits enable entanglement-swapping measurements to be performed deterministically. Both theoretical and experimental research toward realizing photonic GKP qubits is actively underway~\cite{takase2023gottesman, larsen2025integrated, aghaee2025scaling, konno2024logical, yu2026extensible}.

However, in most all-photonic quantum repeater designs based on RGSs with tree encoding~\cite{azuma2015all, fukui2021all, pant2017rate}, at most one end-to-end entangled Bell pair can be extracted per protocol run---even at large multiplexing levels. RGSs that allow the extraction of more than one Bell pair per run have only recently been proposed and require further investigation in terms of resource requirements and practical feasibility~\cite{li2025generalized}. In response, a new all-photonic repeater design has been proposed based on the $[[7,1,3]]$ Steane error correction code using GKP qubits, without relying on tree encoding~\cite{rozpkedek2023all}. This scheme employs an elementary entangled Bell pair composed of eight qubits: seven GKP qubits protected by the Steane code and one bare GKP qubit. During the protocol, the seven protected GKP qubits are stored in a fiber spool inside the repeater, while the bare qubit is transmitted through an optical fiber to a minor node located midway between adjacent repeaters, where it is consumed for the first-stage entanglement swapping. To mitigate errors accumulated during propagation in the fiber spool, teleportation-based error correction (TEC) is applied at short spatial intervals, requiring a substantial number of additional GKP qubits. By employing multiplexing, in which each repeater generates multiple elementary entangled Bell pairs, multiple end-to-end entangled Bell pairs can be established within a single protocol run. In this scheme, the outcomes of the first-stage entanglement swapping---involving bare GKP qubits---are leveraged to optimize the multiplexing in the subsequent second-stage entanglement swapping. 

In a recent development along a related but distinct direction, H\"{a}ussler and van Loock proposed a second-generation memory-based all-photonic repeater architecture using elementary entangled Bell pairs composed of a single photon and a logical qubit protected by one of several candidate quantum error correction codes, together with TECs~\cite{haussler2025long}. A notable advantage of their architecture is that it can operate with substantially longer repeater spacings while transmitting only single photons through the optical fibers. At the same time, in the standard classification of quantum repeaters, second-generation architectures rely on quantum memories to store successfully generated entangled links over multiple protocol runs until neighboring links are also established. Consistent with this classification, in the scheme of H\"{a}ussler and van Loock, this memory-based feature arises because the first-stage entanglement swapping at minor nodes is probabilistic. This distinguishes their second-generation architecture from third-generation architectures~\cite{azuma2015all, fukui2021all, pant2017rate, rozpkedek2023all}, which are designed to generate end-to-end entangled Bell pairs within a single protocol run. We provide a more detailed comparison between our proposed scheme and their scheme in Subsection~\ref{subsec:Loock}.

Nevertheless, Rozp\k{e}dek \textit{et al.}'s scheme~\cite{rozpkedek2023all}, along with all previous third-generation all-photonic proposals~\cite{azuma2015all, fukui2021all, pant2017rate}, still suffers from the constraint of small repeater spacing, whereas memory-based repeaters can typically operate with a repeater spacing of tens of kilometers~\cite{jiang2007optimal}. In addition, Rozp\k{e}dek \textit{et al.}'s scheme suffers from a substantial GKP qubit resource overhead due to the reliance on TEC for storage in fiber spools. Here, we propose an architecture that employs elementary entangled Bell pairs composed of two logically protected GKP qubits---each encoded using the $[[7,1,3]]$ Steane code---resulting in a total of 14 GKP qubits per pair. This extends Rozp\k{e}dek \textit{et al.}'s scheme, in which only one qubit was encoded while the other remained bare. In our approach, the GKP code is further concatenated with the $[[7,1,3]]$ Steane code for both qubits involved in the first- and second-stage entanglement swapping, thereby enhancing protection against both small continuous-variable shift errors and larger shifts that manifest as discrete-variable errors. While the Steane code alone can detect up to two-qubit errors and correct one-qubit errors, combining it with the GKP code enables probabilistic correction of up to three-qubit errors. Furthermore, our architecture employs a multi-reflection mirror-based optical cavity as a free-space photonic memory module to store the seven protected GKP qubits inside each repeater. TEC for storage in fiber spools, along with the associated GKP qubits, is no longer required. We also extend the multiplexing strategy from~\cite{rozpkedek2023all} that leverages the outcomes of the first-stage entanglement swapping---now involving logical qubits encoded in seven physical GKP qubits---to maximize the efficiency of the subsequent inner-leaves entanglement swapping. 

We show that our architecture enables quantum communication over a distance of 1,000 km, with an equidistant repeater spacing of 9 km and a multiplexing level of 15, where the multiplexing level refers to the number of end-to-end entangled Bell pairs successfully established per protocol run. This repeater spacing is significantly longer than those of other third-generation all-photonic quantum repeater schemes. It is approximately 1.8--3.6 times longer than that of Rozp\k{e}dek \textit{et al.}'s scheme~\cite{rozpkedek2023all} and 4.5 times longer than that of Fukui \textit{et al.}'s scheme~\cite{fukui2021all}, both of which are based on GKP qubits. It is also approximately 6.0 times longer than that of Pant \textit{et al.}'s scheme~\cite{pant2017rate} and 2.3 times longer than that of Azuma \textit{et al.}'s scheme~\cite{azuma2015all}, both of which are based on single photons rather than GKP qubits.

Moreover, we show that only 3,800 GKP qubits need to be available at each repeater station at the beginning of each protocol run to realize our architecture with 9 km equidistant repeater spacing and a multiplexing level of 15. In the construction of elementary entangled Bell pairs, we introduce two novel heuristic methods: one that allows up to weight-2 errors but requires a larger number of GKP qubits, and another that tolerates weight-3 errors while significantly reducing the qubit requirements. To estimate this initial GKP qubit requirement, we interpret the state transitions as a Markov chain, enabling us to determine the resource requirements both efficiently and accurately, without relying on Monte Carlo simulations as in previous works~\cite{rozpkedek2023all, pant2017rate}. In particular, according to previous studies, Rozp\k{e}dek \textit{et al.}'s scheme requires approximately $10^3$--$10^4$ GKP qubits per repeater to achieve a communication distance of over 1,000 km~\cite{rozpkedek2023all}. The reliance on TEC for storage in fiber spools adds an additional $\sim 10^{4}$ GKP qubits. Pant \textit{et al.}'s scheme requires approximately $3.3 \times 10^6$ single photons per repeater~\cite{pant2017rate}, while Azuma \textit{et al.}'s scheme demands approximately $10^{24}$ single photons per repeater, as estimated by Pant \textit{et al.}~\cite{pant2017rate}, both accounting for photons wasted during failed cluster-state preparation attempts. Without accounting for photons wasted during failed cluster-state preparation attempts and consumed during construction in Azuma \textit{et al.}'s scheme, the required number decreases to 16,400 single photons per repeater~\cite{azuma2015all}. Under the same assumption, our scheme requires only 420 GKP qubits. All of these comparisons are made with previously proposed all-photonic quantum repeater schemes, in which photonic graph states are constructed by sending individual photonic qubits through optical fusion circuits, rather than by using emitters that directly generate entangled cluster states.

In conclusion, these results demonstrate that our architecture achieves both the longest repeater spacing and the lowest initial photonic-qubit requirement among proposed third-generation all-photonic quantum repeater schemes, offering a significant advantage toward the realization of a future quantum internet. We also expect that our architecture is naturally compatible with all-photonic quantum switches for generating Greenberger--Horne--Zeilinger (GHZ) states among multiple end users.

Parts of this work were previously presented in a shorter conference version~\cite{shiina2026extending}.

The paper is structured as follows. Section~\ref{sec:code} introduces the GKP code and the $[[7,1,3]]$ Steane code. Section~\ref{sec:scheme} presents the overall architecture of our all-photonic quantum repeater along with its core design principles. Section~\ref{sec:construction} describes two heuristic construction methods for the elementary entangled Bell pairs: one that introduces at most weight-2 errors, and another that allows at most weight-3 errors while significantly reducing the number of required GKP qubits. Section~\ref{sec:outer} details the outer-leaves swapping process, while Section~\ref{sec:inner} focuses on the inner-leaves swapping process. Section~\ref{sec:e2e} derives an analytical expression for the end-to-end error probability and introduces key performance metrics. Section~\ref{sec:simulation} details the setup of our numerical simulations and specifies the parameter values used. Section~\ref{sec:results} presents the performance of our architecture. Section~\ref{sec:discussion} compares our architecture with three other classes of all-photonic quantum repeater schemes and examines the sensitivity of the performance to key simulation parameters. Lastly, Section~\ref{sec:last} summarizes our work and suggests directions for future research.

\section{GKP Code and \texorpdfstring{$[[7,1,3]]$}{[[7,1,3]]} Steane Code}
\label{sec:code}
\subsection{GKP Error Correction Code}
\label{subsec:GKP}

In our architecture, we use GKP qubits as the fundamental building blocks, each protected by the GKP error correction code. Throughout this paper, every vertex of a graph state---a type of multipartite entangled state defined by a graph, where vertices represent qubits and edges represent entanglement---corresponds to a GKP qubit. GKP qubits are known for their robustness against photon loss, as quantum information is encoded in the quantum state of a bosonic harmonic oscillator, which is composed of a superposition of many photon-number states.

In the single-mode GKP code based on a square lattice, we use the simultaneous $+1$ eigenspace of the following two stabilizer operators as the logical computational subspace:
\begin{equation}
\begin{split}
\hat{S}_q = e^{i2\sqrt{\pi}\hat{q}},\;\;\; \hat{S}_p = e^{-i2\sqrt{\pi}\hat{p}}.
\end{split}
\end{equation}
The simultaneous eigenstates of these stabilizers in the $q$-basis are partitioned into two disjoint subsets, which serve as the logical $\ket{0}$ and $\ket{1}$ states, respectively:
\begin{equation}
\begin{split}
\ket{0_{\mathrm{GKP}}} &= \sum_{n\in \mathbb{Z}} \ket{q = 2n\sqrt{\pi}}, \\
\ket{1_{\mathrm{GKP}}} &= \sum_{n\in \mathbb{Z}} \ket{q = (2n + 1)\sqrt{\pi}}.
\end{split}
\end{equation}
Similarly, in the $p$-basis, the simultaneous eigenstates are divided into two disjoint subsets that define the logical $\ket{+}$ and $\ket{-}$ states, respectively:
\begin{equation}
\begin{split}
\ket{+_{\mathrm{GKP}}} &= \sum_{n\in \mathbb{Z}} \ket{p = 2n\sqrt{\pi}}, \\
\ket{-_{\mathrm{GKP}}} &= \sum_{n\in \mathbb{Z}} \ket{p = (2n + 1)\sqrt{\pi}}.
\end{split}
\end{equation}

It is physically impossible to realize such ideal GKP states with an infinitely sharp Dirac comb structure, as represented in Eqs.~(2), since this would require infinite squeezing. To model more realistic GKP states, we apply a Gaussian envelope operator $\exp[-\Delta \hat{n}]$ to the ideal states, where $\hat{n} = \hat{a}^\dagger \hat{a}$ is the photon number operator and $\Delta$ characterizes the width of each peak in the comb structure~\cite{noh2020fault}. To further account for imperfections due to finite squeezing, we adopt the twirling approximation, in which an ideal GKP state is assumed to pass through a Gaussian random displacement channel with standard deviation $\sigma_{\mathrm{GKP}}$~\cite{conrad2021twirling, fukui2018high}. In our model, in addition to finite GKP squeezing, photon loss in the fiber and other optical components is the dominant source of effective displacement noise. The amount of squeezing, denoted by $s$, is related to the standard deviation $\sigma_{\mathrm{GKP}}$, which characterizes the spread of the Dirac comb wave function around the ideal peak positions. This relationship is given by
\begin{equation}
\begin{split}
s = -10\log_{10} (2\sigma^2_{\mathrm{GKP}}).
\end{split}
\end{equation}
Throughout this paper, we fix the standard deviation to $\sigma_{\rm{GKP}} = 0.12$, which corresponds to a squeezing level of $s = 15.4\,\mathrm{dB}$.

The stabilizer measurements associated with the two operators in Eq.~(1) allow us to measure displacements in the $q$ and $p$ quadratures modulo $\sqrt{\pi}$. Therefore, shift errors with magnitudes less than $\sqrt{\pi}/2$, half the resolution of the stabilizers, can be properly corrected after such stabilizer measurements by interpreting the outcomes as originating from the nearest peak. On the other hand, shift errors with magnitudes between $\sqrt{\pi}/2$ and $3\sqrt{\pi}/2$ lead to logical bit-flip errors, as they are misidentified as being closer to the wrong peak. During the construction of elementary entangled Bell pairs, we employ discard windows centered at $\pm \sqrt{\pi}/2$ to mitigate such errors, as discussed further in Subsection~\ref{subsec:post}.

\subsection{\texorpdfstring{$[[7, 1, 3]]$}{[[7, 1, 3]]} Steane Error Correction Code}
In our architecture, we employ the $[[7,1,3]]$ Steane code in both the inner-leaves and outer-leaves swapping processes, as discussed in Sections~\ref{sec:outer} and~\ref{sec:inner}. The $[[7,1,3]]$ Steane code is a well-established quantum error-correcting code that encodes one logical qubit into seven physical qubits~\cite{steane1996error}. The $X$-type and $Z$-type stabilizer generators can be written in the following matrix form, respectively:
{
\setlength{\arraycolsep}{1.8pt}
\begin{equation}
\begin{pmatrix}
\opI & \opI & \opI & \opX & \opX & \opX & \opX\\
\opI & \opX & \opX & \opI & \opI & \opX & \opX\\
\opX & \opI & \opX & \opI & \opX & \opI & \opX
\end{pmatrix},
\;
\begin{pmatrix}
\opI & \opI & \opI & \opZ & \opZ & \opZ & \opZ\\
\opI & \opZ & \opZ & \opI & \opI & \opZ & \opZ\\
\opZ & \opI & \opZ & \opI & \opZ & \opI & \opZ
\end{pmatrix}.
\end{equation}
}
\noindent This code alone can detect up to two qubit errors and correct any single-qubit error. Here, $I$ denotes the identity operator.

\begin{figure*}[t]
  \centering
  \includegraphics[width=\textwidth]{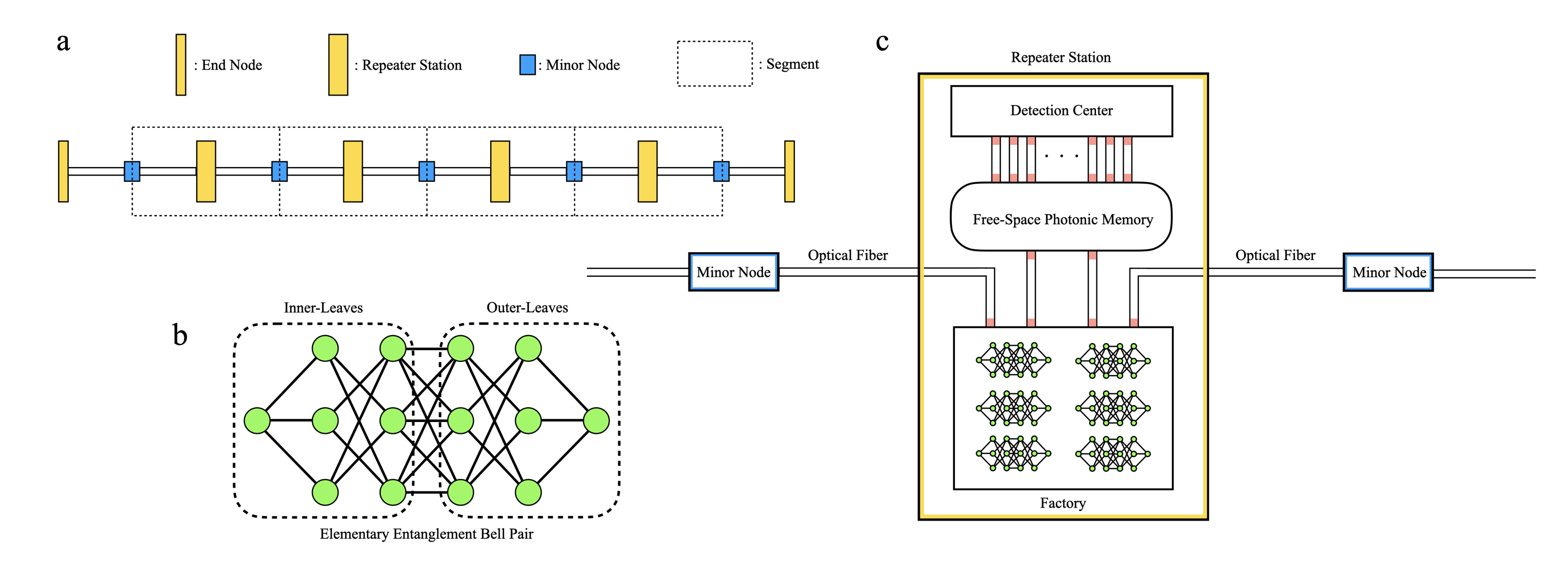}
    \caption{
    Overview of the proposed all-photonic quantum repeater architecture. (a) Schematic of the full communication chain. Repeater stations (yellow) are placed at equal intervals between two end users (also yellow, drawn at half the width of a repeater station), with minor nodes (blue) positioned at the midpoints between each pair of repeater stations. Each segment (dashed box) contains one repeater station and two halves of minor nodes. (b) Graph structure of an elementary entangled Bell pair. The graph represents the minimum-edge graph state that implements the $[[7,1,3]]$ Steane code. Each state is partitioned into two symmetric components, the inner-leaves and the outer-leaves. (c) Internal configuration of a repeater station, consisting of a factory module that produces a sufficient number of elementary entangled Bell pairs, a free-space photonic memory module that serves as a temporary buffer for the inner-leaves, and a detection center module that performs Bell measurements. Optical fibers are used to transmit the outer-leaves to adjacent minor nodes, while the outcomes of the outer-leaves swapping are returned to the original repeater station as analog information, mainly through a separate, non-blocking classical communication channel; in principle, the same optical fiber path could also be used. The light-red square boxes represent the connectors between the optical fibers and the quantum chips or the photonic memories.
}
  \label{fig:1}
\end{figure*}

The logical codewords of the $[[7,1,3]]$ Steane code are given by~\cite{nielsen2010quantum}:
\begin{equation}
\begin{split}
\ket{0_L} = \frac{1}{\sqrt{8}} \Bigl[ &\ket{0000000} + \ket{1010101} + \ket{0110011} \\ 
+ &\ket{1100110}  + \ket{0001111} + \ket{1011010} \\ 
+ &\ket{0111100}  + \ket{1101001} \Bigr],
\end{split}
\end{equation}
\begin{equation}
\begin{split}
\ket{1_L} = \frac{1}{\sqrt{8}} \Bigl[ &\ket{1111111} + \ket{0101010} + \ket{1001100} \\ 
+ &\ket{0011001}  + \ket{1110000} + \ket{0100101} \\ 
+ &\ket{1000011}  + \ket{0010110} \Bigr].
\end{split}
\end{equation}

\section{All-Photonic Quantum Repeater Architecture}
\label{sec:scheme}
We propose a new all-photonic quantum repeater architecture for long-distance quantum communication between two end users, Alice and Bob, as illustrated in Fig.~\ref{fig:1}(a)--(c). Rather than performing all entanglement-swapping operations simultaneously, our architecture separates them into first- and second-stage entanglement swapping. Repeater stations are placed at equal intervals along the communication channel. In addition, minor nodes are positioned at the midpoints between each pair of adjacent repeater stations (Fig.~\ref{fig:1}(a)). Each repeater station comprises three core modules: a factory module that generates elementary entangled Bell pairs; a free-space photonic memory module, which serves as a temporary buffer for GKP qubits designated for second-stage entanglement swapping; and a detection center module, which performs continuous-variable (CV) Bell measurements on GKP qubits. In our setting, these measurements are realized by interfering two modes at a 50:50 beam splitter, followed by homodyne detection of the $q$ quadrature on one output and the $p$ quadrature on the other, thereby projecting the incoming modes onto the CV Bell basis. In contrast, each minor node is equipped only with beam splitters and homodyne detectors, making it significantly simpler and more lightweight than the repeater stations. We define a segment as the region consisting of one repeater station and two neighboring half minor nodes---one on each side---whose length is equal to the distance between two adjacent repeater stations.

First, the factories in all repeater stations between Alice and Bob generate a sufficient number of elementary entangled Bell pairs. The corresponding graph structure (Fig.~\ref{fig:1}(b)) represents the minimum-edge graph state that implements two code blocks of the $[[7,1,3]]$ Steane code. This property was verified using the analysis tools from~\cite{sharma2025minimizing}. Since each edge corresponds to a CZ gate, minimizing the number of edges directly reduces the physical resources required. Furthermore, each graph state is naturally partitioned into two symmetric components, referred to as the \textit{inner-leaves} and the \textit{outer-leaves}. Each of these components consists of seven physical qubits protected by the $[[7,1,3]]$ Steane code.

Next, the outer-leaves are transmitted to adjacent minor nodes through the connecting optical fibers, where the first-stage entanglement swapping---referred to as \textit{outer-leaves swapping}---is performed, as explained in more detail in Section~\ref{sec:outer}. The outcomes of this procedure are subsequently employed during the second-stage entanglement swapping. As a result of outer-leaves swapping, neighboring repeater stations share entangled links. Meanwhile, the inner-leaves are transmitted to the free-space photonic memory, where they are temporarily stored until the outcomes of the outer-leaves swapping, conveyed as analog information, are returned to the repeater station. In this work, we mainly consider the case in which this analog information is returned through a separate, non-blocking classical communication channel. This avoids interference between the outgoing outer-leaves and the returning analog information and enables pipelined operation, thereby improving the rate per unit time. In principle, the analog information could instead follow the same optical fiber path as the outer-leaves, so that a single optical fiber may suffice. In this case, however, the rate per unit time is reduced because the return of the analog information can block the transmission of the next set of outer-leaves. A more detailed discussion of the rate per unit time for these two cases is provided in Subsection~\ref{subsec:Loock}. The internal configuration of a repeater station is summarized in Fig.~\ref{fig:1}(c).

Finally, once the analog information has returned to the original repeater station, it is used to select the optimal pairs of inner-leaves for the second-stage entanglement swapping---referred to as \textit{inner-leaves swapping}---in order to maximize the total secret-key (or entanglement) rate per protocol run, as explained in more detail in Section~\ref{sec:inner}. The selected inner-leaves pairs are subsequently transferred from the free-space photonic memory to the detection center, where CV Bell measurements are performed to complete the inner-leaves swapping.

\begin{figure}[ht]
  \centering
  \includegraphics[width=1\linewidth]{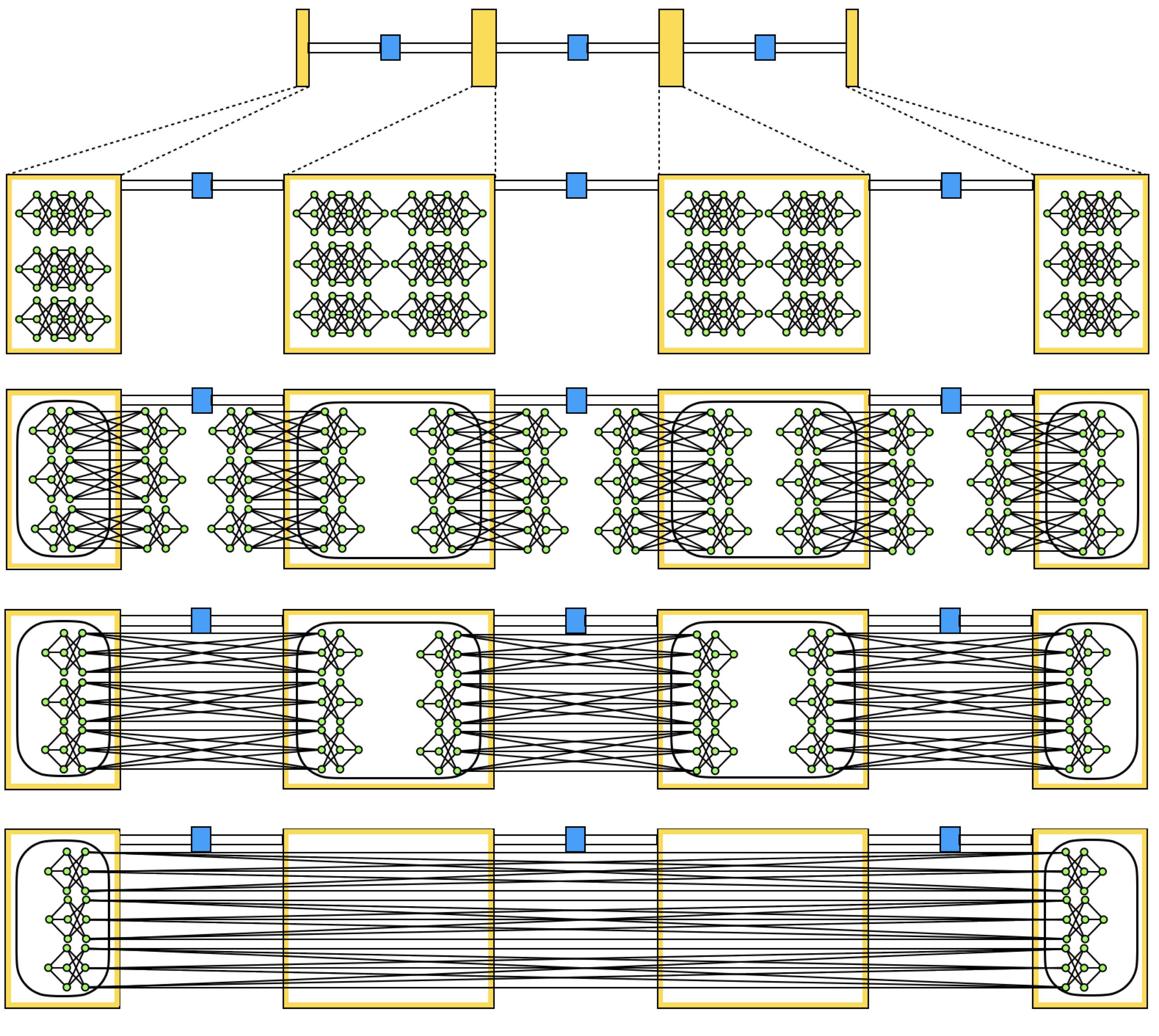}
  \caption{
    Illustration of the full protocol procedure for end-to-end entanglement generation. The top layer depicts the first step, where each factory within the repeater stations (yellow boxes) generates a sufficient number of elementary entangled Bell pairs. In the second layer, the outer-leaves are transmitted to adjacent minor nodes (blue squares) via optical fibers. At each minor node, outer-leaves swapping is performed, and the resulting network of entangled links between neighboring repeater stations is shown in the third layer. The analog information from these measurements is sent back to the originating repeater stations, while the inner-leaves remain buffered in photonic memories. In the bottom layers, the optimal pairs of inner-leaves are selected based on the received analog information and subsequently processed via inner-leaves swapping, thereby establishing long-distance entanglement between the end users. This illustration shows the case of multiplexing level $k=3$.
}
  \label{fig:2}
\end{figure}

In summary, our architecture consists of three main procedures: (1) the construction of elementary entangled Bell pairs, (2) outer-leaves swapping, and (3) inner-leaves swapping, as illustrated in Fig.~\ref{fig:2}.

\subsection{Multiplexing Capability}
\label{subsec:multiplexing}

Our architecture offers a significant advantage in that it enables multiplexing in a straightforward manner. This allows multiple end-to-end (e2e) entangled Bell pairs to be generated within a single protocol run. Specifically, to obtain $k$ e2e entangled Bell pairs (i.e., at a multiplexing level of $k$), each repeater station must generate $2k$ elementary entangled Bell pairs---sending half to the right adjacent minor node and the other half to the left. Fig.~\ref{fig:2} illustrates the case of multiplexing level $k=3$.

In Section~\ref{sec:results}, we show that the total secret-key (or entanglement) rate per protocol run increases with the multiplexing level $k$, but so does the number of required GKP qubits. In that section, from the perspective of our defined cost metric, we investigate the optimal multiplexing level, denoted $k_{\mathrm{opt}}$.

\subsection{Comparison with Related All-Photonic Quantum Repeater Schemes}
\label{subsec:comparison}

Our architecture is a two-way all-photonic quantum repeater scheme~\cite{azuma2015all} and is also classified as a third-generation quantum repeater scheme~\cite{muralidharan2016optimal}. 

We can also simplify our scheme by removing any form of memory, such as a free-space photonic memory or a fiber spool, partitioning each elementary entangled Bell pair into two outer-leaves, and performing all entanglement swapping simultaneously, as in Azuma \textit{et al.}'s scheme~\cite{azuma2015all}. We compare the performance of our architecture with that of this memory-free scenario in Subsection~\ref{subsec:memoryless}.

We also compare our architecture with a scheme that employs elementary entangled Bell pairs composed of one logically protected inner-leave and one bare outer-leaf. More specifically, we examine two variations of this scheme in Subsection~\ref{subsec:LP}: one using a free-space photonic memory as a temporary buffer, and the other using an optical fiber, which corresponds to the scenario proposed by Rozp\k{e}dek \textit{et al.}~\cite{rozpkedek2023all}.

Finally, we compare our architecture with the second-generation memory-based all-photonic repeater architecture~\cite{haussler2025long} in terms of the rate per unit time in Subsection~\ref{subsec:Loock}. Since the first-stage entanglement swapping in their architecture is probabilistic, multiple protocol runs are generally required to generate a single end-to-end entangled Bell pair, whereas our architecture is designed to generate end-to-end entangled Bell pairs within a single protocol run. We also discuss how the rate per unit time depends on whether the analog information is returned through a separate classical communication channel or through the same optical fiber path as the outer-leaves.

\subsection{Free-Space Photonic Memory}
\label{subsec:Mirror Room}

The use of fiber spools as photonic memories has been considered in previous schemes~\cite{rozpkedek2023all, pant2017rate, haussler2025long}. However, fiber spools suffer from significant loss, which either requires the implementation of local teleportation-based error correction approximately every 250 m~\cite{rozpkedek2023all}, leading to substantial GKP qubit resource overhead, or necessitates very dense repeater spacing~\cite{pant2017rate}, both of which are undesirable in practice. In our setting, each qubit that constitutes the inner-leaves waits for the outcome of the corresponding outer-leaves swapping by bouncing off a mirror once every two meters inside a structure referred to as the \textit{free-space photonic memory}. The term “photonic memory” is used broadly to denote any device capable of temporarily storing photons, such as fiber spools or cavity-based systems, where quantum information remains encoded in photonic modes without being transferred to matter systems. In this work, we employ the term “free-space photonic memory” to describe any system that realizes this scenario of repeated reflections at meter-scale intervals. Physically, the free-space photonic memory can be regarded as a type of multi-reflection mirror-based optical cavity; however, its physical scale is significantly larger than that of conventional optical cavities typically used in laboratory settings.

We consider multi-pass reflection cells as promising candidates for realizing such a free-space photonic memory. There are many reflection cell architectures, but one of the most promising candidates is the Robert cell~\cite{robert2007simple}, which is a modified version of the Herriott cell~\cite{herriottFoldedOpticalDelay1965}. The Robert cell is designed to reflect light multiple times between three high-reflectivity mirrors (reflectivity $\geq 99.95\%$), with reflection spots filling up nearly the entire surface area of the mirrors. Thus, compared to alternative cell designs, the Robert cell can create a longer effective optical path within a smaller footprint. The Robert cell has already been used in free-space photonic memory architectures~\cite{arnold2023free}. In this implementation, the total optical path length reaches approximately 340 times the physical length of the cavity, largely limited by the surface area of the mirrors. Furthermore, the system operates in free space at room temperature. The high-reflectivity mirror coatings operate well over a wavelength range of tens of nanometers, making the setup compatible with large-bandwidth, short-pulse quantum states such as GKP qubits.

Expanding upon the prior demonstration of a Robert cell with 340 reflections, a 3-inch-diameter mirror, a cell length of 1.1 m, and a total delay of 1.25~$\mu$s~\cite{arnold2023free}, a suitable configuration can be designed to meet the storage requirements of the repeater spacing considered in this work. The required storage time corresponds to approximately 45 $\mu$s. Scaling to a 12-inch (300 mm) mirror increases the available mirror surface area by a factor of 16, providing a lower bound of $16 \times 340 = 5{,}440$ reflections under a comparable spot pattern. Robert cells support a range of discrete configurations with varying spot densities; more densely packed configurations can exceed this lower bound while still avoiding spot overlap. A configuration with a physical length of 2.033 m, a mirror diameter of 300 mm, and a mirror radius of curvature of 5 m supports 6{,}678 reflections, providing 13.8 km of optical path length and a storage time of 46 $\mu$s. Using the mirror reflectivity discussed in the next paragraph, this yields a total transmission of $T = 0.9999918^{6678} \approx 94.7\%$. % This configuration is consistent with the $L_\text{cavity} = 2$ m and $\eta_m = 0.999995$ parameters used in simulation (Table~\ref{tab:params}).

We also consider the high-reflectivity end mirrors (ETMs) used in Advanced LIGO as another candidate for the free-space photonic memory. The demonstrated performance of the ETMs indicates that the efficiency of mirror reflection assumed in our simulations is realistic with near-term technology. For the Advanced LIGO ETMs at $1064 \, \mathrm{nm}$, the measured losses are $T \approx 3 \, \mathrm{ppm}$, $S \approx 4.9  \, \pm \, 1.5 \, \mathrm{ppm}$, and $A \approx 0.27 \, \pm \, 0.07 \, \mathrm{ppm}$, where $T$, $S$, and $A$ denote the power transmission, scattering, and absorption, respectively. These values imply a reflectivity of $R = 1 - (T + S + A) \approx 0.9999918$ ($99.99918\%$)~\cite{pinard2016mirrors}. Comparable reflectivities are also commercially available~\cite{FiveNineSpecs}. In each arm of Advanced LIGO, the separation between the input and end mirrors is approximately $4 \, \mathrm{km}$ (design length $3994.5 \, \mathrm{m}$)~\cite{aasi2015advanced}. These optics formed part of the instrument that enabled the first direct observation of gravitational waves~\cite{abbott2016observation}.

\subsection{Amplification}
\label{subsec:amplification}

Although noiseless amplification of an unknown quantum state is forbidden by the no-cloning theorem~\cite{wootters1982single}, noisy amplification of the quadrature variables is physically allowed. For GKP qubits, such noisy amplification can be exploited as a loss-compensation technique to enhance communication performance. In particular, three amplification techniques are commonly employed: postamplification, preamplification, and classical computation (CC) amplification~\cite{fukui2021all}.

Postamplification and preamplification rely on physical phase-insensitive amplifiers and therefore inevitably introduce Gaussian noise. For a given loss-compensation task, postamplification generally leads to larger effective noise than preamplification. In contrast, CC amplification implements the effect of phase-sensitive amplification through classical postprocessing by rescaling the measurement outcomes, effectively restoring one quadrature at the expense of the other. This method minimizes the added noise when the two modes involved in the Bell measurement experience symmetric loss, as is often the case in two-way protocols.

In our architecture, CC amplification is employed in both the outer-leaves and inner-leaves swapping procedures. During the construction of elementary entangled Bell pairs, CC amplification is used in nearly all steps. However, in the \textit{refreshment} process---discussed in Section~\ref{sec:construction}---we use preamplification instead, due to the symmetric-loss requirement for CC amplification.

\section{Construction of Elementary Entangled Bell Pairs}
\label{sec:construction}
\begin{figure*}[t]
  \centering
  \includegraphics[width=\textwidth]{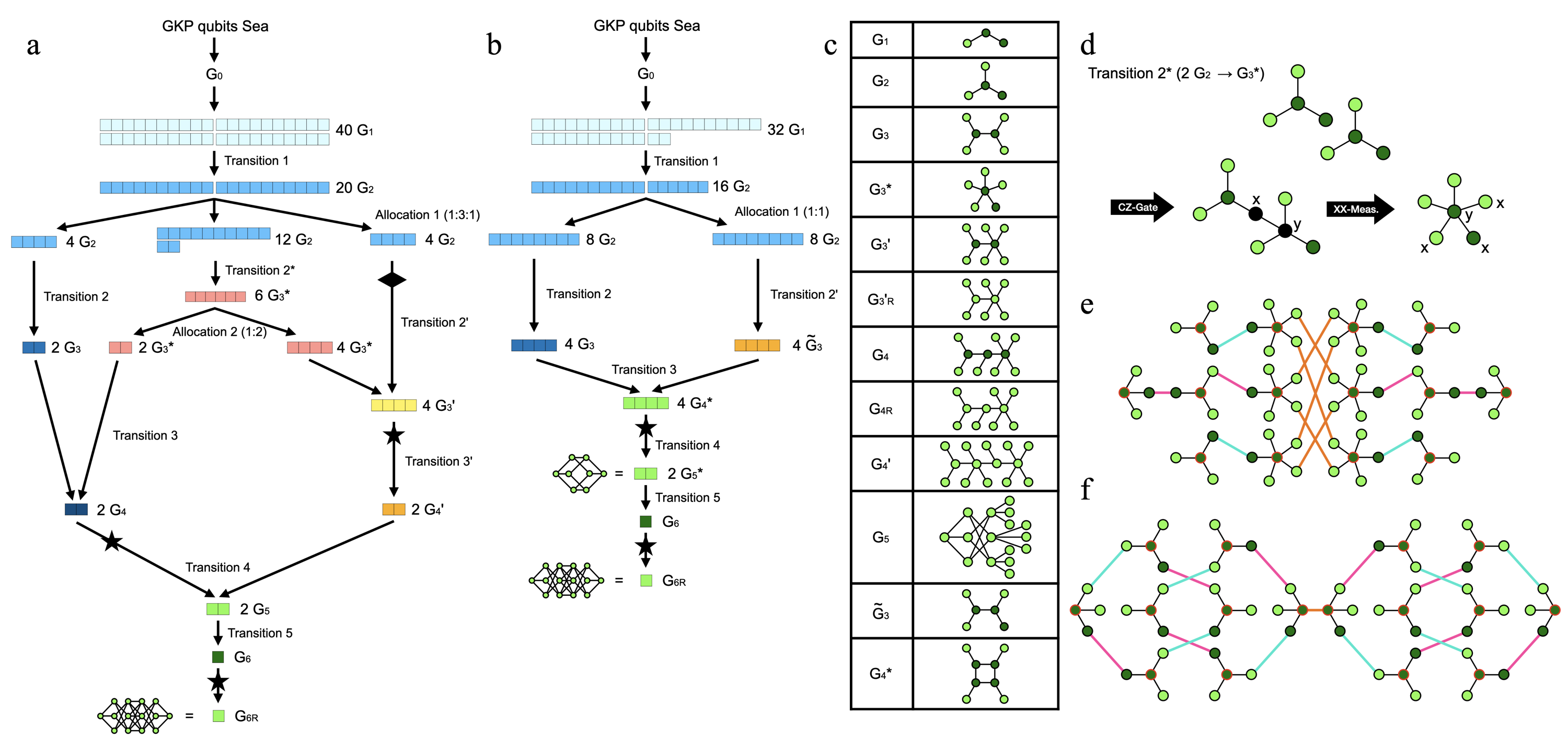}
  \caption{Illustration of the two construction methods for elementary entangled Bell pairs. (a) and (b) show the construction flow of graph states in UW2 and UW3, respectively. The numbers shown for each state indicate the required count of that state in the case where all measurements are assumed to be deterministic, i.e., when no discard windows are used. The filled star denotes the \textit{refreshment} process. The filled rhombus denotes a step in which a graph state passes through an additional optical switch, enabling the implementation of CC amplification by ensuring that both modes involved in the Bell measurement experience symmetric loss. (c) A list of the intermediate states. Light-green filled circles depict qubits with $((\Delta q)^2,\,(\Delta p)^2) = (\sigma_{\rm GKP}^2,\,\sigma_{\rm GKP}^2)$, whereas dark-green filled circles depict qubits with $((\Delta q)^2,\,(\Delta p)^2) = (\sigma_{\rm GKP}^2,\,2\sigma_{\rm GKP}^2)$. (d) Example of how graph states are combined via a CZ gate and an XX measurement, and how correlated errors propagate. Black filled circles depict qubits with $((\Delta q)^2,\,(\Delta p)^2) = (\sigma_{\rm GKP}^2,\,3\sigma_{\rm GKP}^2)$. (e) and (f) show the decomposition of the target elementary entangled Bell pair in UW2 and UW3, respectively. Qubits marked with a red circle denote core qubits, while those without a red circle denote arm qubits. In (e), after this decomposition is prepared, the pink connections form $G_4$ states and the cyan connections correspond to $G_3'$ states. The orange connections indicate the final step in constructing the $G_6$ state. In (f), after preparing this decomposition, the pink connections form $G_3$ and $\tilde{G_3}$ states. Following these, the cyan connections correspond to $G_4^*$ states. The orange connections indicate the final step in constructing the $G_6$ state.
}
  \label{fig:3}
\end{figure*}

In this section, we present two methods for generating elementary entangled Bell pairs within the factory module of a repeater. Before introducing these methods, we first describe the post-selected measurement common to both methods in Subsection~\ref{subsec:post}. Subsection~\ref{subsec:uw2} introduces the \textit{up-to-weight-2 error construction method} (UW2), which restricts correlated errors to weight~2 or less but requires a relatively large number of GKP qubits. Subsection~\ref{subsec:uw3} presents the \textit{up-to-weight-3 error construction method} (UW3), which allows correlated errors up to weight~3 while significantly reducing the number of required GKP qubits. Next, Subsection~\ref{subsec:correlated} investigates the correlated errors responsible for the differences between UW2 and UW3. Finally, Subsection~\ref{subsec:recycling} introduces a new concept, GKP qubit recycling, in which unmeasured GKP qubits are extracted from graph states that would otherwise be discarded and reused as resources for refreshment.

\subsection{Post-Selected Measurement}
\label{subsec:post}

We begin by describing the post-selected measurements employed in both methods. These measurements are used only during the construction of elementary entangled Bell pairs and are not used in the outer-leaves swapping process (Section~\ref{sec:outer}) or the inner-leaves swapping process (Section~\ref{sec:inner}), which will be discussed later. Therefore, once the required elementary entangled Bell pairs have been prepared, no additional post-selection reduces the multiplexing level during either swapping process.

These post-selected measurements employ discard windows that are placed between the regions corrected to logical 0 and 1 in the $q$ quadrature and to logical $+$ and $-$ in the $p$ quadrature during GKP error correction~\cite{fukui2021all}. Specifically, for the Gaussian distribution centered at zero, we employ discard windows centered at $\pm \sqrt{\pi}/2$, corresponding to the nearest boundaries between neighboring logical regions. We do not include contributions from regions associated with more distant periodically repeated boundaries, since the Gaussian tails at those displacements are negligibly small.

Using larger discard windows reduces the bit- and phase-flip error probabilities. However, it also increases the probability that measurement outcomes fall into the discard regions, resulting in greater consumption of GKP qubits. Therefore, the window size must be chosen by balancing the bit- and phase-flip error probabilities against the required number of GKP qubits. The optimal window size was determined in Subsection~\ref{subsec:window}.

\subsection{Up-to-weight-2 Error Construction Method}
\label{subsec:uw2}

The process begins with each repeater preparing the required number of GKP qubits~\cite{gottesman2001encoding}. Each GKP qubit is initialized in the so-called \textit{qunaught} state~\cite{walshe2020continuous}, which can be expressed (up to normalization) as
\begin{equation}
\begin{split}
\ket{\oslash} \propto \sum_{n \in \mathbb{Z}} e^{-in\sqrt{2\pi}\hat{p}} \ket{0}_q 
= \sum_{n \in \mathbb{Z}} e^{in\sqrt{2\pi}\hat{q}} \ket{0}_p.
\end{split}
\end{equation}
Here, we note that this state exhibits a periodicity of $\sqrt{2\pi}$ in both the $q$ and $p$ quadratures, corresponding to the lattice spacing of the GKP grid. For convenience, we collectively refer to the ensemble of such qunaught states within a factory module as the \textit{GKP qubit sea}, from which the qubits required for our methods are drawn. Notably, the required size of the GKP qubit sea depends on the chosen method.

Subsequently, two GKP qubits prepared in the qunaught state interfere at a 50:50 beam splitter to generate a Bell pair~\cite{fukui2024resource}. We then apply a Hadamard gate---which corresponds to a Fourier transform for GKP qubits~\cite{gottesman2001encoding} 
and can be implemented optically by a $\pi/2$ phase delay~\cite{baranes2023free}---to one of the qubits in the Bell pair, thereby converting it into a graph state. The resulting two-qubit graph state is denoted by $G_0$. The GKP qubits constituting the $G_0$ state have identical initial GKP squeezing in both the $q$ and $p$ quadratures, i.e., $((\Delta q)^2, \, (\Delta p)^2) = (\sigma^2_{\rm{GKP}}, \, \sigma^2_{\rm{GKP}})$, where $(\Delta q)^2$ and $(\Delta p)^2$ denote the quadrature variances in the $q$ and $p$ coordinates, respectively. Here, $\sigma_{\rm GKP}$ is the standard deviation determined by the initial squeezing level of the GKP states.

%________________G1_______________
First, we consider constructing the $G_1$ state from a single $G_0$ state and an additional GKP qubit. The $G_1$ state is locally equivalent to a three-qubit GHZ state under local complementation on the central qubit~\cite{hein2004multiparty}. We apply a Control-Z (CZ) gate between one of the two GKP qubits of the $G_0$ state and an additional GKP qubit prepared in the $\ket{+}$ state, thereby creating an edge between them in the graph-state representation. Experimentally, this CZ gate can be deterministically implemented using either inline or offline squeezers together with linear optics~\cite{walshe2025linear, tzitrin2021fault}. In our scheme, however, we apply such a deterministic implementation only for the $G_0 \!\to\! G_1$ transition. This CZ gate transforms the initial quadratures of the two GKP qubits, $(q_1,\,p_1)$ and $(q_2,\,p_2)$, into $(q_1,\,p_1 - q_2)$ and $(q_2,\,p_2 - q_1)$, respectively. Consequently, their GKP squeezing changes from $(\sigma_{\rm GKP}^2,\,\sigma_{\rm GKP}^2)$ to $(\sigma_{\rm GKP}^2,\,2\sigma_{\rm GKP}^2)$ after the gate. In Fig.~\ref{fig:3}, we depict a qubit with $((\Delta q)^2,\,(\Delta p)^2) = (\sigma_{\rm GKP}^2,\,\sigma_{\rm GKP}^2)$ as a light-green filled circle, and a qubit with $((\Delta q)^2,\,(\Delta p)^2) = (\sigma_{\rm GKP}^2,\,2\sigma_{\rm GKP}^2)$ as a dark-green filled circle.

%________________G2_______________
Second, we consider the construction of the $G_2$ state from two $G_1$ states, which we also refer to as the \textit{one-core-three-arms} state. In this notation, the \textit{core qubit} ultimately remains to constitute the elementary entangled Bell pair, whereas the \textit{arm qubits} act as ancillary resources that are consumed during the construction process. We note, however, that in the case of the $G_2$ state, the core qubit may also be consumed as an exception.

From this point on, all subsequent measurements are probabilistic, with their success probabilities depending on the quadrature variances $((\Delta q)^2,\,(\Delta p)^2)$ and the size of the discard windows, as well as on the efficiency of the homodyne detector and on the number of optical switches $x$ through the factor $\eta_s^x$, where $\eta_s$ denotes the efficiency of each optical switch that is applied to the graph states after measurements with discard windows. The optical switch serves to route the graph states into successful ones that proceed to the next transition and unsuccessful ones that are discarded.

The basic idea common to both construction methods is to perform the more difficult measurements (on qubits represented as dark-green filled circles in Fig.~\ref{fig:3}) earlier in the construction process, while leaving the easier measurements (on qubits represented as light-green filled circles in Fig.~\ref{fig:3}) for later transitions. This strategy minimizes potential losses by reducing the cost associated with failures at more advanced transitions, where the cumulative investment up to that point is higher.

Here, we apply a CZ gate between the dark-green arm of a $G_1$ state and the core qubit of another $G_1$ state. 
Experimentally, this CZ gate followed by an XX measurement---i.e., a joint measurement of the $X \otimes X$ operator---can be realized by interfering the qubits on a beam splitter and subsequently performing two homodyne detections, one in the $p$ basis and the other in the $q$ basis. The choice of these measurement bases is motivated by the fact that the beam splitter implements a CNOT rather than a CZ gate. Specifically, by applying the operator $I \otimes H$ to the two qubits before measurement, the quadratures of the two GKP qubits, $(q_1,\,p_1)$ and $(q_2,\,p_2)$, are transformed into $(q_1,\,p_1)$ and $(-p_2,\,q_2)$, respectively. The subsequent CNOT gate further maps them to $(q_1,\,p_1-q_2)$ and $(q_1-p_2,\,q_2)$. Finally, by measuring the first qubit in the $p$ basis and the second qubit in the $q$ basis, the desired CZ operation followed by an XX measurement is implemented.

Although this operation is not a bare CZ gate, its effect on the quadrature variances is equivalent: they change from $(\sigma_{\rm GKP}^2,\,2\sigma_{\rm GKP}^2)$ to $(\sigma_{\rm GKP}^2,\,3\sigma_{\rm GKP}^2)$, which is depicted as a black filled circle in Fig.~\ref{fig:3}. At this stage, an optical switch is employed to discard states whose measurement outcomes fall within a discard window. In our scheme, the optical switch is always used after a CZ gate followed by an XX measurement, except immediately after a \textit{refreshment} process, which will be discussed later.

%________________G3_______________
Third, we aim to construct the $G_3$, $G_3^*$, and $G_3'$ states, which we also refer to as the \textit{two-core-four-arms state}, the \textit{one-core-five-arms state}, and the \textit{two-core-six-arms state}, respectively. The $G_3$ state is obtained by applying a CZ gate between a dark-green arm of one $G_2$ state and a dark-green arm of another $G_2$ state, followed by an XX measurement. The $G_3^*$ state is constructed by applying a CZ gate between a dark-green arm of one $G_2$ state and the core qubit of another $G_2$ state, again followed by an XX measurement. Finally, the $G_3'$ state is built by applying a CZ gate between two dark-green arms---one from the $G_3^*$ state and the other from the $G_2$ state---and then performing an XX measurement. %More detailed descriptions of all transitions are provided in Appendix~A.

%________________G4_______________
Fourth, we construct the $G_4$ and $G_4'$ states, which we also refer to as the \textit{three-core-seven-arms state} and the \textit{four-core-ten-arms state}, respectively. The $G_4$ state is obtained by applying a CZ gate between a light-green arm of the $G_3$ state and a dark-green arm of the $G_3^*$ state, followed by an XX measurement. To obtain the $G_4'$ state, the $G_3'$ state first undergoes a \textit{refreshment} process, represented as a filled star in Fig.~\ref{fig:3}, which resets the accumulated effect of optical switches and restores the qubits to their initial quadrature variances, i.e., the light-green qubits. This \textit{refreshment} process is a local, teleportation-based error-correction step~\cite{namiki2016role, rozpkedek2021quantum, rozpkedek2023all}. Specifically, for a graph state consisting of $n$ qubits, we prepare $n$ Bell pairs (not graph states, i.e., not $G_0$ states), apply CNOT gates between each qubit of the target graph state and one qubit of a Bell pair, and then perform joint $XZ$ measurements (i.e., measurements of the $X \otimes Z$ operator). However, this process has a relatively low success probability and therefore is not employed frequently. We denote the $G_3'$ state after this \textit{refreshment} process as ${G_3}'_R$, where the subscript $R$ indicates that the state has undergone a refreshment process. This notation will be used consistently for other refreshed states in the subsequent constructions. The $G_4'$ state is then constructed by applying a CZ gate between a light-green arm of one ${G_3}'_R$ state and a light-green arm of another ${G_3}'_R$ state, followed by an XX measurement.

%_________________G5______________
Fifth, we construct the $G_5$ state from the ${G_4}_R$ and $G_4'$ states. The ${G_4}_R$ state is obtained by applying a \textit{refreshment} process to the $G_4$ state. We also refer to the $G_5$ state as the \textit{seven-core-nine-arms state}.

%_________________G6_____________
Finally, we realize the $G_6$ state, which is the elementary entangled Bell pair we aim to generate, by combining two $G_5$ states. By subsequently applying a \textit{refreshment} process to the $G_6$ state, we obtain ${G_6}_R$, which constitutes the final stage of UW2.

\subsection{Up-to-weight-3 Error Construction Method}
\label{subsec:uw3}

This method follows the same procedure as UW2 up to the construction of the $G_2$ state. In UW3, the $G_3$, $G_3^*$, and $G_3'$ states are prepared from the $G_2$ states, whereas here we prepare the $G_3$ and $\tilde{G_3}$ states. The $\tilde{G_3}$ state has the same structure as the \textit{two-core-four-arms state}, except that one of the four arms is replaced by a dark-green arm. It is obtained by applying a CZ gate between a dark-green arm of one $G_2$ state and a light-green arm of another $G_2$ state---rather than a dark-green arm as in the construction of the $G_3$ state---followed by an XX measurement. This $\tilde{G_3}$ state is prepared to appropriately position the correlation errors within the $G_5^{*}$ state that will be constructed later. Importantly, the qubits measured for obtaining $G_6$ from the $G_5^{*}$ state are free of correlation errors.

Next, we construct the $G_4^{*}$ state, referred to as the \textit{four-core-four-arms state}, by combining a $G_3$ state with a $\tilde{G_3}$ state. Subsequently, the $G_5^{*}$ state, which exhibits a cubic structure, is obtained by combining two $G_4^{*}$ states after undergoing the \textit{refreshment} process.

Finally, we realize the $G_6$ state, which is the elementary entangled Bell pair we aim to generate, by combining two $G_5^{*}$ states. As in UW2, a subsequent \textit{refreshment} process applied to the $G_6$ state yields ${G_6}_R$, which constitutes the final stage of UW3.

\subsection{Correlated Errors}
\label{subsec:correlated}

In this subsection, we examine the origin of the differences between UW2 and UW3, which arise from differences in the propagation of correlated errors. As an illustrative example, we focus on the construction of the $G_3^*$ state (Fig.~\ref{fig:3}(d)). After applying a CZ gate between a dark-green arm of one $G_2$ state and the core qubit of another $G_2$ state, the two qubits involved in the CZ operation become black qubits with quadrature variances $(\sigma_{\rm GKP}^2,\,3\sigma_{\rm GKP}^2)$.

Next, each qubit is measured in the $X$ basis. We denote by $x$ the logical error originating from the arm qubit measurement, and by $y$ the logical error originating from the core qubit measurement. It should be noted that the error $y$ propagates only to the core of the $G_3^*$ state, whereas the error $x$ propagates to three of the arms of the $G_3^*$ state. In this way, the correlated errors propagate through the system.

To address the suppression of correlated errors, we consider decomposing the target elementary entangled Bell pair, as shown in Figs.~\ref{fig:3}(e) and~\ref{fig:3}(f). Fig.~\ref{fig:3}(e) corresponds to UW2. As in the example of the $G_3^*$ state construction, weight-3 (W3) correlated errors can occur. However, by designing one of the three arms to connect to another logical block, the correlated errors within a single logical block are restricted to weight-2 (W2). Fig.~\ref{fig:3}(f) corresponds to UW3. Up to the $G_5^{*}$ state, only W2 correlated errors are present. However, during the final step of combining the two $G_5^{*}$ states, W3 correlated errors can arise and propagate into each logical block.

\subsection{GKP Qubit Recycling}
\label{subsec:recycling}

We introduce a GKP qubit recycling procedure for graph states rejected after measurements with discard windows. As discussed above, the optical switch routes graph states into successful ones that proceed to the next transition and unsuccessful ones that are normally discarded. Here, instead of discarding the unsuccessful graph state, we consider extracting unmeasured GKP qubits from it and recycling them as resources for the refreshment process.

As an illustrative example, we consider the construction of the $G_3^*$ state shown in Fig.~\ref{fig:3}(d).  Suppose that the measurement outcome associated with the arm qubit, which can induce the logical error denoted by $x$, falls within a discard window. In this case, the resulting $G_3^*$ state is no longer accepted as a reliable graph state for the next transition and is therefore routed away by the optical switch. However, unlike a failed single-photon fusion operation, the GKP measurement itself has still been performed, and an unreliable but physically available $G_3^*$ state remains. We then perform a $Z$-basis measurement on the central GKP qubit of this unreliable $G_3^*$ state. We note that the quadrature variance relevant to this measurement is $(\Delta q)^2=\sigma_{\mathrm{GKP}}^2$, since a $Z$-basis measurement corresponds to a measurement in the $q$ quadrature. This measurement removes the measured qubit from the graph state and isolates the remaining five GKP qubits. After appropriate reinitialization into qunaught states, these isolated GKP qubits can be used to prepare Bell pairs for the refreshment process.

We also introduce a \textit{sequential refreshment strategy}. In this strategy, the measurements involved in the refreshment process are performed sequentially rather than simultaneously. After each successful measurement, the procedure proceeds to the next measurement. If a measurement fails, the Bell pairs that have not yet been used remain available as refreshment resources and can be carried over to subsequent refreshment attempts.

The recycling procedure and the sequential refreshment strategy are expected to supply a significant fraction of the GKP qubits required for refreshment. However, further investigation is needed to quantify this resource reduction. In particular, the construction method may need to be optimized while explicitly taking both the recycling procedure and the sequential refreshment strategy into account.

\section{Outer-Leaves Swapping}
\label{sec:outer}
For a multiplexing level of $k$, each repeater station constructs $2k$ elementary entangled Bell pairs. Among these $2k$ outer-leaves, $k$ outer-leaves are sent to the right adjacent minor node, while the remaining $k$ outer-leaves are sent to the left. Meanwhile, the corresponding $2k$ inner-leaves are stored within the free-space photonic memory. This process is illustrated in the second layer of Fig.~\ref{fig:2}.

Each minor node simultaneously receives $2k$ outer-leaves from both the right and left repeater stations, as it is located precisely midway between the two adjacent repeaters. For clarity, we focus on a single pair of outer-leaves. The minor node then performs a CV Bell measurement, implemented by interfering the two modes at a 50:50 beam splitter, followed by homodyne detection of the $q$ quadrature on one output and the $p$ quadrature on the other. Since each outer-leaves cluster consists of seven physical qubits, we obtain 14 measurement outcomes, denoted $q_i$ and $p_i$ with $i \in \{1,\dots,7\}$.

First, for the GKP error correction, each $q_i$ and $p_i$ is adjusted to the nearest multiple of $\sqrt{\pi}$. Consequently, shift errors with magnitudes smaller than $\sqrt{\pi}/2$ can be properly corrected after this step.

Importantly, the error likelihoods are simultaneously computed from the measured values of $q_i$ and $p_i$. We denote the error likelihoods as $z(x_i)$, where $x_i$ represents either $q_i$ or $p_i$, and the following formula converts $x_i$ into $z(x_i)$:
\begin{equation}
z(x_i) =
\frac{
    \sum\limits_{n \in \mathbb{Z}} \exp\!\left[-\frac{(x_i - (2n+1)\sqrt{\pi})^2}{2\sigma^2}\right]
}{
    \sum\limits_{n \in \mathbb{Z}} \exp\!\left[-\frac{(x_i - n\sqrt{\pi})^2}{2\sigma^2}\right]
},
\label{eq:zxi}
\end{equation}
where $\sigma$ is the standard deviation of the Gaussian random displacement noise applied immediately before the GKP measurement~\cite{noh2020fault}. 

Next, for the $[[7,1,3]]$ Steane error correction, we compute two syndromes: one from the measured values of $q_i$ and the other from the measured values of $p_i$.

For illustration, suppose that the syndrome $(0,1,0)$ is obtained. Using this syndrome together with the parity-check matrix of the $[[7,1,3]]$ Steane code given below:
\begin{equation}
\begin{pmatrix}
0 & 0 & 0 & 1 & 1 & 1 & 1\\
0 & 1 & 1 & 0 & 0 & 1 & 1\\
1 & 0 & 1 & 0 & 1 & 0 & 1
\end{pmatrix},
\end{equation}
we can identify the corresponding error pattern.

Continuing with the same example, when we consider weight-1 (W1) errors, we find that the second column corresponds to the syndrome $(0,1,0)$, and thus identify that the second qubit has an error. Similarly, for weight-2 (W2) errors, we find that the sum of the first and third columns corresponds to the same syndrome $(0,1,0)$, which means that both the first and third qubits have errors. In this way, a single syndrome can correspond to eight different error pattern candidates. For syndromes other than $(0,0,0)$, the candidates consist of one W1 error, three W2 errors, and four weight-3 (W3) errors. For the syndrome $(0,0,0)$, the candidates consist of the no-error case (weight-0, W0) together with seven W3 errors. Therefore, for every syndrome, there are always eight candidate error patterns.

\begin{table}[t]
\centering
\caption{An example of eight error pattern candidates corresponding to the syndrome $(0,1,0)$, categorized into weight-1 (W1), weight-2 (W2), and weight-3 (W3) errors.}
\label{tab:syndrome010}

\footnotesize
\renewcommand{\arraystretch}{1.0}
\setlength{\tabcolsep}{0pt}

\begin{tabular}{c|c}
\toprule
\makebox[1.5cm][c]{Index} & \makebox[5cm][c]{Error Pattern} \\
\midrule
\midrule

\multicolumn{2}{l}{\textbf{Weight-1 Error (W1)}} \\
\midrule
\numcell{1} & \RowStrut \imgcell{FIGURES/FIG501.png} \\
\noalign{\vskip 0.15cm}

\midrule
\multicolumn{2}{l}{\textbf{Weight-2 Error (W2)}} \\
\midrule
\numcell{2} & \RowStrut \imgcell{FIGURES/FIG502.png} \\
\numcell{3} & \RowStrut \imgcell{FIGURES/FIG503.png} \\
\numcell{4} & \RowStrut \imgcell{FIGURES/FIG504.png} \\
\noalign{\vskip 0.15cm}

\midrule
\multicolumn{2}{l}{\textbf{Weight-3 Error (W3)}} \\
\midrule
\numcell{5} & \RowStrut \imgcell{FIGURES/FIG505.png} \\
\numcell{6} & \RowStrut \imgcell{FIGURES/FIG506.png} \\
\numcell{7} & \RowStrut \imgcell{FIGURES/FIG507.png} \\
\numcell{8} & \RowStrut \imgcell{FIGURES/FIG508.png} \\
\noalign{\vskip 0.15cm}

\bottomrule
\end{tabular}
\end{table}

Here, the error likelihoods $z(x_i)$ become useful. Since there are seven measured values for each quadrature, we compute a total error likelihood $\xi^j$ for each candidate index $j \in \{1,\dots,8\}$:
\begin{equation}
  \xi^j(x) = \prod_{m \in \mathcal{E}^j} z(x_m) \;
             \prod_{n \in \mathcal{N}^j} \bigl(1 - z(x_n)\bigr),
\end{equation}
where $x$ represents either $q$ or $p$, and $\mathcal{E}^j$ and $\mathcal{N}^j$ denote, respectively, the sets of qubits in error and no-error for the $j$-th candidate, satisfying $|\mathcal{E}^j| + |\mathcal{N}^j| = 7$.

However, it is not necessary to determine which of the eight candidate error patterns actually occurred. Based on the logical codewords of the $[[7,1,3]]$ Steane code, the same correction operation applies to both W1 errors and all W3 errors for syndromes other than $(0,0,0)$, and to all W3 errors for the syndrome $(0,0,0)$, since these errors are equivalent up to a stabilizer.

Therefore, it is appropriate to compare the sum of $\xi^j$ corresponding to W1 and W3 errors, denoted as $\Xi_{\mathrm{I}}$, with the sum of $\xi^j$ corresponding to W2 errors, denoted as $\Xi_{\mathrm{II}}$, for syndromes other than $(0,0,0)$:
\begin{equation}
\begin{split}
    \Xi_{\mathrm{I}} &= \sum_{j \in \{W1,\,W3\}} \xi^{j}, \\
    \Xi_{\mathrm{II}} &= \sum_{j \in \{W2\}} \xi^{j}.
\end{split}
\end{equation}
Based on this comparison between $\Xi_{\mathrm{I}}$ and $\Xi_{\mathrm{II}}$, we perform the $[[7,1,3]]$ Steane error correction.

For the syndrome $(0,0,0)$, $\Xi_{\mathrm{I}}$ is denoted as the likelihood $\xi^{1}$ corresponding to the no-error case, and $\Xi_{\mathrm{II}}$ as the sum of $\xi^j$ corresponding to all W3 errors:
\begin{equation}
\begin{split}
    \Xi_{\mathrm{I}} &= \xi^{1}, \\
    \Xi_{\mathrm{II}} &= \sum_{j \in \{W3\}} \xi^{j}.
\end{split}
\end{equation}
In the same manner, we perform the $[[7,1,3]]$ Steane error correction based on the comparison between $\Xi_{\mathrm{I}}$ and $\Xi_{\mathrm{II}}$.

\section{Inner-Leaves Swapping}
\label{sec:inner}
While the outer-leaves undergo their swapping procedure---comprising transmission to the minor nodes, the CV Bell measurements performed there, and the return of the outcomes represented as analog information---the inner-leaves remain stored in the free-space photonic memory. When the repeater spacing is set to $L_0$, the one-way travel time to a minor node is $L_0/2c_{\mathrm{fiber}}$, where $c_{\mathrm{fiber}}$ denotes the speed of light in the optical fiber. Since the classical outcomes must also return to the repeater station, an additional $L_0/2c_{\mathrm{fiber}}$ is required, resulting in a total waiting time of $L_0/c_{\mathrm{fiber}}$ for the inner-leaves.

The inner-leaves transferred from the factory to the free-space photonic memory are stored by bouncing every two meters. If the efficiency of a single mirror reflection is denoted by $\eta_m$, then after $t$ reflections the total efficiency is $\eta_m^t$, 
where the number of reflections is estimated as $t = 1000 L_0 c /(2 c_{\mathrm{fiber}})$, with the factor $1000$ converting the repeater spacing $L_0$ from kilometers to meters.

After the analog information from the outer-leaves swapping is returned, we perform inner-leaves swapping using the ranking-based strategy~\cite{rozpkedek2023all}. In this strategy, all the best-ranked entangled links between each pair of adjacent repeater stations are connected to establish an end-to-end entangled Bell pair, all the second-best-ranked entangled links are connected to establish another, and so on for the remaining ranks. This strategy has been shown to achieve the maximum secret-key (or entanglement) rate~\cite{rozpkedek2023all}.

Here, we rank them based on the secret-key fraction, $r$, as discussed in Section~\ref{sec:e2e}. However, in this context, $r$ represents the local-level secret-key fraction, corresponding to the entangled links between neighboring repeater stations, as illustrated in the third layer of Fig.~\ref{fig:2}. To estimate $r$, we introduce a pseudo-error probability defined as
\begin{equation}
    P_{\mathrm{pseudo}} = 1-\frac{\Xi_{\mathrm{selected}}}{\sum\limits_{j}\xi^{j}} 
      = 1-\frac{\Xi_{\mathrm{selected}}}{\Xi_{\mathrm{I}}+\Xi_{\mathrm{II}}},
\end{equation}
where $\Xi_{\mathrm{selected}}$ denotes the $\Xi$ actually applied in the outer-leaves swapping, i.e., $\max\{\Xi_{\mathrm{I}},\,\Xi_{\mathrm{II}}\}$. More specifically, $P_{\mathrm{pseudo}}$ is obtained separately for the $q$ and $p$ quadratures, and the resulting two pseudo-error probabilities are then used as inputs to the calculation of $r$.

Finally, based on the ranking defined above, we send the appropriate pairs of inner-leaves from the free-space photonic memory to the detection center, where, similarly to the outer-leaves swapping, a CV Bell measurement is performed under the protection of both the GKP code and the $[[7,1,3]]$ Steane code.

Following this process, Alice and Bob share $k$ end-to-end entangled Bell pairs. Each Bell pair can be independently used either for secret-key extraction or as an entanglement resource.

\section{End-to-End Error and Performance Metrics}
\label{sec:e2e}
This section formalizes the end-to-end performance of the proposed all-photonic quantum repeater architecture. Thus far, our analysis has focused on a single segment, examining the following components in detail:  (1) the construction of elementary entangled Bell pairs (Section~\ref{sec:construction}), (2) outer-leaves swapping (Section~\ref{sec:outer}), and (3) inner-leaves swapping (Section~\ref{sec:inner}). From this point onward, we extend our analysis to the complete end-to-end connection between two distant parties, Alice and Bob.

When we restrict our attention to errors arising solely from (1) the construction of elementary entangled Bell pairs and (2) outer-leaves swapping within a segment, we denote the corresponding error probability by $P_{\mathrm{outer}}$. This quantity represents the probability that a bit-flip or phase-flip error remains after both GKP error correction and the $[[7,1,3]]$ Steane code. Similarly, when considering only the errors originating from (1) the construction of elementary entangled Bell pairs and (3) inner-leaves swapping within a segment, we denote the corresponding error probability by $P_{\mathrm{inner}}$.

For each quadrature, the probability that an error occurs within a segment, denoted by $P_{\mathrm{segment}}$, is given by
\begin{equation}
P_{\mathrm{segment}} = P_{\mathrm{outer}}(1 - P_{\mathrm{inner}}) + (1 - P_{\mathrm{outer}})P_{\mathrm{inner}}.
\end{equation}

When there are $n$ repeaters between Alice and Bob, the communication channel consists of $n$ full segments, along with two additional half-segments at the end nodes. The end-to-end (e2e) error probability, denoted by $P_{\mathrm{e2e}}$, can then be expressed as
\begin{equation}
\begin{split}
P_{\mathrm{e2e}} =\;& (1 - P_{\mathrm{outer}})\cdot \frac{1 - (1 - 2P_{\mathrm{segment}})^n}{2} \\
&+ P_{\mathrm{outer}}\cdot \frac{1 + (1 - 2P_{\mathrm{segment}})^n}{2},
\end{split}
\end{equation}
where $\{1 - (1 - 2P_{\mathrm{segment}})^n\}/2$ corresponds to the probability that the logical parity across the $n$ segments is flipped (i.e., an odd number of segment-level errors occur), and $\{1 + (1 - 2P_{\mathrm{segment}})^n\}/2$ corresponds to the probability that the parity is preserved (i.e., an even number of errors occur). The two half-segments at the end nodes are jointly treated as a single contribution represented by $P_{\mathrm{outer}}$.

Moreover, when the number of repeaters $n$ is large or when $P_{\mathrm{outer}}$ is sufficiently small, the influence of the half-segments can be neglected. In this case, the end-to-end error probability simplifies to
\begin{equation}
P_{\mathrm{e2e}} = \frac{1 - (1 - 2P_{\mathrm{segment}})^n}{2}.
\end{equation}

Using $P_{\mathrm{e2e},\,q}$ and $P_{\mathrm{e2e},\,p}$, which are obtained separately for the $q$ and $p$ quadratures, the total secret-key (or entanglement) rate per protocol run at multiplexing level $k$ is given by
\begin{equation}
R_{\mathrm{protocol\,run}} = \sum_{l=1}^{k} r\bigl(P_{\mathrm{e2e},\,q}^{\,l},\,P_{\mathrm{e2e},\,p}^{\,l}\bigr),
\end{equation}
where the index $l \in \{1,\dots,k\}$ labels the multiplexed channels, and the function $r$ is defined as
\begin{equation}
r\bigl(P_{\mathrm{e2e},\,q}^{\,l},\,P_{\mathrm{e2e},\,p}^{\,l}\bigr) = \max\bigl\{r_1^{\,l},\,r_2^{\,l},\,0\bigr\}.
\end{equation}
Here, $r_1$ represents both the asymptotic one-way six-state secret-key fraction and the achievable one-way entanglement generation rate, whereas $r_2$ corresponds to both the asymptotic six-state secret-key fraction when advantage distillation is employed and the achievable two-way entanglement generation rate~\cite{bennett1996mixed, renner2008security}.

Moreover, we define the secret-key (or entanglement) rate per channel per protocol run as
\begin{equation}
\begin{split}
R_{\mathrm{channel}} 
&= \frac{1}{k} \, R_{\mathrm{protocol\,run}} \\ 
&= \frac{1}{k} \sum_{l=1}^{k} r\bigl(P_{\mathrm{e2e},\,q}^{\,l},\,P_{\mathrm{e2e},\,p}^{\,l}\bigr).
\end{split}
\end{equation}

On the other hand, we estimate the number of GKP qubits that need to be available at each repeater station at the beginning of each protocol run using the method that tracks the state transitions of a Markov chain. Here, the state transition refers to the process by which multiple graph states are transformed into other graph states through a sequence of optical fusion operations and measurements, and this process is probabilistic owing to the use of post-selection in GKP measurements~\cite{fukui2018high}. Each repeater station must simultaneously generate $2k$ elementary entangled Bell pairs at a multiplexing level of $k$. The GKP qubits required for the refreshment process are expected to be supplied to a significant extent through a combination of the GKP qubit recycling procedure and the sequential refreshment strategy introduced in Subsection~\ref{subsec:recycling}. The refreshment resources are scheduled to become available with a slight delay relative to the initial GKP qubits. Therefore, the GKP qubits used for refreshment are not included in the initial GKP qubit requirement defined here.

We define two quantities representing this initial GKP qubit requirement per repeater station per protocol run: one corresponding to a 0.999 success probability for all repeater stations to prepare the required number of elementary entangled Bell pairs simultaneously, and the other corresponding to a 0.9999 success probability. We denote them as \textit{Initial number of GKP qubits (>0.999)} and \textit{Initial number of GKP qubits (>0.9999)}, respectively.

Finally, at the same multiplexing level $k$, by combining the secret-key (or entanglement) rate per channel per protocol run with the initial number of GKP qubits required per repeater station per protocol run, we define the following cost function:
\begin{equation}
\mathrm{Cost}(k) = \frac{(\mathrm{Initial \,\, number \,\, of \,\, GKP \,\, qubits})^2}{R_{\mathrm{channel}}}.
\end{equation}
Here, since scaling up the generation of GKP qubits is more challenging than increasing the rate, we square the initial number of GKP qubits required in the cost definition. Although this choice is not unique, it provides a more realistic measure of the experimental overhead than a simple linear scaling. By plotting the cost as a function of $k$, we can determine the optimal multiplexing level, $k_{\mathrm{opt}}$, as discussed in Subsection~\ref{subsec:Cost}.

\section{Simulation Parameters}
\label{sec:simulation}
We simulate the performance of our all-photonic quantum repeater architecture by analyzing error propagation through noisy channels and imperfect operations, without explicitly tracking the evolution of quantum states. Numerical simulations were performed using the following parameters:

\begin{table*}[ht]
\centering
\caption{Simulation parameters.}
\label{tab:sim-params}
\footnotesize
\renewcommand{\arraystretch}{1.4}
\setlength{\tabcolsep}{6pt}
\begin{tabular}{c|c|c}
\toprule
\makebox[1.4cm][c]{Symbol} & \makebox[5.8cm][c]{Parameter} & \makebox[1.5cm][c]{Value} \\
\midrule
\midrule
$L_0$ & Distance between two adjacent repeater stations & $9~\mathrm{km}$ \\
$\sigma_{\mathrm{GKP}}$ & Standard deviation of the initial Gaussian random displacement noise due to finite GKP squeezing & $0.12$ \\
$\eta_s$ & Efficiency of the optical switch applied to the graph states after post-selection~\cite{fukui2018high} & $0.995$ \\
$\eta_m$ & Efficiency of mirror reflection per bounce inside the free-space photonic memory & $0.999995$ \\
$\eta_d$ & Efficiency of a single homodyne detection & $0.9975$ \\
\multirow{2}{*}{$\eta_c$} & Efficiency of a single connector between an optical fiber & \multirow{2}{*}{$0.99$} \\[-3.4pt]
& and a quantum chip or a free-space photonic memory & \\
$L_{\mathrm{cavity}}$ & Distance between successive bounces inside the free-space photonic memory & $2~\mathrm{m}$ \\
$k$ & Multiplexing level & $15$ \\
\multirow{2}{*}{$\eta$} & Transmittance through an optical fiber, modeled as $\eta = e^{-L/L_{\mathrm{att}}}$ & \multirow{2}{*}{(function)} \\[-3.4pt]
& for fiber length $L~[\mathrm{km}]$, where $L_{\mathrm{att}} = 22~\mathrm{km}$~\cite{azuma2015all} & \\
$v_7$ & Discard window size parameter for measurement type 7 & $0.3$ \\
\noalign{\vskip 0.15cm}
\bottomrule
\end{tabular}
\end{table*}

$L_0$ --- The distance between two adjacent repeater stations, measured in kilometers. Throughout this paper, we typically set $L_0 = 9~\mathrm{km}$.

$\sigma_{\mathrm{GKP}}$ --- The standard deviation of the initial Gaussian random displacement noise. Throughout this paper, we set $\sigma_{\mathrm{GKP}} = 0.12$.

$\eta_s$ --- The efficiency of the optical switch applied to the graph states after measurements with discard windows. In addition, multiple optical switches located in the detection center within each repeater are collectively modeled as a single optical switch in the simulation. This simplification is justified because these switches can be operated with a longer time allowance than those used after measurements with discard windows, and therefore are expected to achieve higher performance. Throughout this paper, we set $\eta_s = 0.995$.

$\eta_m$ --- The efficiency of mirror reflection per bounce. Throughout this paper, we set $\eta_m = 0.999995$.

$\eta_d$ --- The efficiency of a single homodyne detection. Throughout this paper, we set $\eta_d = 0.9975$.

$\eta_c$ --- The efficiency of a single connector between an optical fiber and a quantum chip or a photonic memory. In Fig.~\ref{fig:1}(c), the light-red square boxes represent these connectors. In contrast, the minor nodes do not require such connectors, because, in practice, their operations can be implemented using a fixed optical setup consisting of optical fibers, beam splitters, and homodyne detectors, without requiring fiber-to-chip coupling. This fixed arrangement is possible because there is no need to actively reorder the $k$ outer-leaves at the minor nodes. Throughout this paper, we set $\eta_c = 0.99$.

$L_{\mathrm{cavity}}$ --- The distance between successive bounces inside the free-space photonic memory, measured in meters. Throughout this paper, we set $L_{\mathrm{cavity}} = 2~\mathrm{m}$.

$k$ --- The multiplexing level. For example, at multiplexing level $k = 15$, 15 end-to-end entangled Bell pairs are generated between the two end nodes.

$\eta$ --- The transmittance through an optical fiber. We model the transmittance of an optical fiber with length $L~[\mathrm{km}]$ as $\eta = e^{-L / L_{\mathrm{att}}}$, where the attenuation length is $L_{\mathrm{att}} = 22~\mathrm{km}$~\cite{azuma2015all}.

$v_7$ --- The discard window size parameter for measurement type 7 in the classification summarized in Appendix~\ref{sec:A}. This measurement type has the largest standard deviation of Gaussian random displacement noise among all measurement types. We employ discard windows centered at $\pm \sqrt{\pi}/2$. The regions between $\sqrt{\pi}/2 - v$ and $\sqrt{\pi}/2 + v$, and between $-\sqrt{\pi}/2 - v$ and $-\sqrt{\pi}/2 + v$, define the discard windows. Therefore, the actual size of each discard window is $2v$. The discard window size parameters for all other measurement types are determined relative to $v_7$ so that the error probabilities are equal across all measurement types. We typically set $v_7 = 0.3$. This choice is supported by Fig.~\ref{fig:4} in Section~\ref{subsec:window}.

The parameters employed in all simulations are summarized in Table~\ref{tab:sim-params}. The simulation code is publicly available for reproducibility~\cite{ShiinaCode2026}.

\section{Results}
\label{sec:results}
In this section, we present the performance of our all-photonic quantum repeater architecture. In Subsection~\ref{subsec:window}, we explain the rationale behind the choice of the discard window size parameter. In Subsection~\ref{subsec:Rate}, we show the results for \(R_{\mathrm{protocol\,run}}\) and \(R_{\mathrm{channel}}\). In Subsection~\ref{subsec:Number}, we discuss the initial GKP qubit requirement. Finally, in Subsection~\ref{subsec:Cost}, we evaluate the cost function \(\mathrm{Cost}(k)\) and determine the optimal multiplexing level \(k_{\mathrm{opt}}\).

\subsection{Optimization of the Discard Window}
\label{subsec:window}
As described in Subsection~\ref{subsec:post}, it is important to determine the discard window size parameter by considering the trade-off between the error probability and the required resources.

We present the relationship between the discard window size parameter and the error probability in Fig.~\ref{fig:4}.

\begin{center}
\includegraphics[width=1\linewidth]{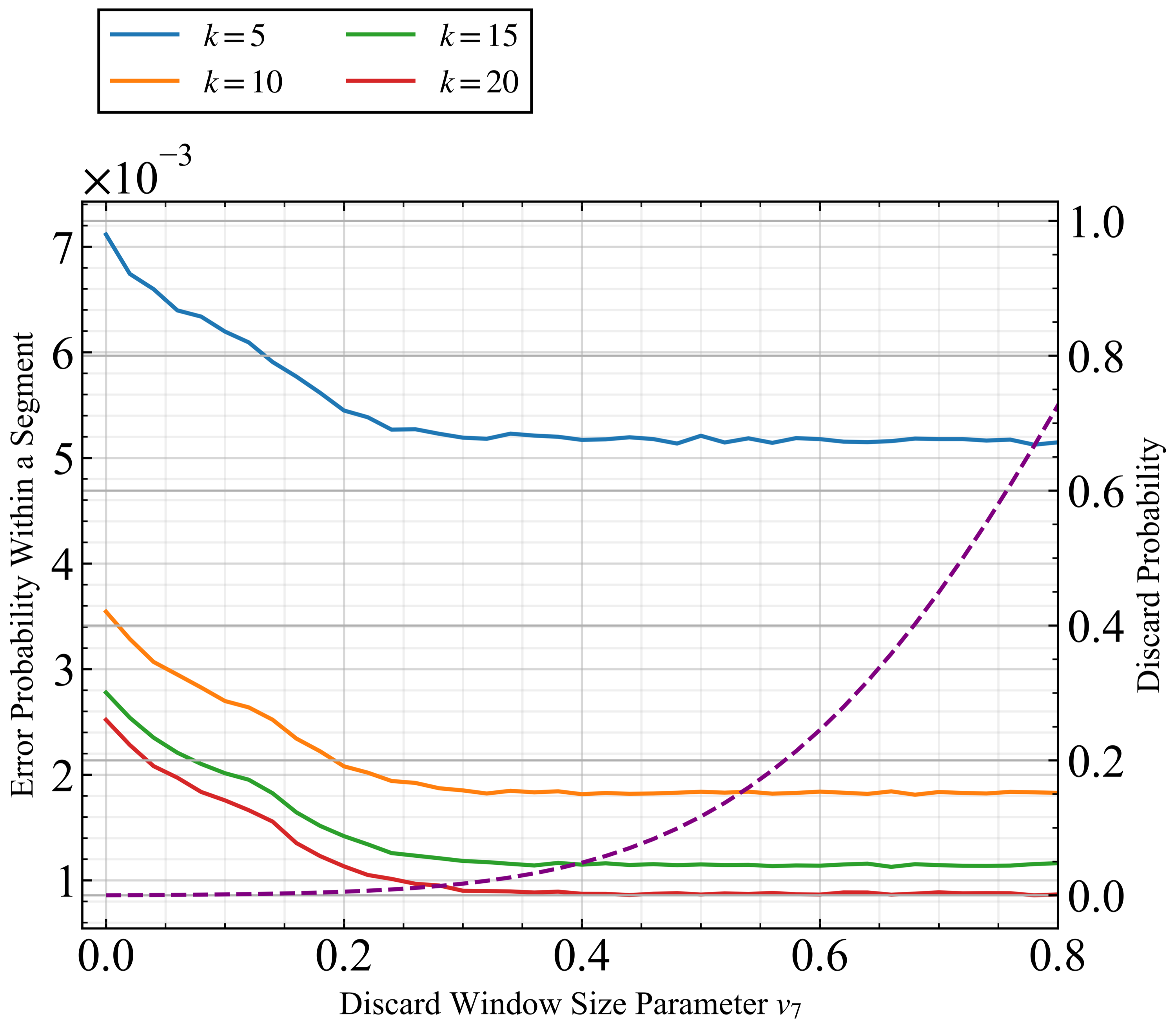}
\captionsetup{hypcap=false}
\captionof{figure}{
        Error probability within a single segment (left vertical axis) and discard probability (right vertical axis) as functions of the discard window size parameter $v_7$. The left vertical axis shows the error probability obtained from Monte Carlo simulations using UW3 for multiplexing levels $k = 5$, $10$, $15$, and $20$, where the error probability corresponding to the best multiplexed channel was selected and plotted. The right vertical axis shows the discard probability, represented by the purple dashed curve, which was obtained analytically from the Gaussian function associated with measurement type~7. Regardless of the multiplexing level, the reduction in the error probability saturates around $v_7 = 0.3$, while the discard probability remains sufficiently low. Therefore, $v_7 = 0.3$ is chosen as the discard window size parameter.
    }
\label{fig:4}
\end{center}

The error probability accounts for the errors accumulated during the following processes: (1) the construction of elementary entangled Bell pairs, (2) outer-leaves swapping, and (3) inner-leaves swapping within a single segment. This probability is shown on the left vertical axis, labeled “Error Probability within a Segment.” Four curves corresponding to $k = 5$, $k = 10$, $k = 15$, and $k = 20$ were obtained from Monte Carlo simulations using UW3. For each curve, the error probability corresponding to the best multiplexed channel was selected and plotted. For UW2, the four curves almost completely overlap with those obtained using UW3.

On the other hand, we refer to the probability that measurement outcomes fall into the discard regions as the \textit{Discard Probability}. This probability is shown on the right vertical axis, labeled “Discard Probability.” The purple dashed curve corresponds to this probability, which was obtained analytically from the Gaussian function associated with measurement type~7.

The horizontal axis represents the discard window size parameter for measurement type~7, $v_7$, as discussed in Section~\ref{sec:simulation}.

As shown in Fig.~\ref{fig:4}, regardless of the multiplexing level, the reduction in the error probability saturates around $v_7 = 0.3$. This saturation can be understood as arising from the fact that, once the construction errors have been sufficiently suppressed by the discard windows, other imperfections, such as loss in the optical fibers, become the dominant limiting factors. On the other hand, the discard probability remains sufficiently low at $v_7 = 0.3$. Based on these observations, choosing $v_7 = 0.3$ is justified. Furthermore, Fig.~\ref{fig:4} indicates that increasing the multiplexing level does not reduce the error probability indefinitely. In other words, increasing the multiplexing level beyond a certain point contributes little to any further reduction in the error probability. For example, the reduction in the error probability when increasing $k$ from 15 to 20 is significantly smaller than that when increasing $k$ from 5 to 10. Since a higher multiplexing level also requires more GKP qubits, this naturally leads us to consider the cost.

%%%%%%%%%%%%%%%%%%%%%%%%%%%%%%%%%%%%%%%%%%%%%%%%%%%%%%%%%%%%%%%%%%%%%%%%%%%%%%%%%%%%
%%%%%%%%%%%%%%%%%%%%%%%%%%%%%%%%%%%%%%%%%%%%%%%%%%%%%%%%%%%%%%%%%%%%%%%%%%%%%%%%%%%%
\subsection{Rate}
\label{subsec:Rate}

First, we show how $R_{\mathrm{protocol\,run}}$ varies with the distance between two adjacent repeater stations in Fig.~\ref{fig:5}. Each data point was obtained from Monte Carlo simulations. Similarly, all data points for \(R_{\mathrm{protocol\,run}}\) and \(R_{\mathrm{channel}}\) reported in the remaining subsections of Section~\ref{sec:results} and throughout Section~\ref{sec:discussion} were obtained from Monte Carlo simulations. The solid curves correspond to UW3, while the dashed curves other than the blue one correspond to UW2. The blue dashed curve represents the PLOB bound~\cite{pirandola2017fundamental}.

With a repeater spacing of $L_0 = 10~\mathrm{km}$, the achievable end-to-end communication distance is limited to approximately $600~\mathrm{km}$. In contrast, reducing the repeater spacing to $L_0 = 9~\mathrm{km}$ enables communication over distances of up to approximately $1{,}500~\mathrm{km}$. For longer communication distances, the repeater spacing would need to be reduced further. However, since $1{,}000~\mathrm{km}$ already covers the relevant communication range for our analysis, we fix the repeater spacing at $9~\mathrm{km}$ in the following sections.

\begin{center}
\includegraphics[width=1\linewidth]{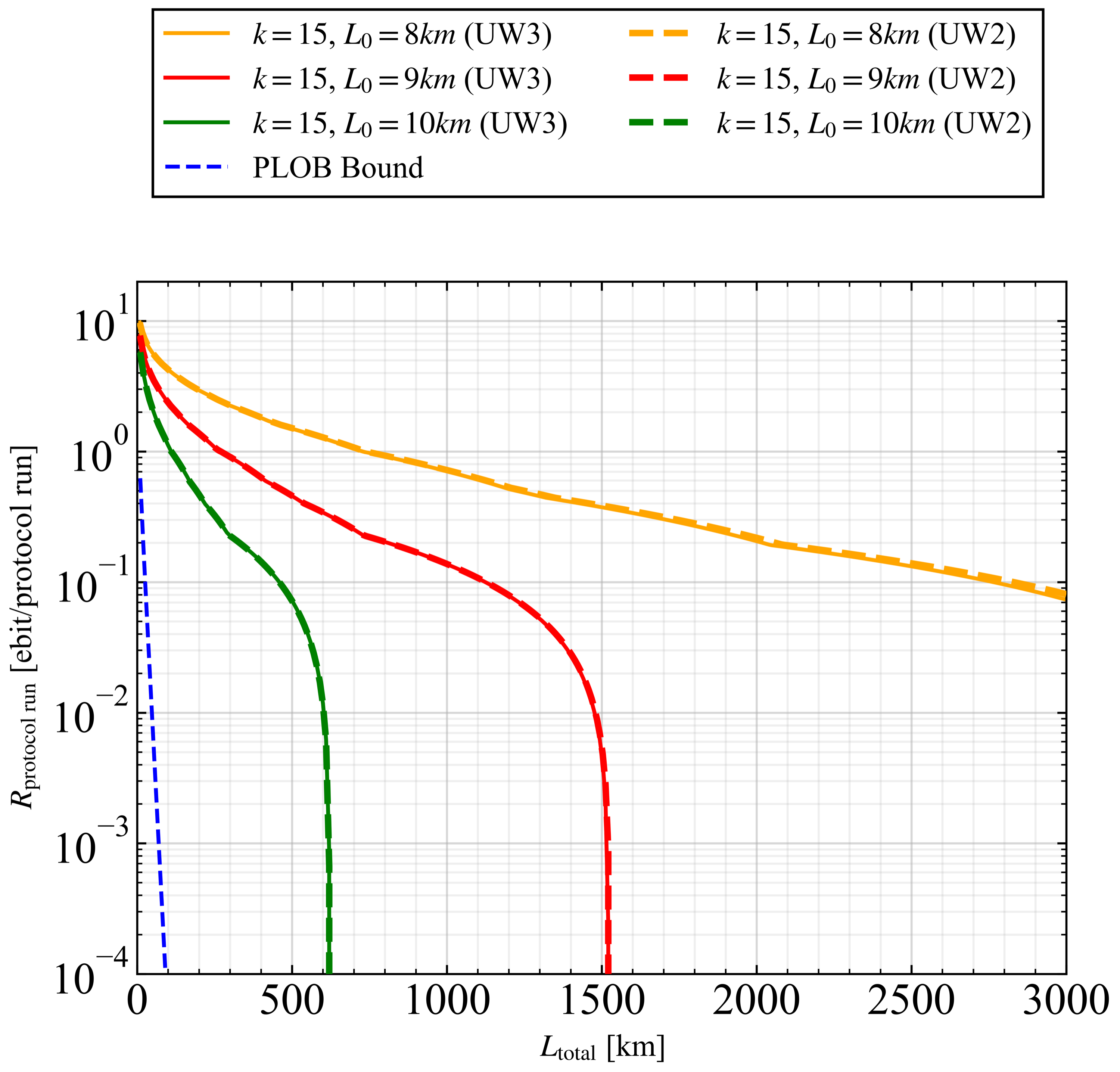}
\captionsetup{hypcap=false}
\captionof{figure}{
        The total secret-key (or entanglement) rate per protocol run, $R_{\mathrm{protocol\,run}}$, as a function of the end-to-end communication distance, $L_{\mathrm{total}}$. All solid curves correspond to UW3, while the dashed curves (excluding the blue one) correspond to UW2. The blue dashed curve represents the PLOB bound~\cite{pirandola2017fundamental}. Each data point was obtained from Monte Carlo simulations. With a repeater spacing of $L_0 = 10~\mathrm{km}$, the achievable end-to-end communication distance is limited to approximately $600~\mathrm{km}$, whereas reducing the repeater spacing to $L_0 = 9~\mathrm{km}$ extends it to approximately $1{,}500~\mathrm{km}$. Since $1{,}000~\mathrm{km}$ already covers the relevant communication range for our analysis, we fix the repeater spacing at $9~\mathrm{km}$ in the following sections. 
    }
    \label{fig:5}
\end{center}

%%%%%%%%%%%%%$R_{\mathrm{protocol\,run}}$_vs_k%%%%%%%%%%%%

Next, with the repeater spacing fixed at $9~\mathrm{km}$, we examine how $R_{\mathrm{protocol\,run}}$ varies with the multiplexing level $k$ for different end-to-end communication distances, taken to be integer multiples of $45~\mathrm{km}$. Each data point was obtained using UW3. For UW2, all curves almost completely overlap with those obtained using UW3. The results for shorter end-to-end communication distances ranging from $45~\mathrm{km}$ to $540~\mathrm{km}$ are shown in Fig.~\ref{fig:6}, while those for longer distances from $585~\mathrm{km}$ to $1{,}035~\mathrm{km}$ are shown in Fig.~\ref{fig:7}.

Fig.~\ref{fig:6} shows that all curves increase approximately linearly with the multiplexing level $k$ for all distances. Similarly, Fig.~\ref{fig:7} shows a general increase in the rate as $k$ increases. The slight nonlinear features, or small kinks, observed in Fig.~\ref{fig:7} may indicate critical points at which the ranking-based strategy becomes more effective. Although this interpretation is not conclusive, one possible explanation is as follows. When the multiplexing level $k$ is small, even the best-ranked entangled links tend to have relatively low quality, resulting in a low end-to-end rate. However, once $k$ exceeds a certain threshold, sufficiently high-quality links begin to appear among the entangled links between neighboring repeater stations. This allows the protocol to connect these high-quality links, leading to a significant improvement in the entanglement rate.

\begin{center}
\includegraphics[width=1\linewidth]{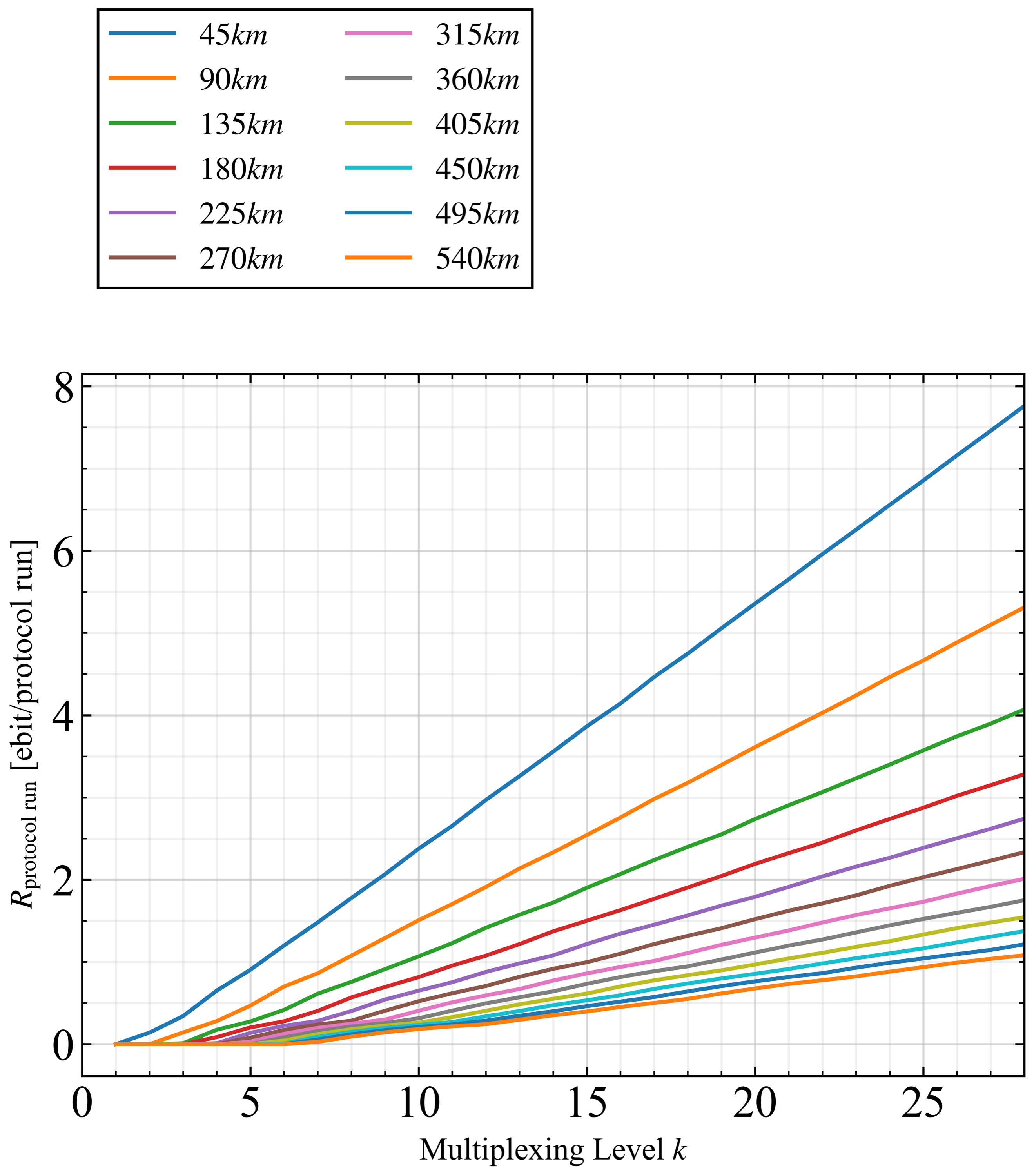}
\captionsetup{hypcap=false}
\captionof{figure}{
	The total secret-key (or entanglement) rate per protocol run, $R_{\mathrm{protocol\,run}}$, as a function of the multiplexing level $k$ for shorter end-to-end communication distances ranging from $45~\mathrm{km}$ to $540~\mathrm{km}$ in integer multiples of $45~\mathrm{km}$. Each data point was obtained from Monte Carlo simulations using UW3. For UW2, all curves almost completely overlap with those obtained using UW3. For all distances shown, the rate increases approximately linearly with $k$.
    }
    \label{fig:6}
\end{center}

\begin{center}
\includegraphics[width=1\linewidth]{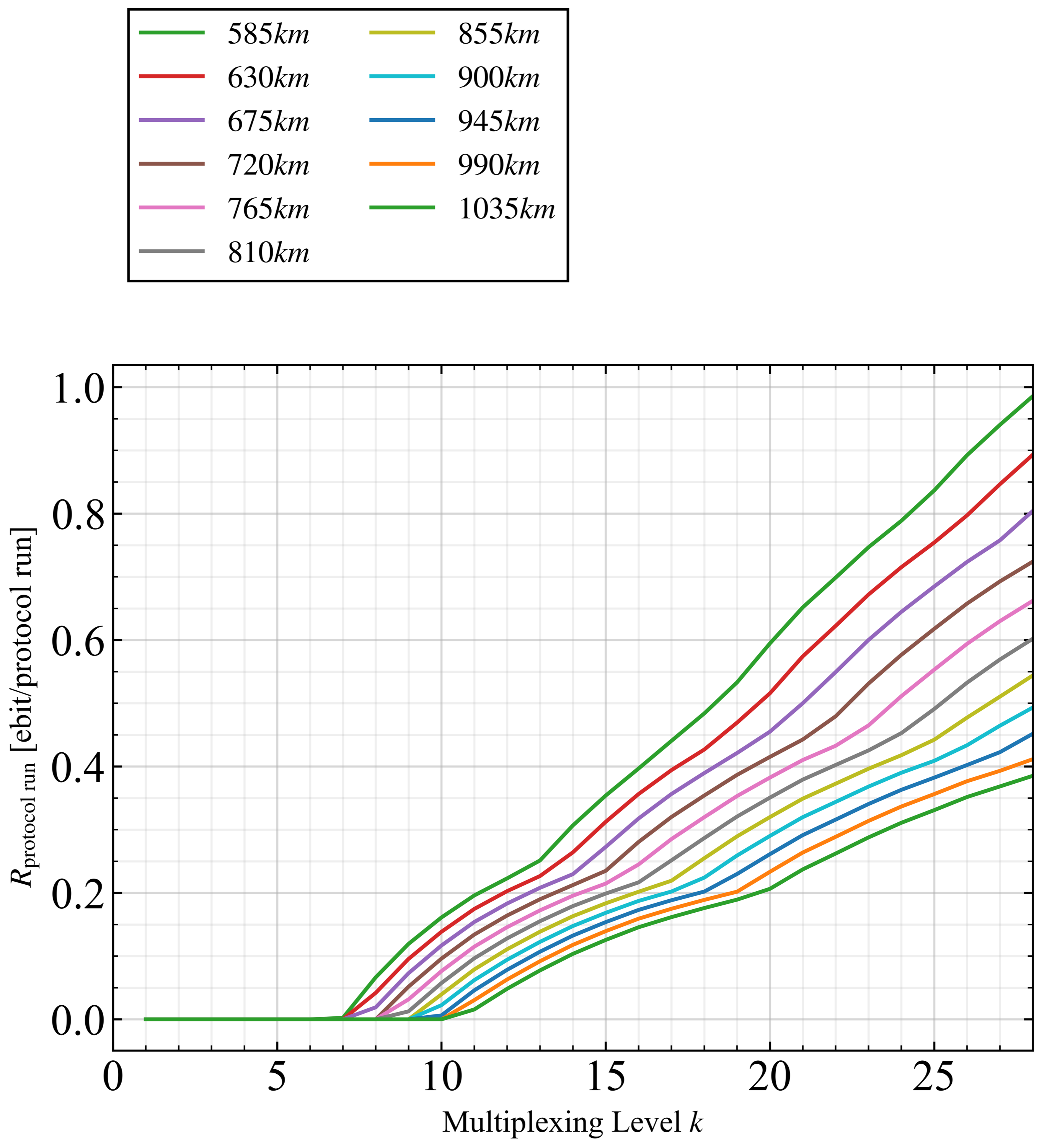}
\captionsetup{hypcap=false}
\captionof{figure}{
        The total secret-key (or entanglement) rate per protocol run, $R_{\mathrm{protocol\,run}}$, as a function of the multiplexing level $k$ for longer end-to-end communication distances ranging from $585~\mathrm{km}$ to $1{,}035~\mathrm{km}$ in integer multiples of $45~\mathrm{km}$. Each data point was obtained from Monte Carlo simulations using UW3. For UW2, all curves almost completely overlap with those obtained using UW3. The rate generally increases with $k$, with slight nonlinear features or small kinks appearing at certain values of $k$.
    }
    \label{fig:7}
\end{center}

%%%%%%%%%%%%%$R_{channel}$_vs_k%%%%%%%%%%%%
Finally, with the repeater spacing fixed at $9~\mathrm{km}$, we investigate how $R_{\mathrm{channel}}$ varies with the multiplexing level $k$ for different end-to-end communication distances, taken to be integer multiples of $45~\mathrm{km}$. Each data point was obtained by dividing the corresponding value of $R_{\mathrm{protocol\,run}}$ in Figs.~\ref{fig:6} and~\ref{fig:7} by the multiplexing level $k$. The results for shorter end-to-end communication distances ranging from $45~\mathrm{km}$ to $540~\mathrm{km}$ are shown in Fig.~\ref{fig:8}, while those for longer distances from $585~\mathrm{km}$ to $1{,}035~\mathrm{km}$ are shown in Fig.~\ref{fig:9}.

Fig.~\ref{fig:8} shows that all curves increase with the multiplexing level $k$ for all distances, but tend to saturate as $k$ becomes large. Similarly, Fig.~\ref{fig:9} shows a general increase in the rate as $k$ increases, with a tendency to saturate as $k$ becomes large, although clear saturation is not yet observed because the range of $k$ remains limited. However, it is natural to expect that $R_{\mathrm{channel}}$ will eventually saturate as $k$ increases further. The nonlinear features observed in the curves of Fig.~\ref{fig:9} may be attributed to the same reason discussed earlier for $R_{\mathrm{protocol\,run}}$.

\begin{center}
\includegraphics[width=1\linewidth]{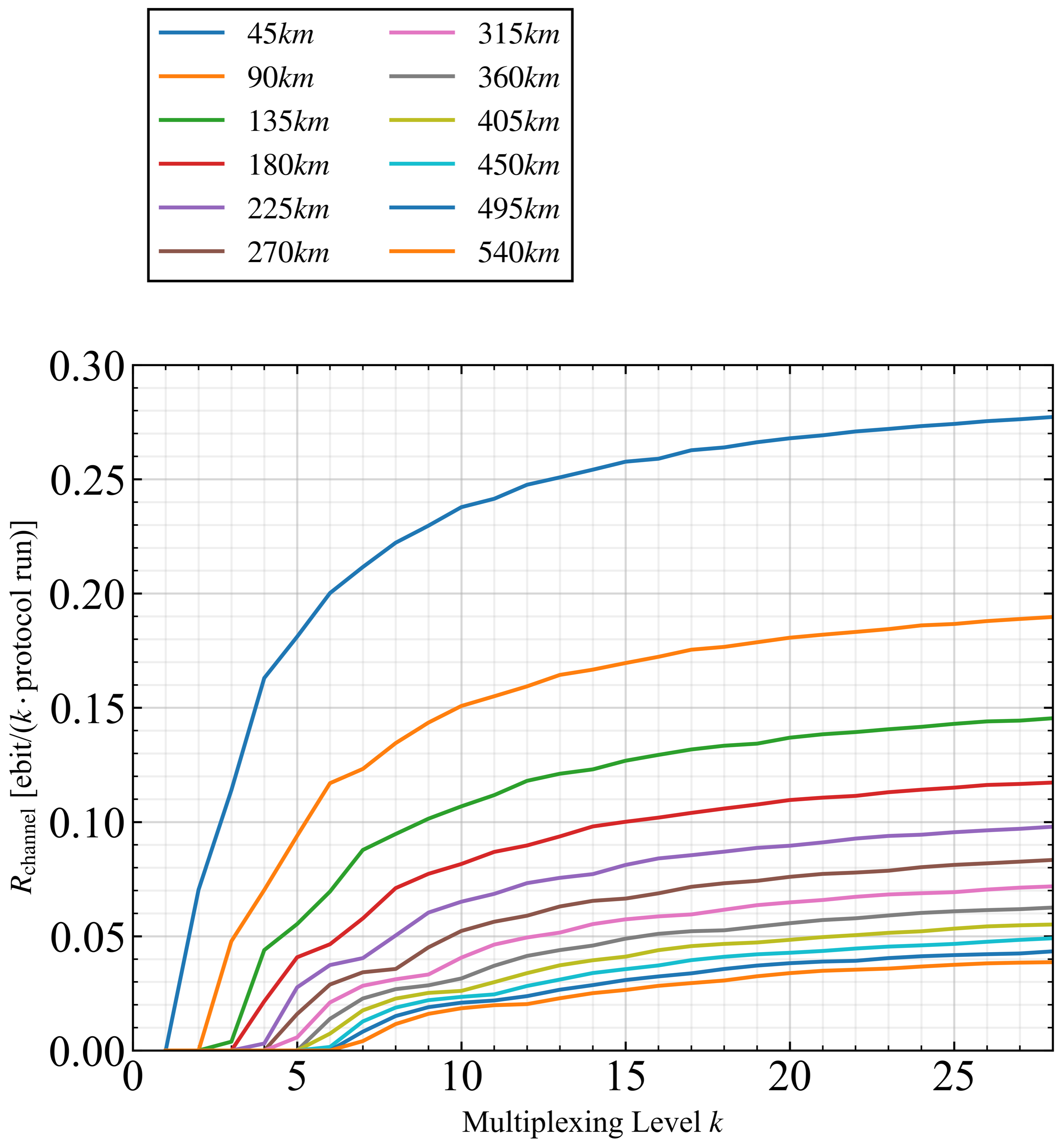}
\captionsetup{hypcap=false}
\captionof{figure}{
        The secret-key (or entanglement) rate per channel per protocol run, $R_{\mathrm{channel}}$, as a function of the multiplexing level $k$ for shorter end-to-end communication distances ranging from $45~\mathrm{km}$ to $540~\mathrm{km}$ in integer multiples of $45~\mathrm{km}$. Each data point was obtained by dividing the corresponding value of $R_{\mathrm{protocol\,run}}$ in Fig.~\ref{fig:6} by the multiplexing level $k$. For all distances shown, the rate increases with $k$ and tends to saturate as $k$ becomes large.
    }
    \label{fig:8}
\end{center}

\begin{center}
\begin{minipage}{1\linewidth}
\centering
\includegraphics[width=1\linewidth]{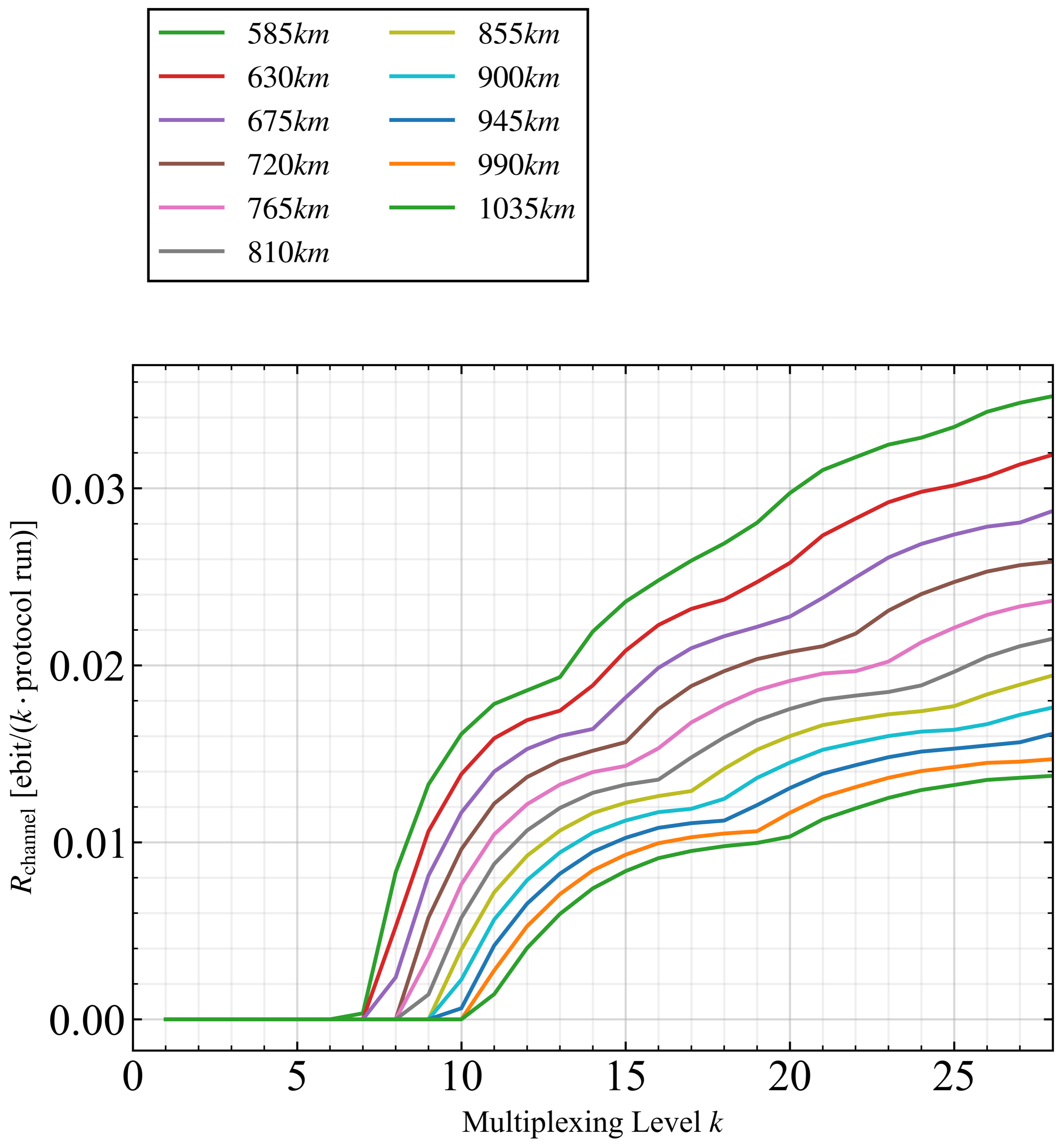}
\captionsetup{hypcap=false}
\captionof{figure}{
        The secret-key (or entanglement) rate per channel per protocol run, $R_{\mathrm{channel}}$, as a function of the multiplexing level $k$ for longer end-to-end communication distances ranging from $585~\mathrm{km}$ to $1{,}035~\mathrm{km}$ in integer multiples of $45~\mathrm{km}$. Each data point was obtained by dividing the corresponding value of $R_{\mathrm{protocol\,run}}$ in Fig.~\ref{fig:7} by the multiplexing level $k$. The rate generally increases as $k$ increases, with a tendency toward saturation for larger $k$, and is expected to eventually saturate as $k$ increases further.
    }
    \label{fig:9}
\end{minipage}
\end{center}

%%%%%%%%%%%%%%%%%%%%%%%%%%%%%%%%%%%%%%%%%%%%%%%%%%%%%%%%%%%%%%%%%%%%%%%%%%%%%%%%%%%%
%%%%%%%%%%%%%%%%%%%%%%%%%%%%%%%%%%%%%%%%%%%%%%%%%%%%%%%%%%%%%%%%%%%%%%%%%%%%%%%%%%%%
\subsection{Initial GKP Qubit Requirement}
\label{subsec:Number}
In this subsection, we present the initial GKP qubit requirement per repeater station per protocol run. First, we show how this requirement varies with the end-to-end communication distance \(L_{\mathrm{total}}\) in Fig.~\ref{fig:10}. Each data point was obtained analytically by tracking the state transitions of a Markov chain.

The solid curves represent the results obtained using UW3, while the dashed curves correspond to those obtained using UW2. For each multiplexing level \(k\), we calculate two quantities representing the initial GKP qubit requirement---\textit{Initial number of GKP qubits (>0.999)} and \textit{Initial number of GKP qubits (>0.9999)}---as defined in Section~\ref{sec:e2e}. Between the two curves with the same color and line style, the lower one corresponds to \textit{Initial number of GKP qubits (>0.999)}, while the upper one corresponds to \textit{Initial number of GKP qubits (>0.9999)}.

The initial GKP qubit requirement per repeater station per protocol run remains relatively stable even as the end-to-end communication distance increases. More importantly, UW2 has a significantly larger initial GKP qubit requirement than UW3. Given this result and the fact that UW2 and UW3 yield the same rate performance, we conclude that UW3 is the more resource-efficient construction method under the baseline parameter values considered here.

\begin{center}
\includegraphics[width=1\linewidth]{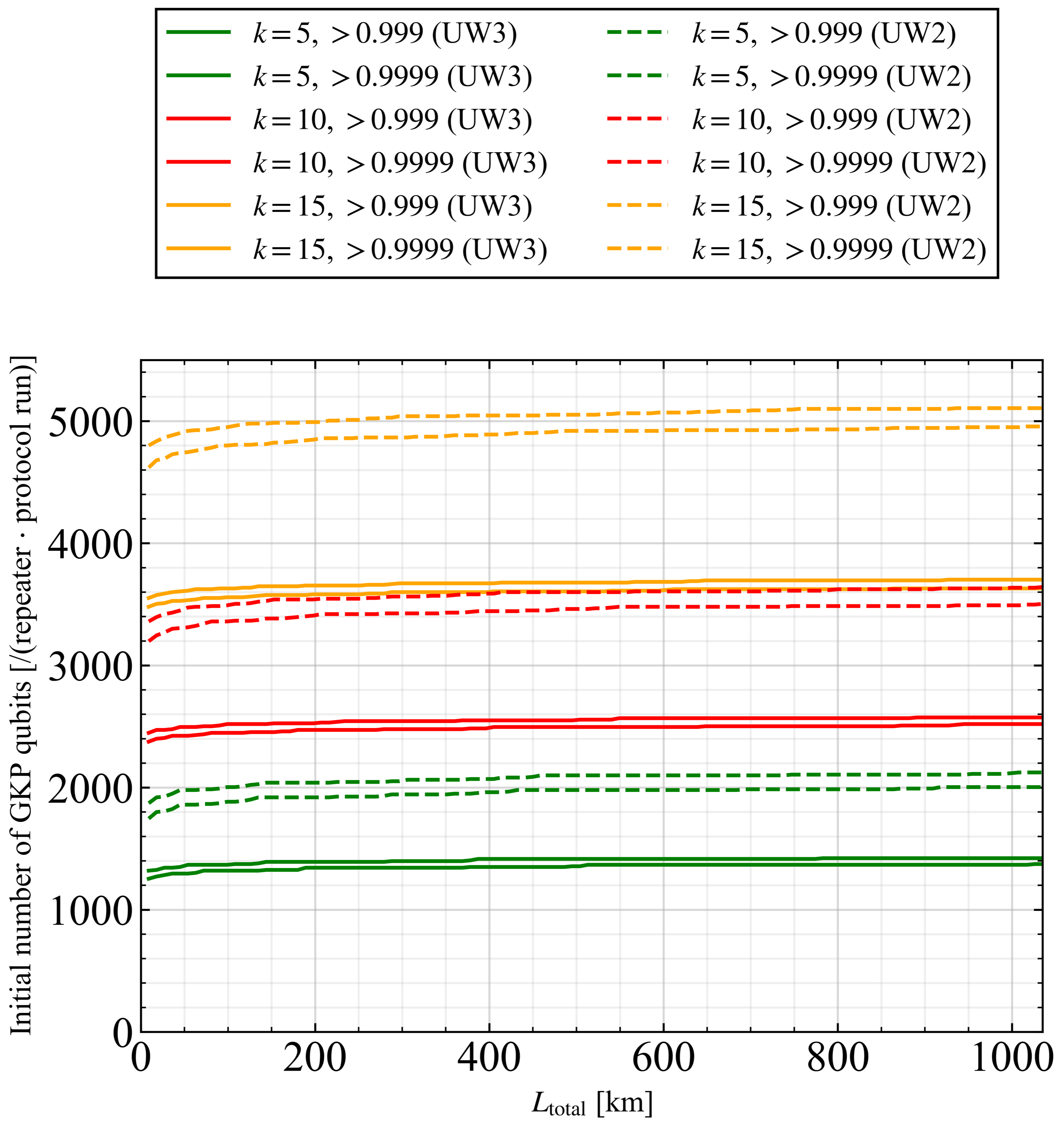}
\captionsetup{hypcap=false}
\captionof{figure}{
         The initial GKP qubit requirement per repeater station per protocol run---\textit{Initial number of GKP qubits (>0.999)} and \textit{Initial number of GKP qubits (>0.9999)}---as a function of the end-to-end communication distance, $L_{\mathrm{total}}$. Each data point was obtained analytically by tracking the state transitions of a Markov chain. The solid and dashed curves correspond to results obtained using UW3 and UW2, respectively. Between the two curves with the same color and line style, the lower one corresponds to \textit{Initial number of GKP qubits (>0.999)}, while the upper one corresponds to \textit{Initial number of GKP qubits (>0.9999)}. The results indicate that the initial GKP qubit requirement per repeater station per protocol run remains nearly constant as $L_{\mathrm{total}}$ increases, and that UW2 has a significantly larger initial GKP qubit requirement than UW3.
    }
    \label{fig:10}
\end{center}

%%%%%%%%%%%%%NoGKP_vs_k%%%%%%%%%%%%
Finally, we examine how the initial GKP qubit requirement varies with the multiplexing level $k$ for two end-to-end communication distances, $45~\mathrm{km}$ and $1{,}035~\mathrm{km}$. Each data point was obtained analytically by tracking the state transitions of a Markov chain. The solid curves correspond to results obtained using UW3, while the dashed curves correspond to those obtained using UW2. The blue and green curves correspond to end-to-end communication distances of $45~\mathrm{km}$ and $1{,}035~\mathrm{km}$, respectively. Between the two curves with the same color and line style, the lower one corresponds to \textit{Initial number of GKP qubits (>0.999)}, while the upper one corresponds to \textit{Initial number of GKP qubits (>0.9999)}.

Fig.~\ref{fig:11} shows that all curves increase linearly with the multiplexing level $k$. Moreover, the curves corresponding to the longer communication distance have larger initial GKP qubit requirements than those for the shorter distance. In addition, the curves obtained using UW2 not only lie above those obtained using UW3 but also exhibit steeper slopes, indicating a larger resource overhead as the multiplexing level increases.

\begin{center}
\begin{minipage}{1\linewidth}
\centering
\includegraphics[width=1\linewidth]{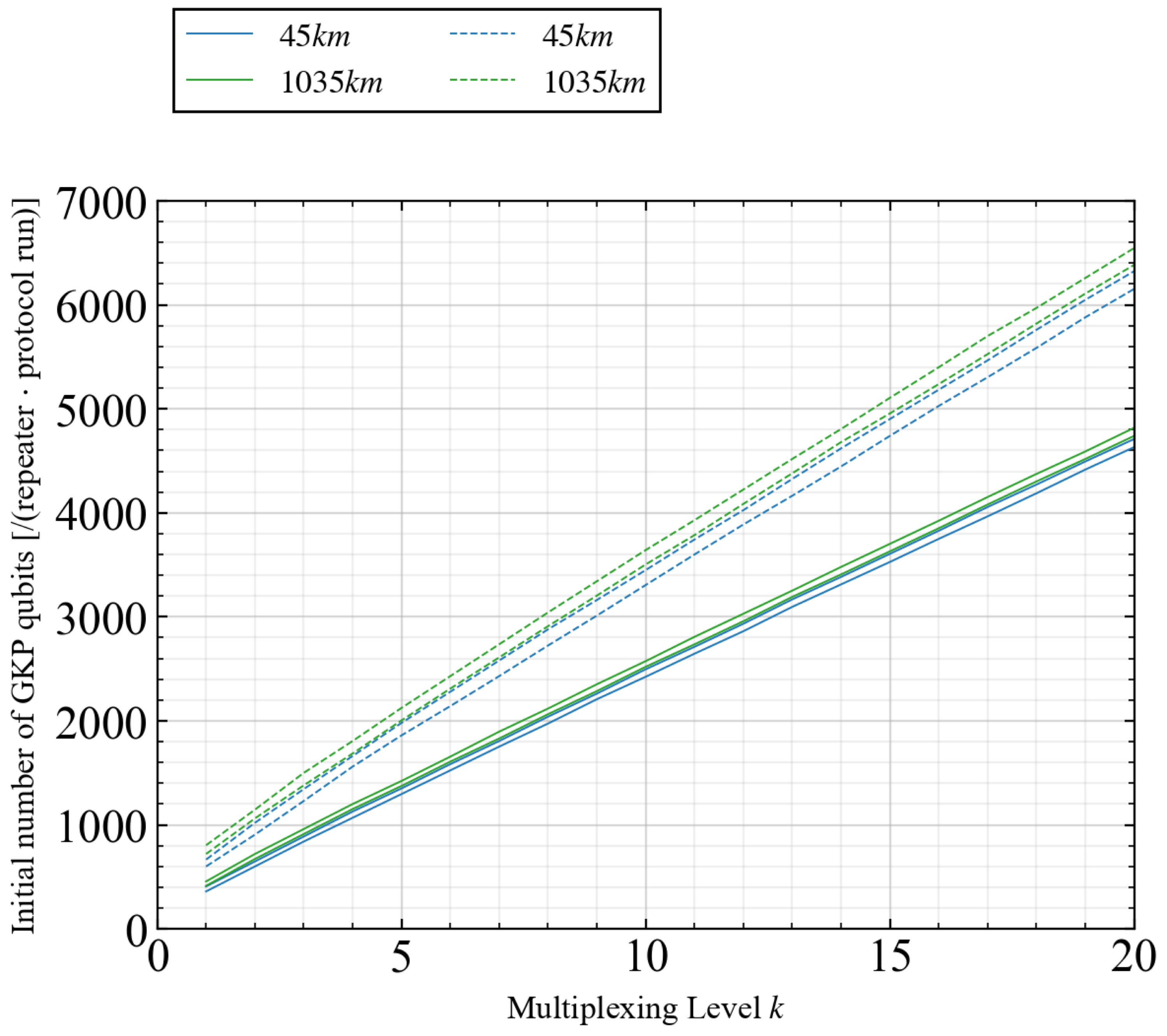}
\captionsetup{hypcap=false}
\captionof{figure}{
        The initial GKP qubit requirement per repeater station per protocol run---\textit{Initial number of GKP qubits (>0.999)} and \textit{Initial number of GKP qubits (>0.9999)}---as a function of the multiplexing level $k$. Each data point was obtained analytically by tracking the state transitions of a Markov chain. The solid and dashed curves correspond to results obtained using UW3 and UW2, respectively. Between the two curves with the same color and line style, the lower one corresponds to \textit{Initial number of GKP qubits (>0.999)}, while the upper one corresponds to \textit{Initial number of GKP qubits (>0.9999)}. All curves increase linearly with the multiplexing level $k$. The curves corresponding to the longer communication distance have larger initial GKP qubit requirements than those for the shorter distance. In addition, the curves obtained using UW2 not only lie above those obtained using UW3 but also exhibit steeper slopes, indicating a larger resource overhead as the multiplexing level increases.
    }
    \label{fig:11}
\end{minipage}
\end{center}

%\clearpage
%%%%%%%%%%%%%%%%%%%%%%%%%%%%%%%%%%%%%%%%%%%%%%%%%%%%%%%%%%%%%%%%%%%%%%%%%%%%%%%%%%%%
%%%%%%%%%%%%%%%%%%%%%%%%%%%%%%%%%%%%%%%%%%%%%%%%%%%%%%%%%%%%%%%%%%%%%%%%%%%%%%%%%%%%
\subsection{Cost Analysis and Optimization Results}
\label{subsec:Cost}
In this subsection, we analyze the cost function and determine the optimal multiplexing level, $k_{\mathrm{opt}}$. We then present the performance results obtained using $k_{\mathrm{opt}}$. 

First, Fig.~\ref{fig:12} shows how the cost varies with the multiplexing level $k$ for different end-to-end communication distances, taken to be integer multiples of $45~\mathrm{km}$. Each data point was obtained by combining the $R_{\mathrm{channel}}$ data used for Figs.~\ref{fig:8} and~\ref{fig:9} with the corresponding initial GKP qubit requirement for each multiplexing level $k$. Unlike Fig.~\ref{fig:11}, which includes only the two cases $L_{\mathrm{total}} = 45~\mathrm{km}$ and $1{,}035~\mathrm{km}$, this figure uses all end-to-end communication distances from $45~\mathrm{km}$ to $1{,}035~\mathrm{km}$ in increments of $45~\mathrm{km}$. It shows the cost obtained using UW3 based on \textit{Initial number of GKP qubits (>0.999)}. Similar trends are observed for UW3 based on \textit{Initial number of GKP qubits (>0.9999)} and for UW2. All curves exhibit a global minimum with respect to $k$. These global minima correspond to the optimal multiplexing levels, $k_{\mathrm{opt}}$. The values of $k_{\mathrm{opt}}$ are indicated in red next to each global minimum.

\begin{center}
\begin{minipage}{1\linewidth}
\centering
\includegraphics[width=1\linewidth]{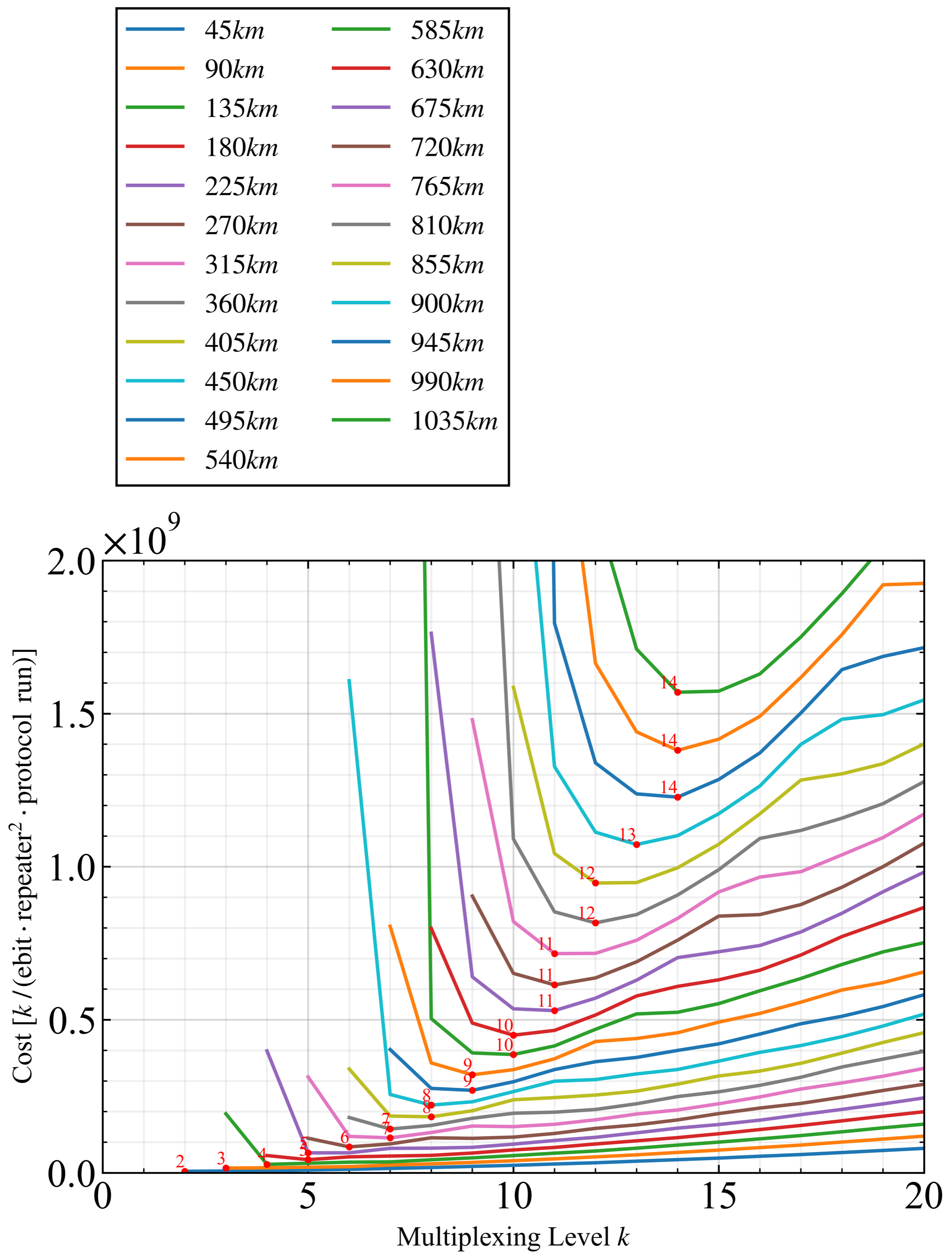}
\captionsetup{hypcap=false}
\captionof{figure}{
        The cost obtained using UW3 based on \textit{Initial number of GKP qubits (>0.999)} as a function of the multiplexing level $k$ for end-to-end communication distances ranging from $45~\mathrm{km}$ to $1{,}035~\mathrm{km}$, taken to be integer multiples of $45~\mathrm{km}$. Each data point was obtained by combining the $R_{\mathrm{channel}}$ data with the corresponding initial GKP qubit requirement. Similar trends are observed for UW3 based on \textit{Initial number of GKP qubits (>0.9999)} and for UW2. All curves exhibit a global minimum with respect to $k$, corresponding to the optimal multiplexing levels $k_{\mathrm{opt}}$. The values of $k_{\mathrm{opt}}$ are indicated in red at each global minimum. Truncated parts of the curves indicate that the achievable rate is zero in the lower-$k$ region, resulting in undefined (infinite) cost values.
    }
    \label{fig:12}
\end{minipage}
\end{center}

%%%%%k_{opt}_vs_L_{total}%%%%%

Next, Fig.~\ref{fig:13} summarizes how the optimal multiplexing level, $k_{\mathrm{opt}}$, varies with the end-to-end communication distance $L_{\mathrm{total}}$. Each data point was obtained by extracting $k_{\mathrm{opt}}$ from Fig.~\ref{fig:12} and from the corresponding graph for UW3 based on \textit{Initial number of GKP qubits (>0.9999)}. The figure shows $k_{\mathrm{opt}}$ obtained using UW3, and a similar trend is observed for UW2. The pink and blue curves correspond to success probabilities of 0.999 and 0.9999, respectively, for all repeater stations to prepare the required number of elementary entangled Bell pairs simultaneously. The difference in line thickness between the two curves is intended solely to improve visual clarity. Both $k_{\mathrm{opt}}$ curves increase monotonically as the end-to-end communication distance increases.

\begin{center}
\begin{minipage}{1\linewidth}
\centering
\includegraphics[width=1\linewidth]{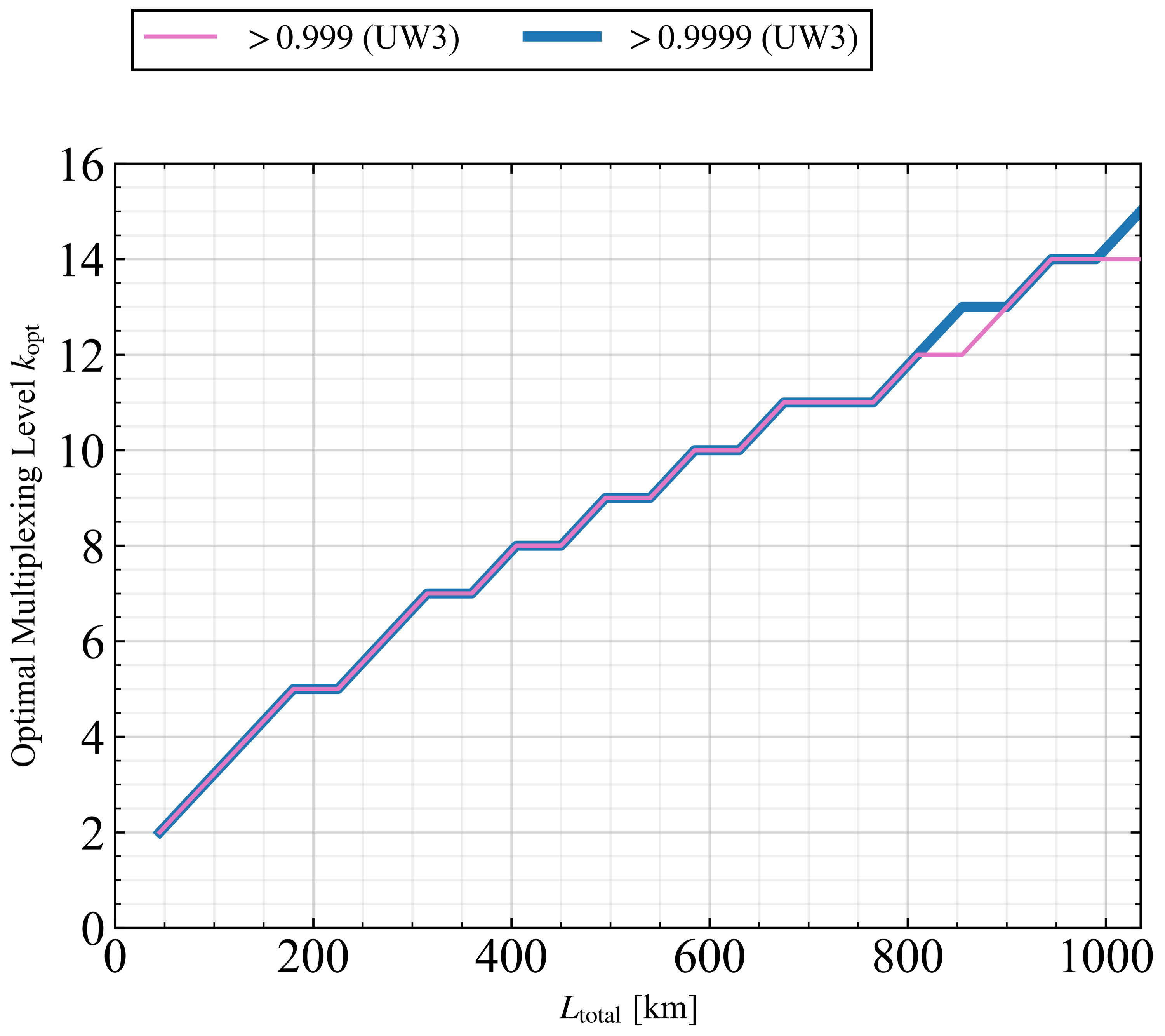}
\captionsetup{hypcap=false}
\captionof{figure}{
        The optimal multiplexing level $k_{\mathrm{opt}}$ as a function of the end-to-end communication distance $L_{\mathrm{total}}$ for UW3. Each data point was obtained by extracting $k_{\mathrm{opt}}$ from Fig.~\ref{fig:12} and from the corresponding graph for UW3 based on \textit{Initial number of GKP qubits (>0.9999)}. The pink and blue curves correspond to success probabilities of 0.999 and 0.9999, respectively, for all repeater stations to prepare the required number of elementary entangled Bell pairs simultaneously. A similar trend is observed for UW2. Different line thicknesses are used solely to improve visual clarity. Both $k_{\mathrm{opt}}$ curves increase monotonically as the end-to-end communication distance increases.
    }
    \label{fig:13}
\end{minipage}
\end{center}

%%%%%Rate_based_on_k_{opt}_vs_L_{total}%%%%%

Based on the optimal multiplexing level $k_{\mathrm{opt}}$, we now present the rate performance results. Fig.~\ref{fig:14} shows how $R_{\mathrm{protocol\,run}}$ based on $k_{\mathrm{opt}}$ varies with the end-to-end communication distance $L_{\mathrm{total}}$. The plotted curves correspond to $R_{\mathrm{protocol\,run}}$ obtained using UW3, and a similar trend is observed for UW2. The pink and blue curves correspond to success probabilities of 0.999 and 0.9999, respectively, for all repeater stations to prepare the required number of elementary entangled Bell pairs simultaneously. The difference in line thickness between the two curves is intended solely to improve visual clarity. For reference, the results for the fixed values $k=5$, $10$, and $15$ are shown as the green, red, and orange curves, respectively. Overall, the $R_{\mathrm{protocol\,run}}$ performance based on $k_{\mathrm{opt}}$ remains relatively stable over the considered range.

\begin{center}
\includegraphics[width=1\linewidth]{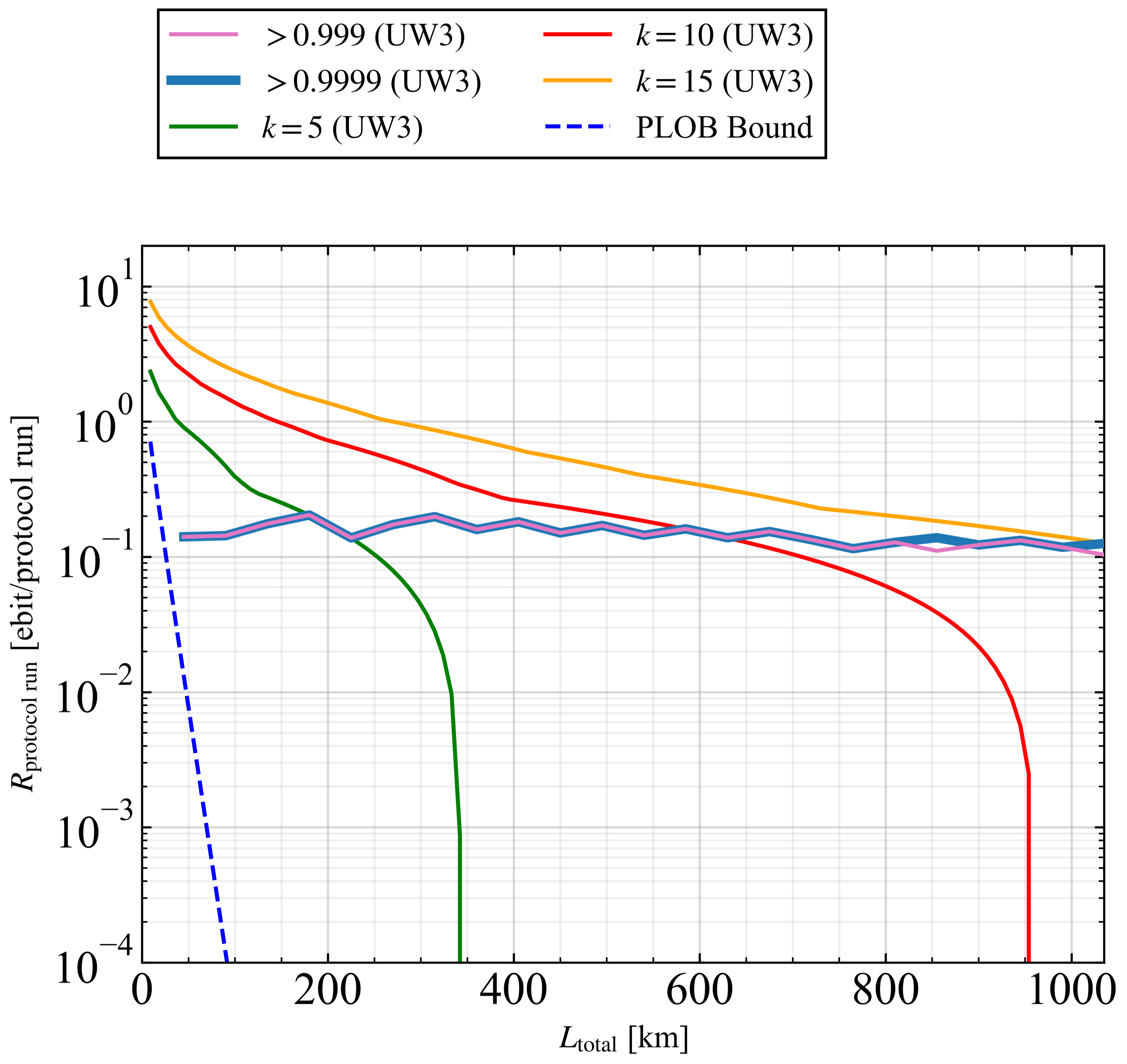}
\captionsetup{hypcap=false}
\captionof{figure}{
        The total secret-key (or entanglement) rate per protocol run, $R_{\mathrm{protocol\,run}}$, obtained using $k_{\mathrm{opt}}$ as a function of the end-to-end communication distance $L_{\mathrm{total}}$ for UW3. The pink and blue curves correspond to success probabilities of 0.999 and 0.9999, respectively, for all repeater stations to prepare the required number of elementary entangled Bell pairs simultaneously. A similar trend is observed for UW2. Different line thicknesses are used solely to improve visual clarity. The performance remains relatively stable over the considered range.
    }
    \label{fig:14}
\end{center}

%%%%%NoGKP_based_on_k_{opt}_vs_L_{total}%%%%%

Fig.~\ref{fig:15} shows how the initial GKP qubit requirement based on $k_{\mathrm{opt}}$ varies with the end-to-end communication distance $L_{\mathrm{total}}$. Each data point was obtained analytically by tracking the state transitions of a Markov chain. The plotted curves correspond to the initial GKP qubit requirement obtained using UW3, and a similar trend is observed for UW2. The pink and blue curves correspond to success probabilities of 0.999 and 0.9999, respectively, for all repeater stations to prepare the required number of elementary entangled Bell pairs simultaneously. For reference, the results for the fixed values $k=5$, $10$, and $15$ are shown as the green, red, and orange curves, respectively. Between the two curves with the same color and line style, the lower one corresponds to \textit{Initial number of GKP qubits (>0.999)}, while the upper one corresponds to \textit{Initial number of GKP qubits (>0.9999)}. The curves obtained using $k_{\mathrm{opt}}$ increase monotonically with the end-to-end communication distance and cross the green, orange, and red curves as the distance increases. Taken together, these results indicate that $k_{\mathrm{opt}}$ achieves a relatively stable $R_{\mathrm{protocol\,run}}$ performance without requiring an unnecessarily large initial number of GKP qubits.

\begin{center}
\includegraphics[width=1\linewidth]{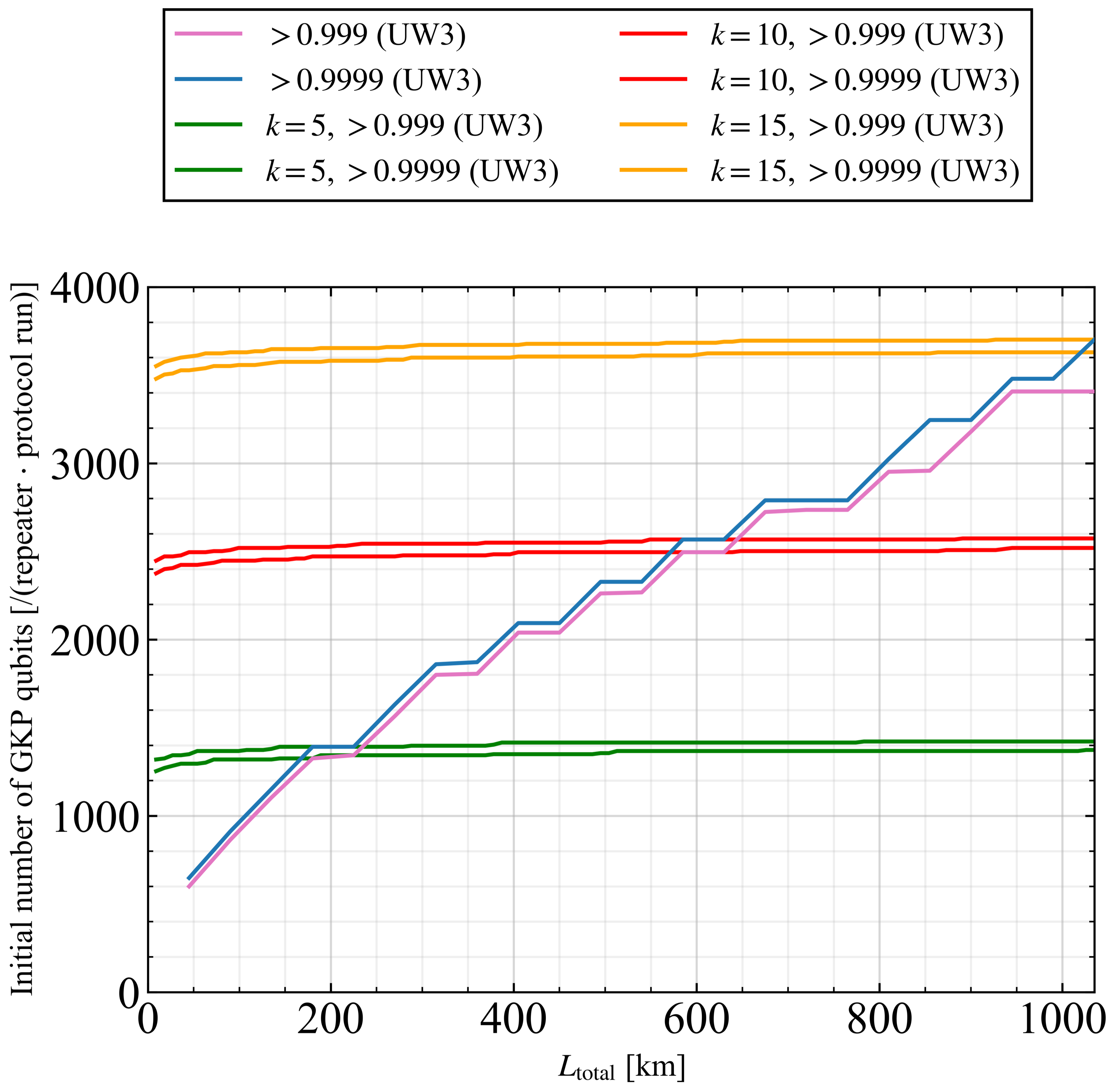}
\captionsetup{hypcap=false}
\captionof{figure}{
        The initial GKP qubit requirement per repeater station per protocol run---\textit{Initial number of GKP qubits (>0.999)} and \textit{Initial number of GKP qubits (>0.9999)}---obtained using $k_{\mathrm{opt}}$ as a function of the end-to-end communication distance $L_{\mathrm{total}}$ for UW3. The pink and blue curves correspond to success probabilities of 0.999 and 0.9999, respectively, for all repeater stations to prepare the required number of elementary entangled Bell pairs simultaneously. A similar trend is observed for UW2. The curves increase monotonically with the end-to-end communication distance and cross the green, orange, and red curves as the distance increases.
    }
    \label{fig:15}
\end{center}

%%%%%Cost_based_on_k_{opt}_vs_L_{total}%%%%%

Finally, Fig.~\ref{fig:16} shows how the cost based on $k_{\mathrm{opt}}$ varies with the end-to-end communication distance $L_{\mathrm{total}}$. The plotted curves correspond to the cost obtained using UW3, and a similar trend is observed for UW2. The pink and blue curves correspond to success probabilities of 0.999 and 0.9999, respectively, for all repeater stations to prepare the required number of elementary entangled Bell pairs simultaneously. For reference, the results for the fixed values $k=5$, $10$, and $15$ are shown as the green, red, and orange curves, respectively. Between the two curves with the same color and line style, the lower one corresponds to \textit{Initial number of GKP qubits (>0.999)}, while the upper one corresponds to \textit{Initial number of GKP qubits (>0.9999)}. The curves obtained using $k_{\mathrm{opt}}$ increase monotonically and rapidly with the end-to-end communication distance and form envelopes of the red and orange curves. Although the green curves lie below the pink and blue curves, Fig.~\ref{fig:14} shows that they are valid only up to an end-to-end communication distance $L_{\mathrm{total}}$ of $350~\mathrm{km}$.

\begin{center}
\includegraphics[width=1\linewidth]{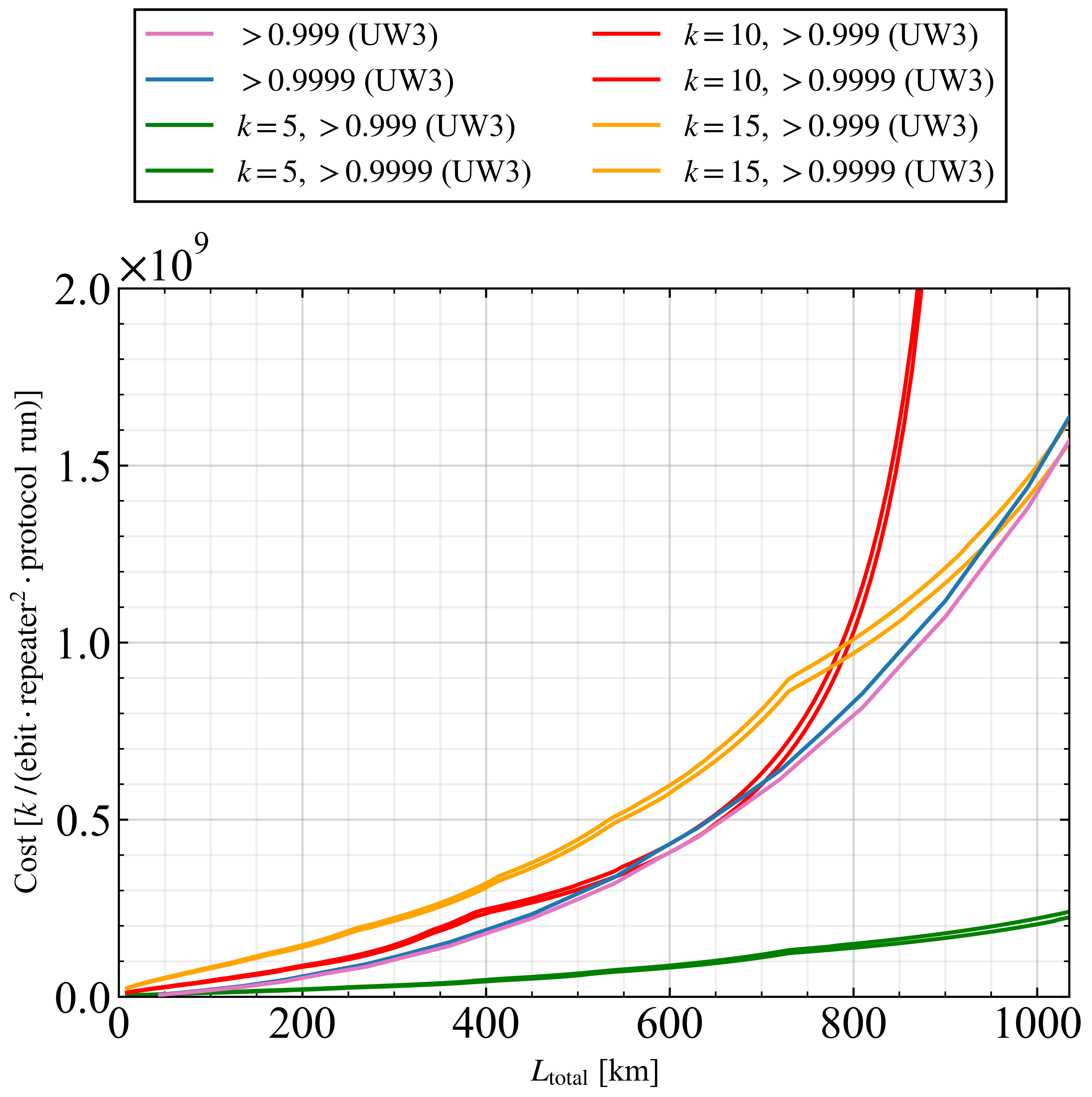}
\captionsetup{hypcap=false}
\captionof{figure}{
        The cost obtained using $k_{\mathrm{opt}}$ as a function of the end-to-end communication distance $L_{\mathrm{total}}$ for UW3. The pink and blue curves correspond to success probabilities of 0.999 and 0.9999, respectively, for all repeater stations to prepare the required number of elementary entangled Bell pairs simultaneously. A similar trend is observed for UW2. The cost curves increase monotonically and rapidly as the distance increases.
    }
    \label{fig:16}
\end{center}

\clearpage

\section{Discussion}
\label{sec:discussion}
In this section, we compare our proposed scheme with three distinct classes of quantum repeater architectures. The first is the memory-less architecture, which does not rely on long-lived quantum memories and therefore avoids the need for dedicated memory hardware and the associated storage loss. This class originated from the pioneering proposal of an all-photonic quantum repeater by Azuma \textit{et al.}~\cite{azuma2015all}, which introduced the concept of long-distance entanglement generation without the use of quantum memories. In contrast to the original approach of Azuma \textit{et al.}, our scheme employs elementary entangled Bell pairs protected by both the GKP code and the $[[7,1,3]]$ Steane code.

The second is the logical--physical architecture. This class has been investigated, for example, by Rozp\k{e}dek et al.~\cite{rozpkedek2023all}, in which each elementary entangled Bell pair consists of one logical qubit encoded using both the GKP code and the $[[7,1,3]]$ Steane code, and one physical GKP qubit.

The third is the second-generation memory-based all-photonic repeater architecture, such as the one proposed by H\"{a}ussler and van Loock~\cite{haussler2025long}. In this architecture, the first-stage entanglement swapping at minor nodes is probabilistic, so multiple attempts are generally required before neighboring entangled links are successfully established. Therefore, in addition to comparing the rate per protocol run, it is also important to compare the rate per unit time.

Subsections~\ref{subsec:memoryless} and~\ref{subsec:LP} compare our scheme with the memory-less and logical--physical architectures, respectively. Subsection~\ref{subsec:SA} examines the sensitivity of our scheme to key simulation parameters. Finally, Subsection~\ref{subsec:Loock} compares our scheme with the second-generation memory-based all-photonic architecture in terms of the rate per unit time.

\subsection{Memory-less Architecture}
\label{subsec:memoryless}

We illustrate this scheme in Fig.~\ref{fig:17}. Compared with our proposed scheme, each repeater requires only half as many elementary entangled Bell pairs. Furthermore, whereas each graph state in our proposed scheme is partitioned into \textit{inner-leaves} and \textit{outer-leaves}, in the memory-less architecture it is instead partitioned into two \textit{outer-leaves}. These \textit{outer-leaves} are then transmitted to adjacent minor nodes through the connecting optical fibers, where \textit{outer-leaves swapping} is performed. All entanglement swapping operations are carried out simultaneously.

\begin{center}
\begin{minipage}{1\linewidth}
\centering
\includegraphics[width=1\linewidth]{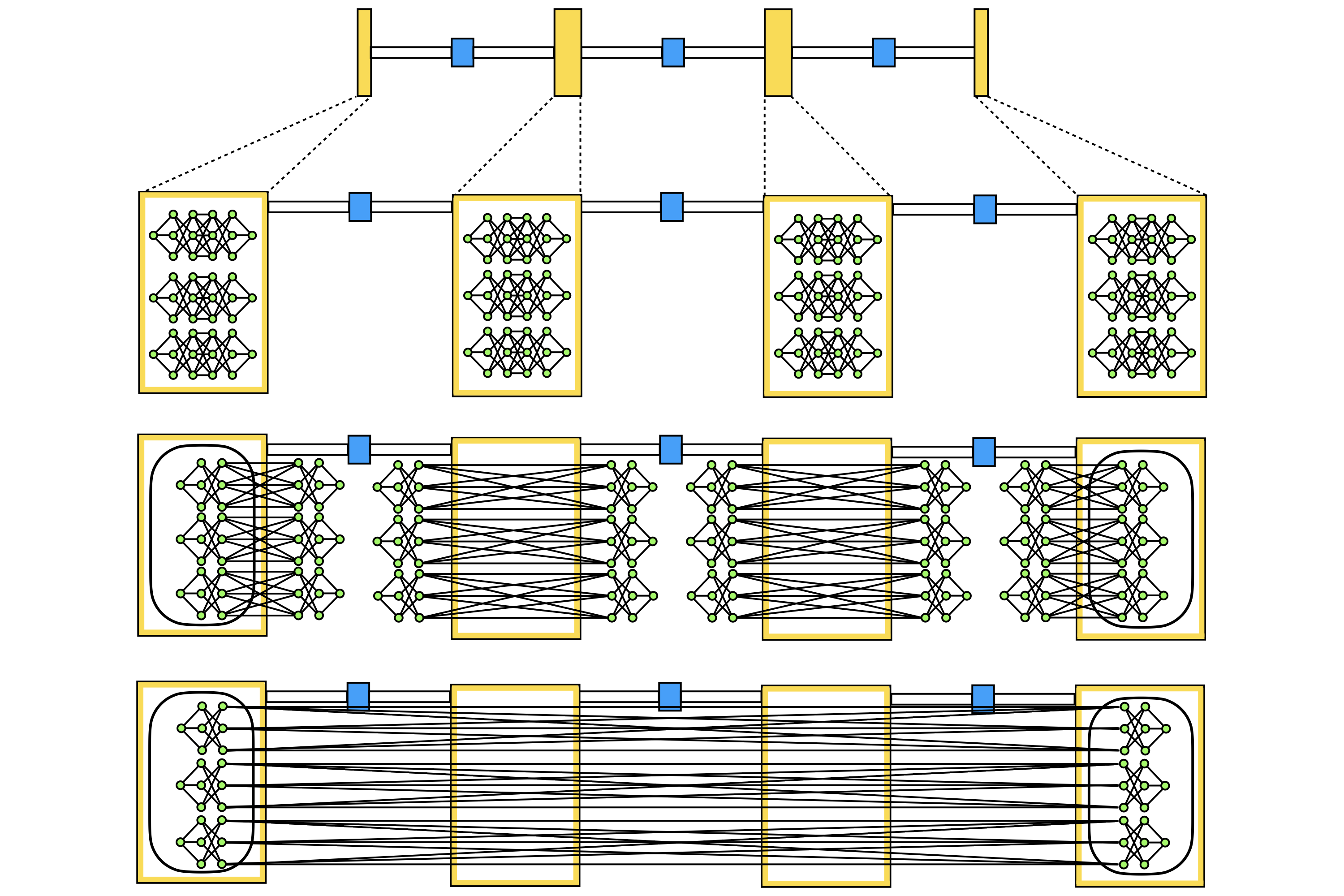}
\captionsetup{hypcap=false}
\captionof{figure}{
    Illustration of the full protocol procedure for the memory-less architecture. The top layer depicts the first step, where each factory within the repeater stations (yellow boxes) generates a sufficient number of elementary entangled Bell pairs. In the second layer, the outer-leaves are transmitted to adjacent minor nodes (blue squares) via optical fibers. At each minor node, outer-leaves swapping is performed. All entanglement swapping operations are carried out simultaneously. This illustration shows the case of multiplexing level $k=3$.
}
  \label{fig:17}
\end{minipage}
\end{center}

%%%%%Rate%%%%%%

We present the rate performance results in Fig.~\ref{fig:18}, which show how $R_{\mathrm{protocol\,run}}$ varies with the distance between two adjacent repeater stations. The solid curves correspond to results obtained using UW3, while the dashed curves other than the blue one correspond to those obtained using UW2. The blue dashed curve represents the PLOB bound~\cite{pirandola2017fundamental}. The multiplexing level is fixed at $k=15$. 

\begin{center}
\begin{minipage}{1\linewidth}
\centering
\includegraphics[width=1\linewidth]{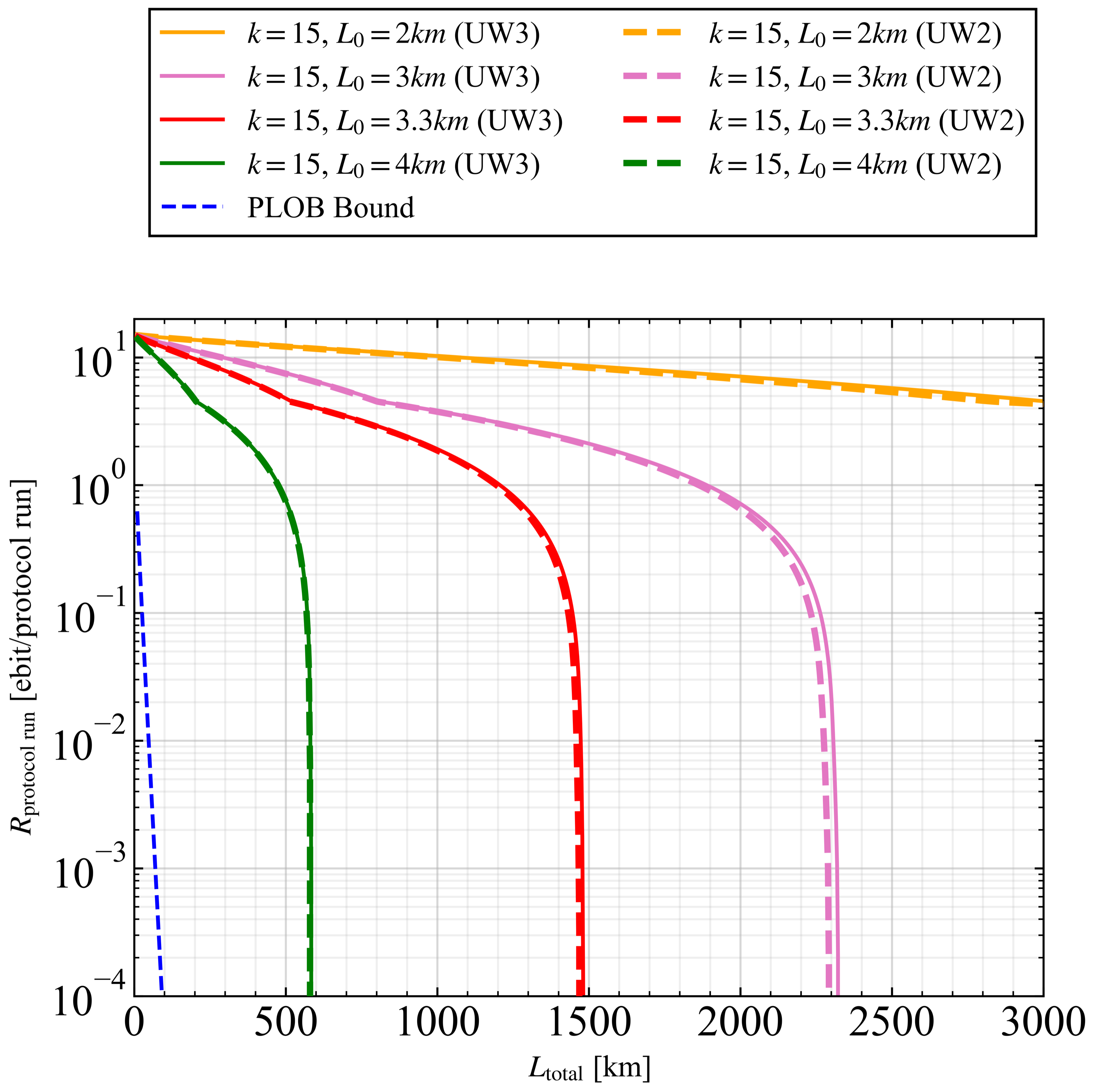}
\captionsetup{hypcap=false}
\captionof{figure}{
        The total secret-key (or entanglement) rate per protocol run, $R_{\mathrm{protocol\,run}}$, as a function of the end-to-end communication distance, $L_{\mathrm{total}}$. All solid curves correspond to results obtained using UW3, while the dashed curves (excluding the blue one) correspond to those obtained using UW2. The blue dashed curve represents the PLOB bound~\cite{pirandola2017fundamental}. Each data point was obtained from Monte Carlo simulations. Here, $v_7=0.3$ is used for all curves; increasing $v_7$ may slightly improve the rate, but it also increases the required number of GKP qubits.
    }
    \label{fig:18}
\end{minipage}
\end{center}

To achieve the same performance as our proposed scheme with $L_{0} = 9~\mathrm{km}$, the repeater spacing $L_{0}$ must be set to $3.3~\mathrm{km}$ or less.

%%%%%Number%%%%%%

We then present the initial GKP qubit requirement per repeater station per protocol run in Fig.~\ref{fig:19}, which shows how this requirement varies with the end-to-end communication distance $L_{\mathrm{total}}$. Each data point was obtained analytically by tracking the state transitions of a Markov chain.

\begin{center}
\includegraphics[width=1\linewidth]{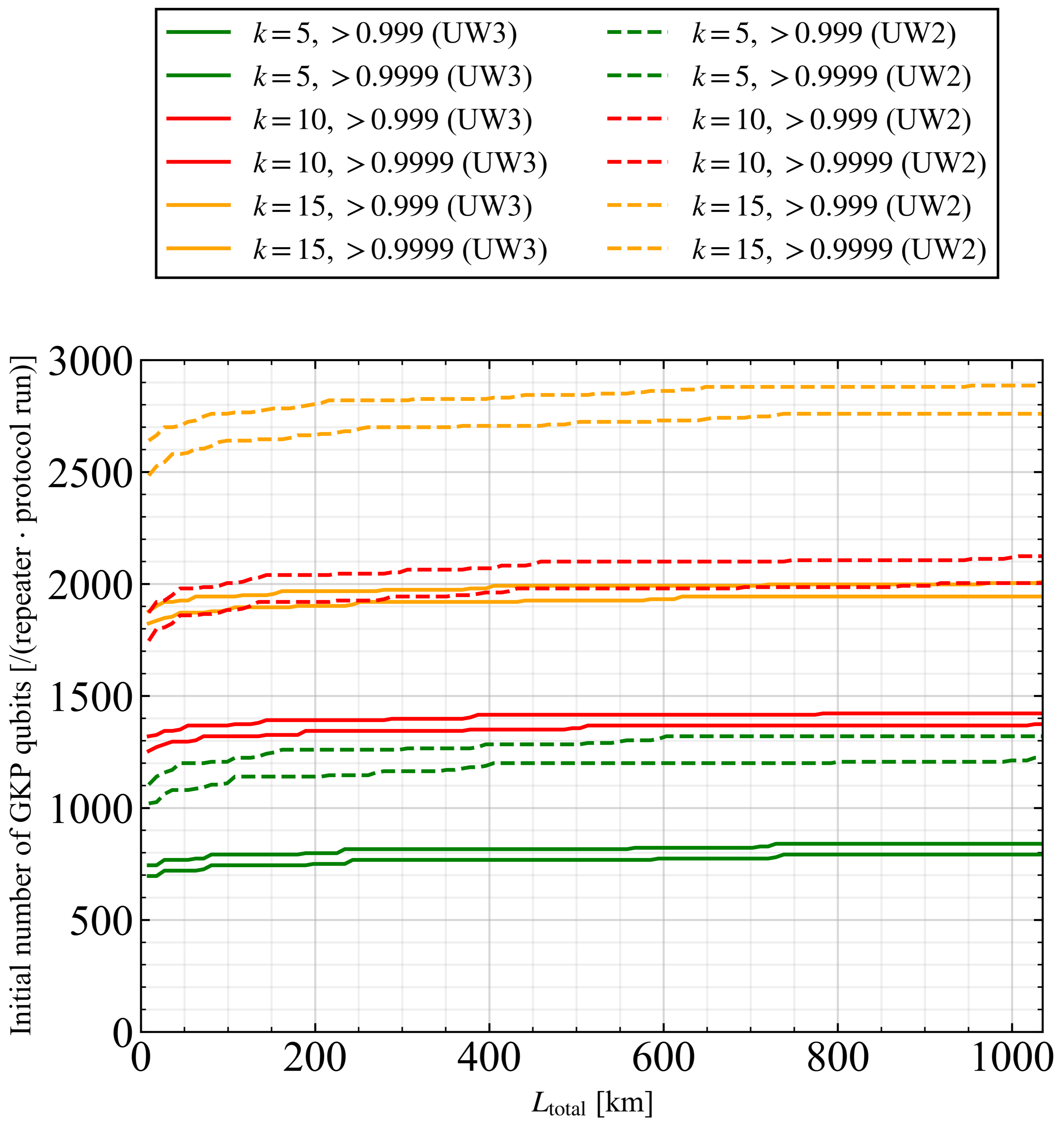}
\captionsetup{hypcap=false}
\captionof{figure}{
        The initial GKP qubit requirement per repeater station per protocol run---\textit{Initial number of GKP qubits (>0.999)} and \textit{Initial number of GKP qubits (>0.9999)}---as a function of the end-to-end communication distance, $L_{\mathrm{total}}$. Each data point was obtained analytically by tracking the state transitions of a Markov chain. The solid and dashed curves correspond to results obtained using UW3 and UW2, respectively. Between the two curves with the same color and line style, the lower one corresponds to \textit{Initial number of GKP qubits (>0.999)}, while the upper one corresponds to \textit{Initial number of GKP qubits (>0.9999)}.
    }
    \label{fig:19}
\end{center}

The solid curves correspond to results obtained using UW3, while the dashed curves correspond to those obtained using UW2. For each multiplexing level $k$, we calculate two quantities representing the initial GKP qubit requirement---\textit{Initial number of GKP qubits (>0.999)} and \textit{Initial number of GKP qubits (>0.9999)}---as defined in Section~\ref{sec:results}. Between the two curves with the same color and line style, the lower one corresponds to \textit{Initial number of GKP qubits (>0.999)}, while the upper one corresponds to \textit{Initial number of GKP qubits (>0.9999)}.

Although each repeater requires only half as many elementary entangled Bell pairs as in Fig.~\ref{fig:10}, the initial GKP qubit requirement remains greater than half of that in Fig.~\ref{fig:10}. This deviation from the naive factor-of-two reduction becomes more pronounced as the multiplexing level $k$ increases or when UW2 is used.

%%%%%Summary%%%%%%

In summary, although the absence of quantum memories provides a significant advantage, the initial GKP qubit requirement remains greater than half of that in our proposed scheme, while the repeater spacing must be reduced to less than half. Therefore, when both the initial GKP qubit requirement and the repeater spacing are taken into account, our proposed scheme offers a clear advantage over the memory-less architecture.

%%%%%%%%%%%%%%%%%%

\subsection{Logical--Physical Architecture}
\label{subsec:LP}

\begin{figure}[ht]
  \centering
  \includegraphics[width=1\linewidth]{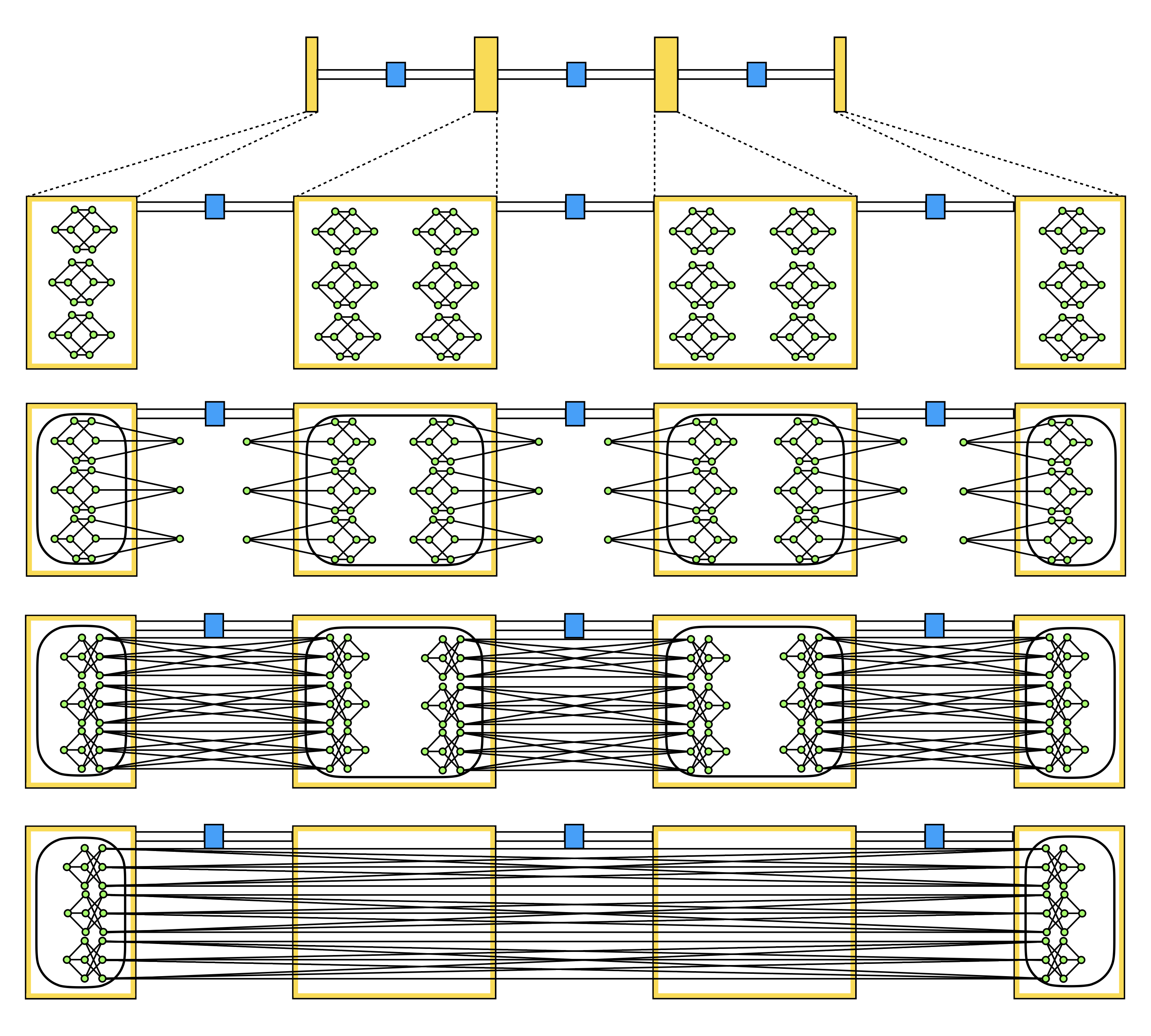}
  \caption{
    Illustration of the full protocol procedure for the logical--physical architecture. The top layer depicts the first step, where each factory within the repeater stations (yellow boxes) generates a sufficient number of elementary entangled Bell pairs. In the second layer, the outer-leaf components are transmitted to adjacent minor nodes (blue squares) via optical fibers. At each minor node, outer-leaf swapping is performed, and the resulting network of entangled links between neighboring repeater stations is shown in the third layer. The analog information from these measurements is sent back to the originating repeater stations. In the bottom layers, the optimal pairs of inner-leaves are selected based on the received analog information and subsequently processed via inner-leaves swapping, thereby establishing long-distance entanglement between the end users. This illustration shows the case of multiplexing level $k=3$.
}
  \label{fig:20}
\end{figure}

We illustrate this scheme in Fig.~\ref{fig:20}. Compared with our proposed scheme, each elementary entangled Bell pair in this architecture consists of two components: an \textit{outer-leaf} component and an \textit{inner-leaves} component, where the \textit{outer-leaf} is a single GKP qubit. This elementary entangled Bell pair is referred to as the $G_5^{*}$ state and is explained in detail in Section~\ref{sec:construction}. Its construction follows UW3 up to an intermediate stage, as depicted in Fig.~\ref{fig:3}(b).

Furthermore, within this logical--physical architecture, two variants can be considered for how the inner-leaves wait for the completion of \textit{outer-leaves swapping} and the return of the associated analog information to the repeater station. The differences in the internal configurations of the repeater station between these two variants are summarized in Fig.~\ref{fig:21}.

The first variant employs the same waiting strategy as our proposed scheme, in which the inner-leaves are stored in a free-space photonic memory, as illustrated in Fig.~\ref{fig:21}(a).

The second variant follows the scheme proposed by Rozp\k{e}dek \textit{et al.}~\cite{rozpkedek2023all}, in which the inner-leaves wait inside spools and teleportation-based error correction (TEC) is applied periodically as they propagate over a fixed distance, as illustrated in Fig.~\ref{fig:21}(b). In this figure, the optical fiber colored blue represents the spool, which is typically a few hundred meters long. Based on the optimization over the candidate TEC spacings of $100, 125, 250, 500,\text{ and }1000~\mathrm{m}$, we fix the TEC spacing at $125~\mathrm{m}$ in the following discussion.

\begin{center}
\begin{minipage}{1\linewidth}
\centering
\includegraphics[width=1\linewidth]{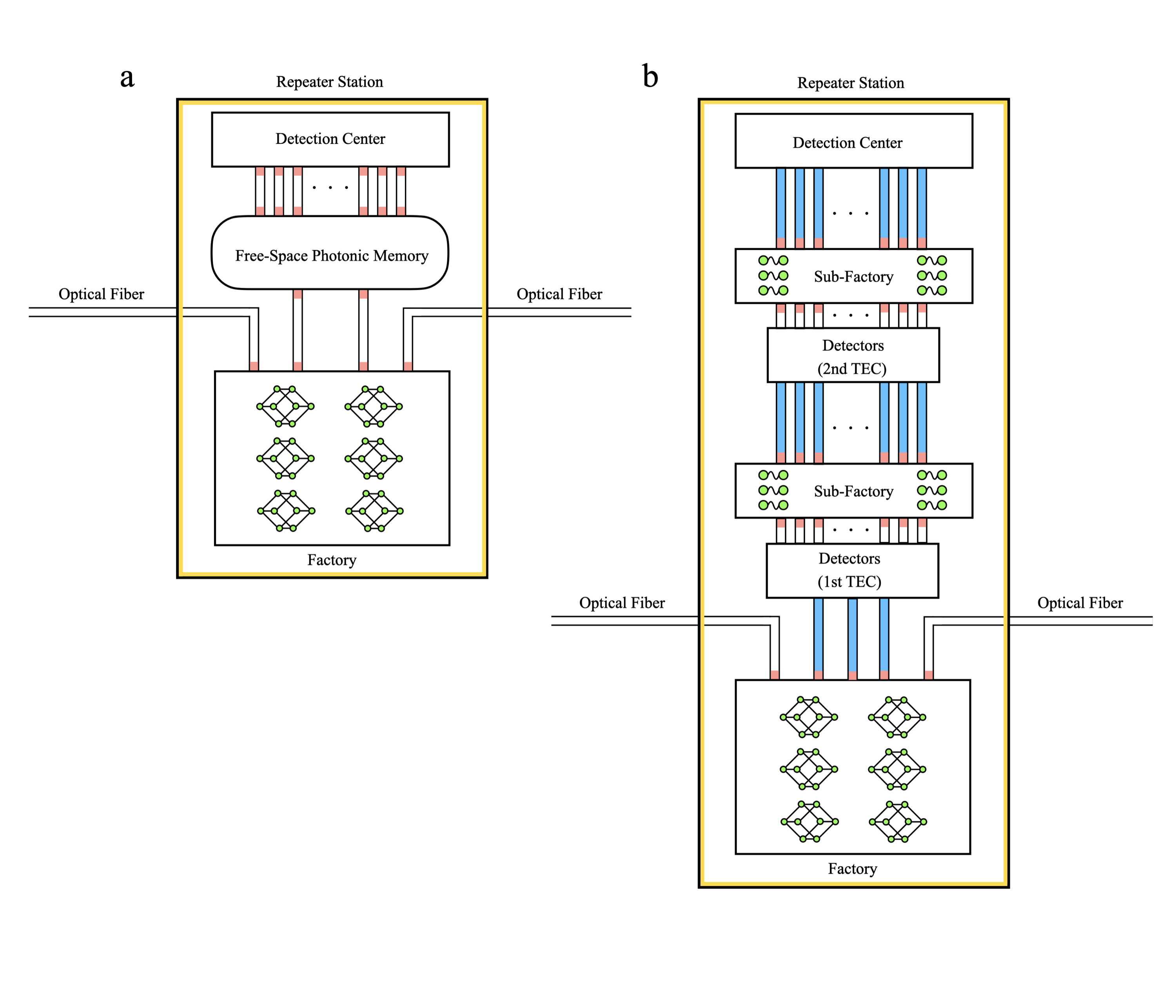}
\captionsetup{hypcap=false}
\captionof{figure}{
    (a) Internal configuration of a repeater station with a free-space photonic memory, consisting of a factory module that produces a sufficient number of elementary entangled Bell pairs, a free-space photonic memory module that serves as a temporary buffer for the inner-leaves, and a detection center module that performs Bell measurements. (b) Internal configuration of a repeater station with spools, consisting of a factory module that produces a sufficient number of elementary entangled Bell pairs, sub-factory modules that produce a sufficient number of Bell pairs for TEC, detector modules composed of beam splitters and homodyne detectors (corresponding to small minor nodes), and a detection center module that performs Bell measurements. Optical fibers colored blue represent the spool, which is typically a few hundred meters long. In both (a) and (b), the light red square boxes represent the connectors between the optical fibers and the quantum chips or the photonic memories.
}
  \label{fig:21}
\end{minipage}
\end{center}

%%%%%Rate%%%%%%

We present the rate performance results in Fig.~\ref{fig:22}, which show how $R_{\mathrm{protocol\,run}}$ varies with the distance between two adjacent repeater stations. The solid curves correspond to the free-space photonic memory variant, while the dashed curves (excluding the blue one) correspond to the spools variant. The blue dashed curve represents the PLOB bound~\cite{pirandola2017fundamental}. The multiplexing level is fixed at $k=15$. 

\begin{center}
\begin{minipage}{1\linewidth}
\centering
\includegraphics[width=1\linewidth]{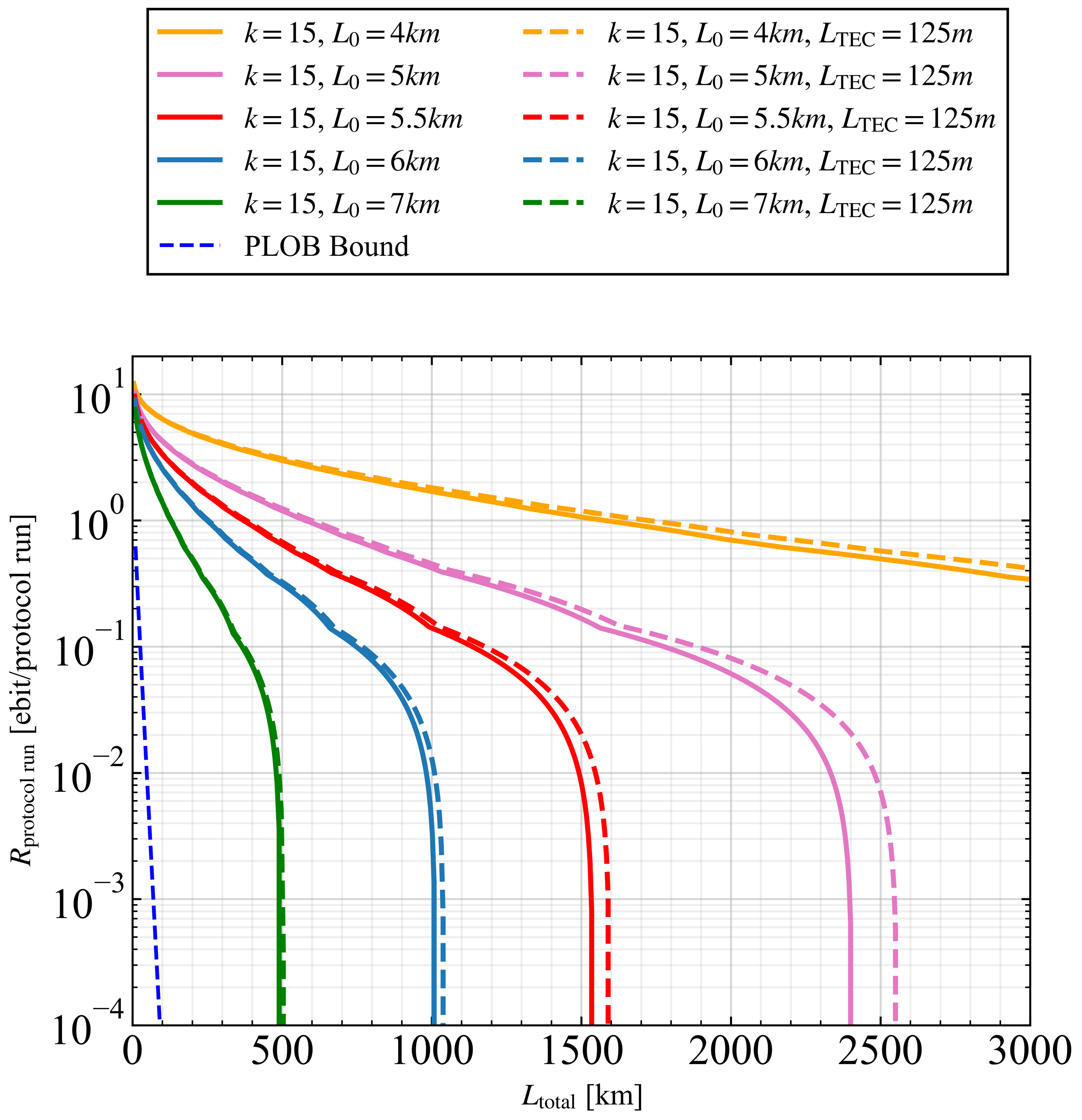}
\captionsetup{hypcap=false}
\captionof{figure}{
        The total secret-key (or entanglement) rate per protocol run, $R_{\mathrm{protocol\,run}}$, as a function of the end-to-end communication distance, $L_{\mathrm{total}}$. All solid curves correspond to the free-space photonic memory variant, while the dashed curves (excluding the blue one) correspond to the spools variant. The blue dashed curve represents the PLOB bound~\cite{pirandola2017fundamental}. Each data point was obtained from Monte Carlo simulations. Here, $v_7=0.3$ is used for all curves; increasing $v_7$ may slightly improve the rate, but it also increases the required number of GKP qubits.
    }
    \label{fig:22}
\end{minipage}
\end{center}

To achieve the same performance as our proposed scheme with $L_{0} = 9~\mathrm{km}$, the repeater spacing $L_{0}$ must be set to $5.5~\mathrm{km}$ or less in both variants. Although the spools variant appears to outperform the free-space photonic memory variant in terms of $R_{\mathrm{protocol\,run}}$ alone, note that the reliance on TEC in spools requires an additional 18,480 GKP qubits in the case of $L_{0} = 5.5~\mathrm{km}$.

%%%%%Number%%%%%%

We then present the initial GKP qubit requirement per repeater station per protocol run in Fig.~\ref{fig:23}, which shows how this requirement varies with the end-to-end communication distance $L_{\mathrm{total}}$. This quantity counts only the GKP qubits required at the beginning of each protocol run to construct the elementary entangled Bell pairs and excludes the additional GKP qubits consumed by TEC in the spools variant. Each data point was obtained analytically by tracking the state transitions of a Markov chain. Since the construction of the elementary entangled Bell pairs is common to both variants, the corresponding curves coincide.

\begin{center}
\begin{minipage}{1\linewidth}
\centering
\includegraphics[width=1\linewidth]{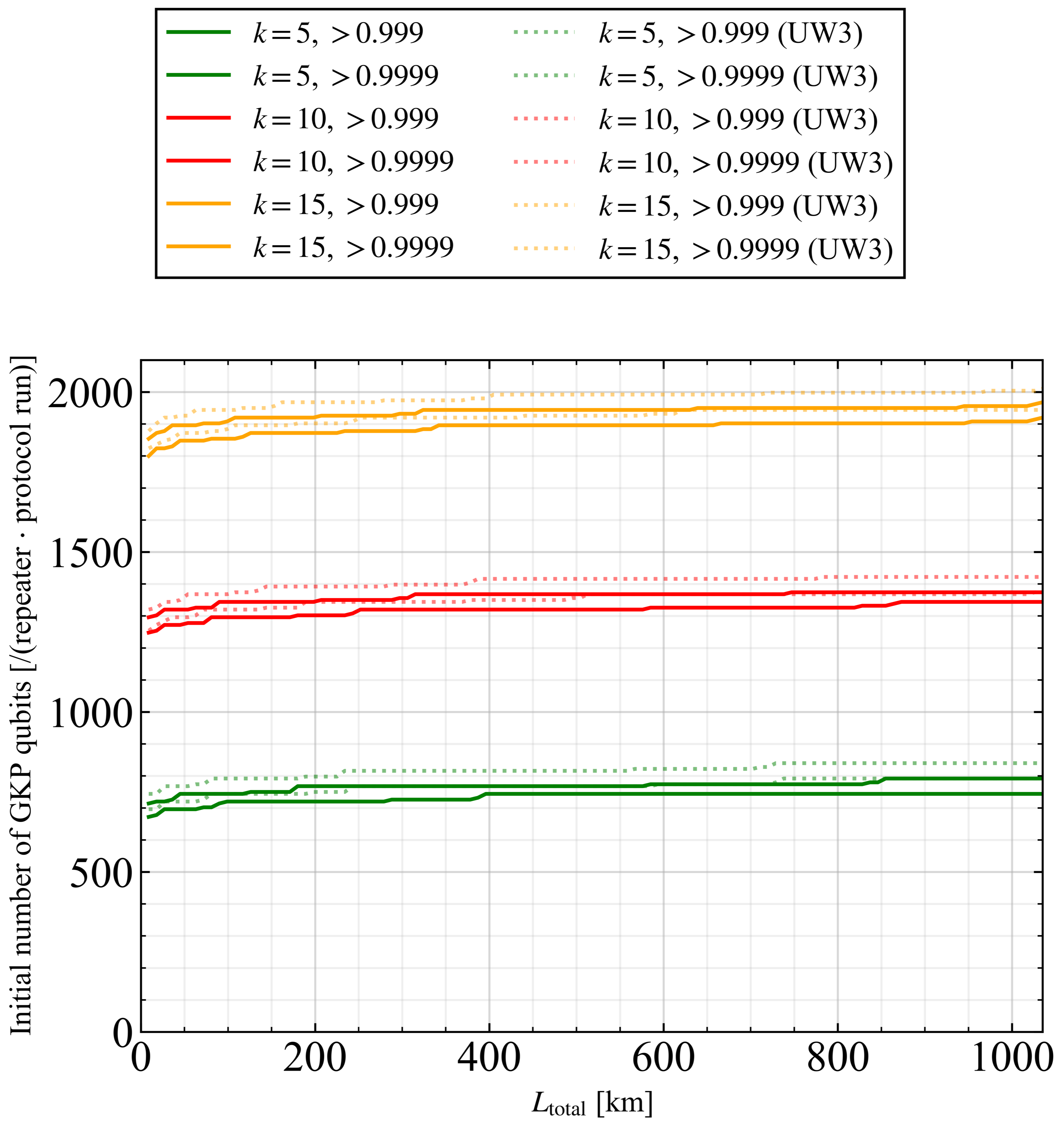}
\captionsetup{hypcap=false}
\captionof{figure}{
        The initial GKP qubit requirement per repeater station per protocol run as a function of the end-to-end communication distance, $L_{\mathrm{total}}$. This quantity counts only the GKP qubits required at the beginning of each protocol run to construct the elementary entangled Bell pairs and excludes the additional GKP qubits consumed by TEC in the spools variant. Each data point was obtained analytically by tracking the state transitions of a Markov chain. The solid and dotted curves correspond to the logical--physical architecture and the memory-less architecture, respectively.
    }
    \label{fig:23}
\end{minipage}
\end{center}

The solid curves represent the results for the logical--physical architecture. For reference, the results for the memory-less architecture are also shown as dotted curves, identical to those in Fig.~\ref{fig:19}. Although the logical--physical architecture has a slightly smaller initial GKP qubit requirement than the memory-less architecture, this requirement remains greater than half of that in Fig.~\ref{fig:10}. This deviation from the naive factor-of-two reduction becomes more pronounced as the multiplexing level $k$ increases or when UW2 is used.

In summary, although the logical--physical architecture has a slightly smaller initial GKP qubit requirement than the memory-less architecture, the initial GKP qubit requirement remains greater than half of that in our proposed scheme. In addition, the achievable repeater spacing still falls short of that of our proposed scheme. Therefore, when both the initial GKP qubit requirement and the repeater spacing are taken into account, our proposed scheme offers an advantage over the logical--physical architecture.

%\clearpage

\subsection{Sensitivity Analysis}
\label{subsec:SA}

In this subsection, we investigate how the secret-key (or entanglement) rate depends on key simulation parameters listed in Table~\ref{tab:sim-params}---namely $\sigma_{\mathrm{GKP}}$, $\eta_s$, and $\eta_m$---to clarify the practical operating regime of the proposed architecture.

Table~\ref{tab:sim-params} summarizes the parameters used in Sections~\ref{sec:results} and~\ref{sec:discussion}. These baseline values are selected based on representative values reported in the literature~\cite{pant2017rate, lenzini2018integrated}. In the following, we quantify how the achievable rate changes when each parameter is varied individually relative to its baseline value.

First, we examine the sensitivity of the rate to $\sigma_{\mathrm{GKP}}$ in Fig.~\ref{fig:x1}. The solid curves correspond to $\sigma_{\mathrm{GKP}} = 0.12$, the dashed curves to $\sigma_{\mathrm{GKP}} = 0.13$, and the dotted curves to $\sigma_{\mathrm{GKP}} = 0.14$. The rate is sensitive to increases in $\sigma_{\mathrm{GKP}}$, with the degradation becoming more pronounced at shorter repeater spacings.

\begin{center}
\begin{minipage}{1\linewidth}
\centering
\includegraphics[width=1\linewidth]{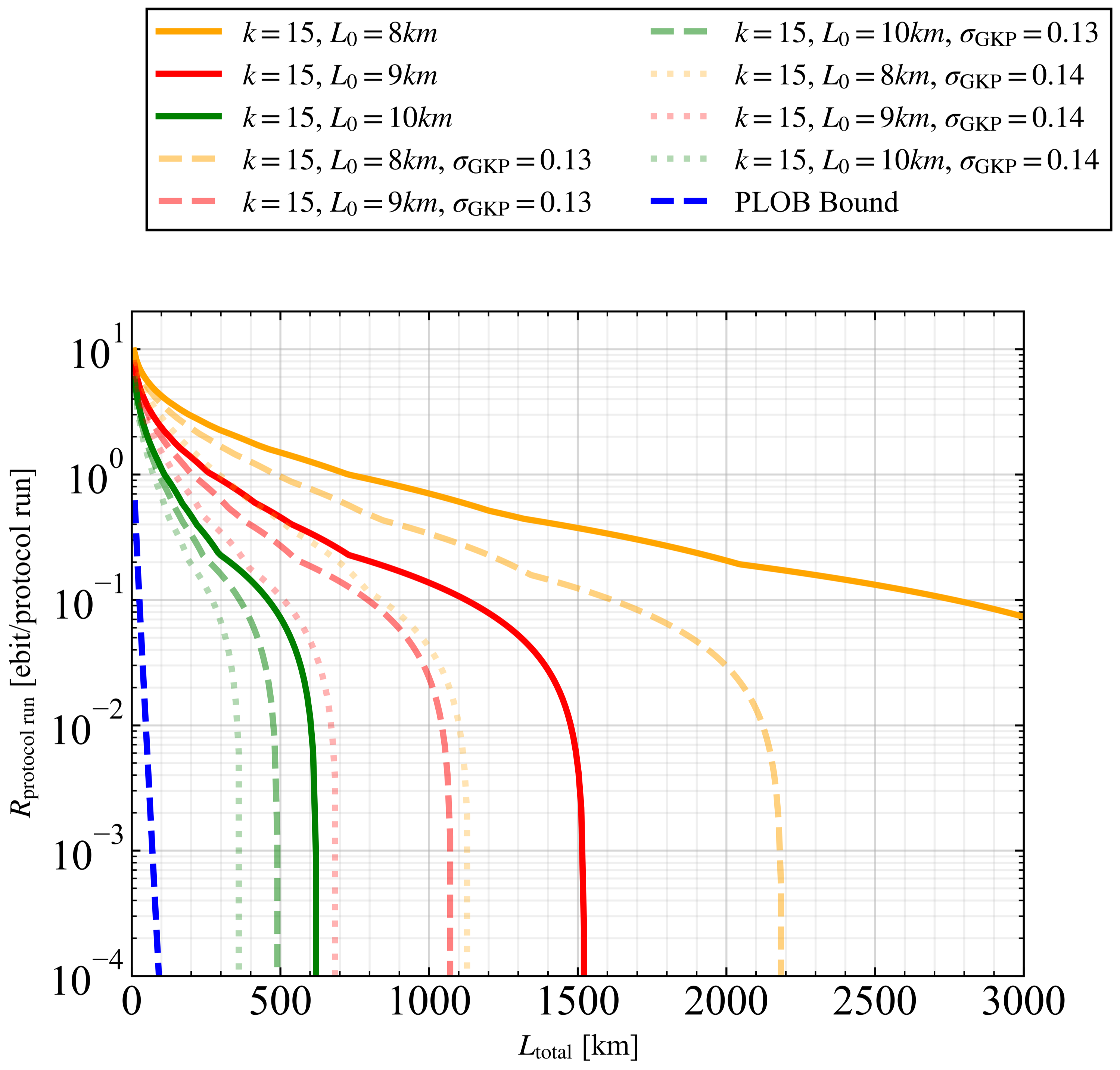}
\captionsetup{hypcap=false}
\captionof{figure}{
Sensitivity analysis of the achievable secret-key (or entanglement) rate with respect to $\sigma_{\mathrm{GKP}}$. The solid, dashed, and dotted curves correspond to $\sigma_{\mathrm{GKP}} = 0.12$, $0.13$, and $0.14$, respectively. The blue dashed curve represents the PLOB bound~\cite{pirandola2017fundamental}. Each data point was obtained from Monte Carlo simulations. Here, $v_7=0.3$ is used for all curves; increasing $v_7$ may slightly improve the rate, but it also increases the required number of GKP qubits.
}
    \label{fig:x1}
\end{minipage}
\end{center}

Next, we examine the sensitivity of the rate to $\eta_{s}$ in Fig.~\ref{fig:x2}. The solid curves correspond to $\eta_{s} = 0.995$, the dashed curves to $\eta_{s} = 0.99$, and the dotted curves to $\eta_{s} = 0.985$. The rate is sensitive to decreases in $\eta_{s}$, with the degradation becoming more pronounced at shorter repeater spacings.

\begin{center}
\begin{minipage}{1\linewidth}
\centering
\includegraphics[width=1\linewidth]{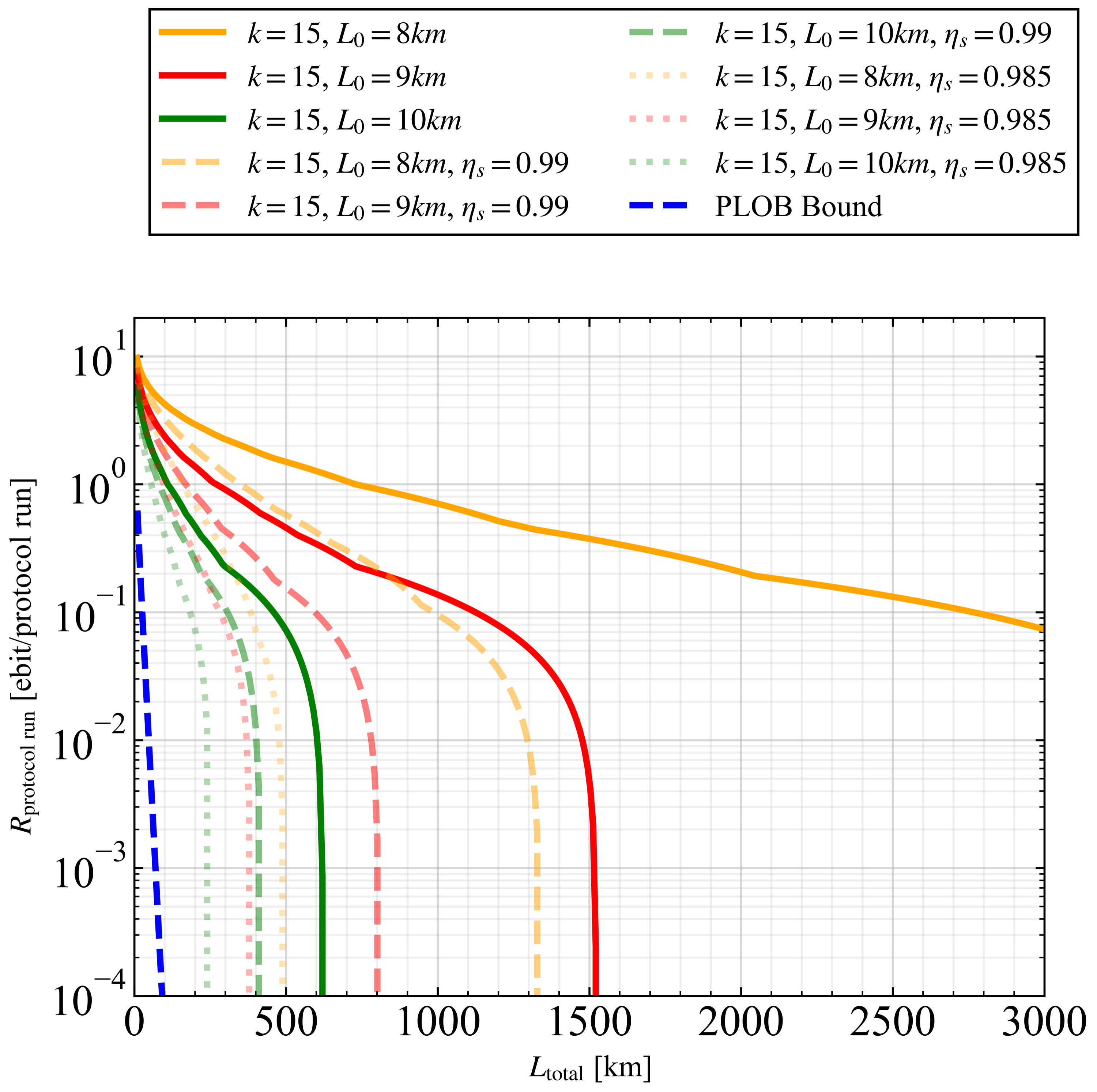}
\captionsetup{hypcap=false}
\captionof{figure}{
Sensitivity analysis of the achievable secret-key (or entanglement) rate with respect to $\eta_{s}$. The solid, dashed, and dotted curves correspond to $\eta_{s} = 0.995$, $0.99$, and $0.985$, respectively. The blue dashed curve represents the PLOB bound~\cite{pirandola2017fundamental}. Each data point was obtained from Monte Carlo simulations. Here, $v_7=0.3$ is used for all curves; increasing $v_7$ may slightly improve the rate, but it also increases the required number of GKP qubits.
}
    \label{fig:x2}
\end{minipage}
\end{center}

Finally, we examine the sensitivity of the rate to $\eta_{m}$ in Fig.~\ref{fig:x3}. The solid, dashed, dash-dotted, and dotted curves correspond to $\eta_{m} = 0.999995$, $0.999994$, $0.999993$, and $0.999992$, respectively. The rate is highly sensitive to decreases in $\eta_{m}$, with the degradation becoming more pronounced at shorter repeater spacings. Motivated by this sensitivity, we additionally plot the relationship between $1-\eta_{m}$ and the required $L_{\mathrm{cavity}}$ in Fig.~\ref{fig:x4} to maintain the baseline performance shown in Section~\ref{sec:results} (i.e., the solid curves).

\begin{center}
\begin{minipage}{1\linewidth}
\centering
\includegraphics[width=1\linewidth]{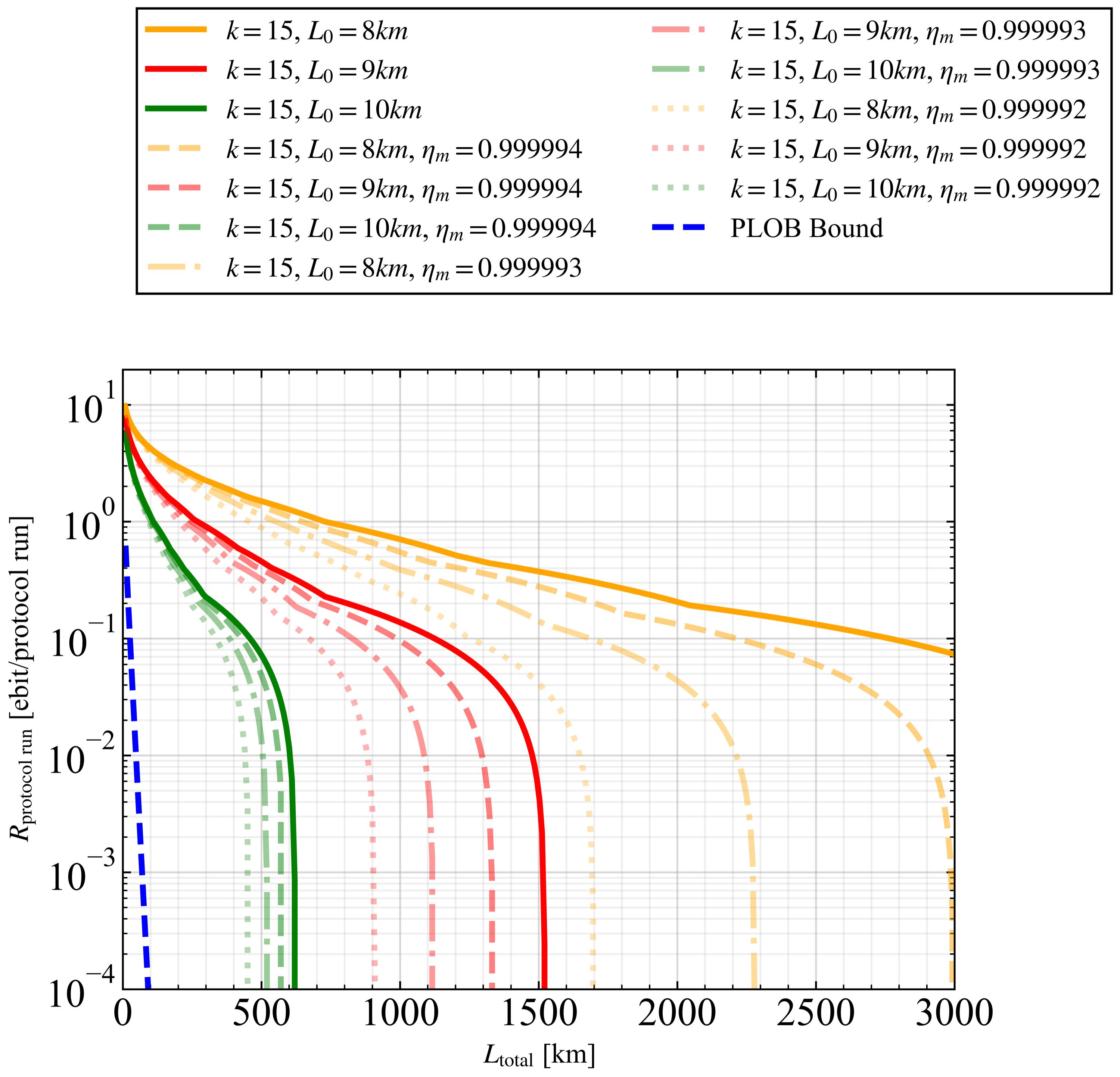}
\captionsetup{hypcap=false}
\captionof{figure}{
Sensitivity analysis of the achievable secret-key (or entanglement) rate with respect to $\eta_{m}$. The solid, dashed, dash-dotted, and dotted curves correspond to $\eta_{m} = 0.999995$, $0.999994$, $0.999993$, and $0.999992$, respectively. The blue dashed curve represents the PLOB bound~\cite{pirandola2017fundamental}. Each data point was obtained from Monte Carlo simulations. Here, $v_7=0.3$ is used for all curves; increasing $v_7$ may slightly improve the rate, but it also increases the required number of GKP qubits.
}
    \label{fig:x3}
\end{minipage}
\end{center}

The common trend observed across Figs.~\ref{fig:x1}--\ref{fig:x3}, namely that the rate becomes less sensitive to variations in the parameters as the repeater spacing increases, is consistent with the expectation that errors from outer-leaves swapping---particularly those accumulated during propagation through the optical fiber---tend to dominate, thereby reducing the relative sensitivity of the rate to these parameters.

Fig.~\ref{fig:x4} shows that the required $L_{\mathrm{cavity}}$ is highly sensitive to $\eta_{m}$. Given that the cavity must be implemented within a repeater station, this result suggests that, under the baseline parameter values considered here, a mirror reflectivity exceeding $0.9999$ is required to maintain the baseline performance.

Consequently, $\eta_m$ emerges as the primary practical constraint. 
Maintaining the baseline performance reported in this work requires sufficiently high $\eta_m$ or a larger $L_{\mathrm{cavity}}$ to reduce the number of bounces for a given waiting time.

\begin{center}
\begin{minipage}{1\linewidth}
\centering
\includegraphics[width=1\linewidth]{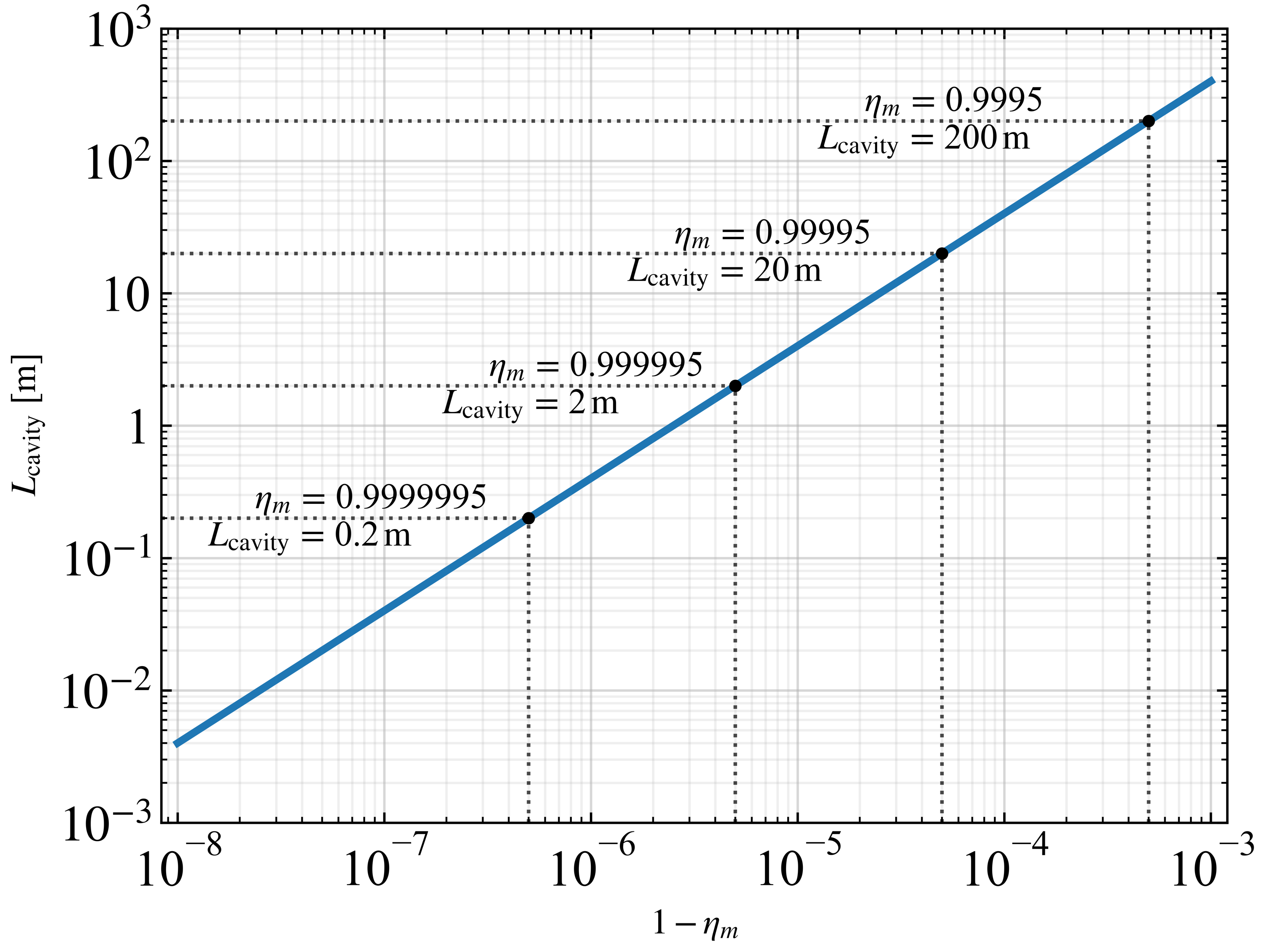}
\captionsetup{hypcap=false}
\captionof{figure}{
        Log--log plot of the relationship between $1-\eta_m$ and the required $L_{\mathrm{cavity}}$ to maintain the baseline performance shown in Section~\ref{sec:results} (i.e., the solid curves in Figs.~\ref{fig:x1}--\ref{fig:x3}). The results highlight the strong sensitivity of the required $L_{\mathrm{cavity}}$ to small deviations in $\eta_m$.
    }
    \label{fig:x4}
\end{minipage}
\end{center}

\subsection{Comparison in Terms of Rate per Unit Time}
\label{subsec:Loock}

The preceding analysis has primarily considered the secret-key (or entanglement) rate per protocol run. However, when comparing architectures with different temporal structures, this quantity alone does not fully characterize the practical communication rate. In particular, our architecture is designed to generate end-to-end entangled Bell pairs within a single protocol run. By contrast, in second-generation memory-based all-photonic repeater architectures, such as the one proposed by H\"{a}ussler and van Loock~\cite{haussler2025long}, the first-stage entanglement swapping at the minor nodes is probabilistic, so multiple attempts are generally required before neighboring entangled links are successfully established. For this reason, a comparison in terms of the rate per unit time is necessary.

Let us denote the secret-key (or entanglement) rate per unit time by $R_{\mathrm{time}}$. This quantity can be written as
\begin{equation}
\begin{split}
R_{\mathrm{time}} = f P R_{\mathrm{protocol\,run}} = f P k R_{\mathrm{channel}},
\end{split}
\end{equation}
where $f$ is the repetition rate, determined by the slowest device or process in the architecture, and $P$ denotes the inverse of the average number of attempts required to establish all entangled links between neighboring repeater stations. In our proposed scheme, the first-stage entanglement swapping, which corresponds to outer-leaves swapping, is deterministic, and therefore $P=1$. Here, $R_{\mathrm{channel}}$ corresponds to the secret-key fraction (SKF), $r$, while $fPk$ corresponds to the raw rate, $R$, in the notation of H\"{a}ussler and van Loock's paper~\cite{haussler2025long}.

For our proposed scheme, we first consider the case in which the analog information from the minor nodes is returned to the repeater stations through a separate, non-blocking classical communication channel. In this case, the return of the analog information does not block the transmission of the next set of outer-leaves, enabling pipelined operation. We therefore assume that the repetition rate is limited by the time required to prepare the initial GKP qubits that need to be available at the beginning of each protocol run. In contrast, the refreshment resources are scheduled to become available with a slight delay relative to the initial GKP qubits, and therefore need not be available at the beginning of the protocol run. These refreshment resources are expected to be supplied to a significant extent through a combination of the GKP qubit recycling procedure and the sequential refreshment strategy introduced in Subsection~\ref{subsec:recycling}. For $L_{\mathrm{total}} = 1{,}000~\mathrm{km}$, $L_0 = 9~\mathrm{km}$, and $k=15$, assuming a repetition rate of $f = 100~\mathrm{kHz}$ yields
\begin{equation}
\begin{split}
R_{\mathrm{time}} = 13.8\times 10^3~\mathrm{ebit/s}.
\end{split}
\end{equation}

We next consider the blocking case in which the return of the analog information from the minor nodes delays the transmission of the next set of outer-leaves. This situation may arise, for example, when the analog information is returned through the same optical fiber path as the outer-leaves and the next outer-leaves transmission is scheduled so as to avoid possible crosstalk between the outgoing quantum signals and the returning classical signals. It may also arise when the available transmission resources in the optical fiber are insufficient to support both directions simultaneously. In this case, the transmission of the next set of outer-leaves must wait until the analog information from the previous outer-leaves swapping has returned to the repeater stations. The repetition rate is therefore limited by the time required for the outer-leaves to travel from the repeaters to the minor nodes and for the corresponding analog information to return from the minor nodes to the repeaters. This assumes that the initial GKP qubits needed at the beginning of each protocol run can be prepared within this communication time. Since this communication time is $\tau_0 = L_0/c_{\mathrm{fiber}}$, the repetition rate is limited to $f = 1/\tau_0 \approx 22.2~\mathrm{kHz}$. Under this assumption, the corresponding rate per unit time is 
\begin{equation}
\begin{split}
R_{\mathrm{time}} = 3.07 \times 10^3~\mathrm{ebit/s}.
\end{split}
\end{equation}

On the other hand, for the scheme of H\"{a}ussler and van Loock, we use the same conditions, $L_{\mathrm{total}} = 1{,}000~\mathrm{km}$ and $L_0 = 9~\mathrm{km}$. In their scheme, however, the multiplexing level $k$ is fixed at one, because each successful protocol run can generate at most one end-to-end entangled Bell pair. Moreover, because the first-stage entanglement swapping is probabilistic, $fPk$ is evaluated using H\"{a}ussler and van Loock's raw-rate expression:
\begin{equation}
\begin{split}
f P k = \left[ \tau_0 \sum_{i=1}^{n} (-1)^{i+1} \binom{n}{i} \frac{1}{1-(1-p)^i} \right]^{-1},
\end{split}
\end{equation}
where $p$ is the success probability for establishing an entangled link between neighboring repeater stations in a single attempt, and $n=L_{\mathrm{total}}/L_0$ is the number of segments.

To give an optimistic estimate for their scheme, we assume that quantum error correction is sufficiently effective that the SKF is close to unity and set $R_{\mathrm{channel}}=1$. Under this assumption, for $L_{\mathrm{total}} \approx 1{,}000~\mathrm{km}$ with $L_0 = 9~\mathrm{km}$ and $n=111$ segments, we obtain
\begin{equation}
\begin{split}
R_{\mathrm{time}} = 1.59 \times 10^3~\mathrm{ebit/s}.
\end{split}
\end{equation}
If the repeater spacing is increased to $L_0 = 50~\mathrm{km}$ or $L_0 = 100~\mathrm{km}$, the corresponding rates become
\begin{equation}
\begin{split}
R_{\mathrm{time}} &= 57.2~\mathrm{ebit/s}, \\
R_{\mathrm{time}} &= 3.56~\mathrm{ebit/s},
\end{split}
\end{equation}
respectively. Although these rates are lower than the rate achieved by our proposed scheme at the same total distance with $L_0=9~\mathrm{km}$, the architecture of H\"{a}ussler and van Loock has an important complementary advantage: it can operate with substantially longer repeater spacings while transmitting only single photons through the optical fibers. In this sense, their work provides a valuable alternative design direction for all-photonic quantum repeaters, particularly for regimes in which long repeater spacing is prioritized over high per-unit-time rates.

\section{Conclusion}
\label{sec:last}
In this work, we propose a novel all-photonic quantum repeater architecture based on elementary entangled Bell pairs protected by both the GKP code and the $[[7,1,3]]$ Steane code. We show that this architecture can probabilistically correct up to three bit-flip errors and three phase-flip errors in a logical qubit encoded in seven physical GKP qubits. To support efficient multiplexing, we introduce a new ranking criterion associated with the outer-leaves, which are protected by both the GKP and $[[7,1,3]]$ Steane codes. This new ranking criterion enables inner-leaves swapping to leverage the outcomes of outer-leaves swapping by ranking the multiplexing channels. We explicitly model an optical switch and incorporate a multi-reflection mirror-based optical cavity as a free-space photonic memory module into our quantum repeater architecture. Additionally, we propose two heuristic construction methods for the elementary entangled Bell pairs, referred to as UW2 and UW3. We confirm that there is no significant difference in their rate performance; however, UW3 offers a clear advantage in terms of the required number of GKP qubits, according to results obtained using a newly developed method for tracking the Markov chain of state transitions. Under the baseline parameter values considered here, our architecture achieves a repeater spacing of 9 km while requiring fewer than 3,800 GKP qubits per repeater station per protocol run.

In future work, it would be valuable to incorporate entanglement distillation into the proposed architecture. Furthermore, the free-space photonic memory opens the possibility of multi-run operation, in which entangled links are buffered and hence ranked across multiple runs of the protocol, potentially leading to an extended rate $R_{\mathrm{multiple\,run}}$. In another direction, integrating free-space photonic memory into the elementary entangled Bell-pair generation process may significantly reduce the overall hardware requirements by enabling temporal multiplexing and reducing the need for parallel generation modules. Finally, our all-photonic quantum repeater architecture can be extended to a quantum switch for generating GHZ states among multiple end users.

\section*{Acknowledgments}
We thank K.~Azuma, P.~Dhara, A.~Jha, P.~Kwiat, R.~Nehra, K.~Seshadreesan, L.~Jiang, and S.~Guha for valuable discussions. We also thank A. Garcia Flores for helpful feedback on the manuscript. We acknowledge support from the NSF (Grant No. ERC-1941583). K.~Goodenough acknowledges the support from the Alexander von Humboldt Foundation. 

\section*{Author Contributions}
R.~Shiina led the project, designed the repeater architecture and the construction methods for the elementary entangled Bell pairs, developed the simulation code, performed the numerical simulations and analytical modeling of the resource overhead, analyzed the results, prepared the figures, and wrote the manuscript. K.~Goodenough proposed the method for estimating the initial number of GKP qubits by interpreting the state transitions as a Markov chain, wrote a draft version of the corresponding code, and provided feedback on the manuscript. N.~Arnold advised on the integration, design, and experimental feasibility of the free-space photonic memory and contributed to the writing of the corresponding section. F.~Rozp\k{e}dek conceived and supervised the project, conceived the design of the repeater architecture, contributed to the development of the MATLAB simulation code, and provided feedback on the manuscript. A large language model was used to assist with grammar and clarity improvements in the manuscript.

\bibliographystyle{ieeetr}
\bibliography{Z.bib}

\clearpage
\onecolumn
\appendix
\section{Measurement Types Used in the Construction Process}
\label{sec:A}
In the construction of elementary entangled Bell pairs described in Section~\ref{sec:construction}, we use 11 measurement types, classified according to the quadrature variances of the measured qubits and the number of optical switches involved. To keep the main text focused on the overall construction methods, we summarize these measurement types in Table~\ref{tab:measurement_types}. The type numbers are assigned according to the measurement configuration; therefore, the same configuration is assigned the same type number in both UW2 and UW3. These type numbers are also used consistently throughout our simulation code.

\begin{table}[ht]
\centering
\caption{Measurement types used in the construction of elementary entangled Bell pairs.}
\label{tab:measurement_types}
\footnotesize
\renewcommand{\arraystretch}{1.8}
\setlength{\tabcolsep}{6pt}

\begin{tabular}{c|c|c}
\toprule
\makebox[2cm][c]{Type} & \makebox[6cm][c]{Variance $\sigma^2$} & \makebox[6cm][c]{Discard window size parameter $v$} \\
\midrule
\midrule
$1$ & $3\sigma^2_{\mathrm{GKP}} + \frac{1-\eta_d}{\eta_d}$ & $0.143$ \\
$2$ & $3\sigma^2_{\mathrm{GKP}} + \frac{1-\eta_s\eta_d}{\eta_s\eta_d}$ & $0.198$ \\
$3$ & $3\sigma^2_{\mathrm{GKP}} + \frac{1-\eta_s^2\eta_d}{\eta_s^2\eta_d}$ & $0.251$ \\
$4$ & $2\sigma^2_{\mathrm{GKP}} + \frac{1-\eta_s\eta_d}{\eta_s\eta_d}$ & $0.0315$ \\
$5$ & $2\sigma^2_{\mathrm{GKP}} + \frac{1-\eta_s^2\eta_d}{\eta_s^2\eta_d}$ & $0.0934$ \\
$6$ & $3\sigma^2_{\mathrm{GKP}} + 1 - \eta_s^2 + \frac{1-\eta_d}{\eta_d}$ & $0.250$ \\
$7$ & $3\sigma^2_{\mathrm{GKP}} + 1 - \eta_s^3 + \frac{1-\eta_d}{\eta_d}$ & $0.3^\dagger$ \\
$8$ & $2\sigma^2_{\mathrm{GKP}} + 1 - \eta_s^2 + \frac{1-\eta_d}{\eta_d}$ & $0.0919$ \\
$9$ & $2\sigma^2_{\mathrm{GKP}} + 1 - \eta_s^3 + \frac{1-\eta_d}{\eta_d}$ & $0.149$ \\
$10$ & $3\sigma^2_{\mathrm{GKP}} + 1 - \eta_s + \frac{1-\eta_d}{\eta_d}$ & $0.198$ \\
$11$ & $2\sigma^2_{\mathrm{GKP}} + 1 - \eta_s + \frac{1-\eta_d}{\eta_d}$ & $0.0310$ \\
\noalign{\vskip 0.15cm}
\bottomrule
\end{tabular}

\vspace{0.2ex}
\makebox[15.2cm][l]{$^\dagger$ The value $0.3$ is fixed.}

\end{table}

Measurement type 7 has the largest standard deviation of Gaussian random displacement noise among all measurement types. Therefore, we use its discard window size parameter, denoted by $v_7$, as the reference value. The discard window size parameters for all other measurement types are determined relative to $v_7$ so that the error probabilities are equal across all measurement types.

Measurement types 6--11 correspond to the measurements used during the refreshment process. During this process, we use preamplification rather than CC amplification, because the symmetric-loss requirement for CC amplification is not satisfied.

In addition, no optical switch is applied immediately after a refreshment process. Once the refreshment process succeeds, the subsequent transition corresponds to a relatively easy measurement with a narrow discard window, namely measurement type 4. In this case, the benefit of routing the resulting graph states with an additional optical switch is small, and we therefore omit the optical switch for this transition.

\section{Detailed Construction Procedure for UW2 and UW3}
\label{sec:B}
\begin{figure*}[t]
    \centering
    \includegraphics[width=0.9\textwidth]{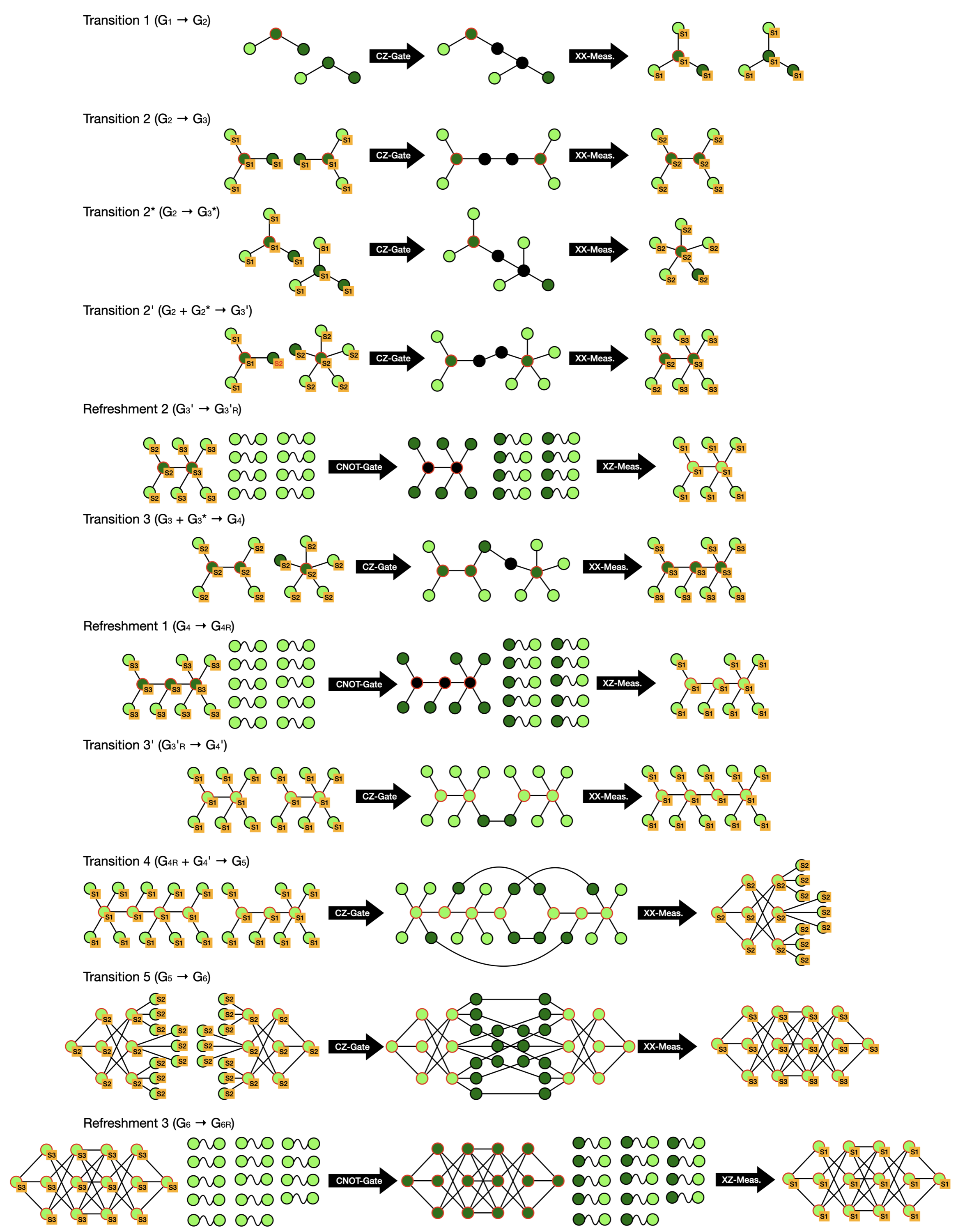}
    \caption{
    Detailed construction procedure for UW2. 
    }
    \label{fig:detailed_UW2}
\end{figure*}

\begin{figure*}[t]
    \centering
    \includegraphics[width=0.9\textwidth]{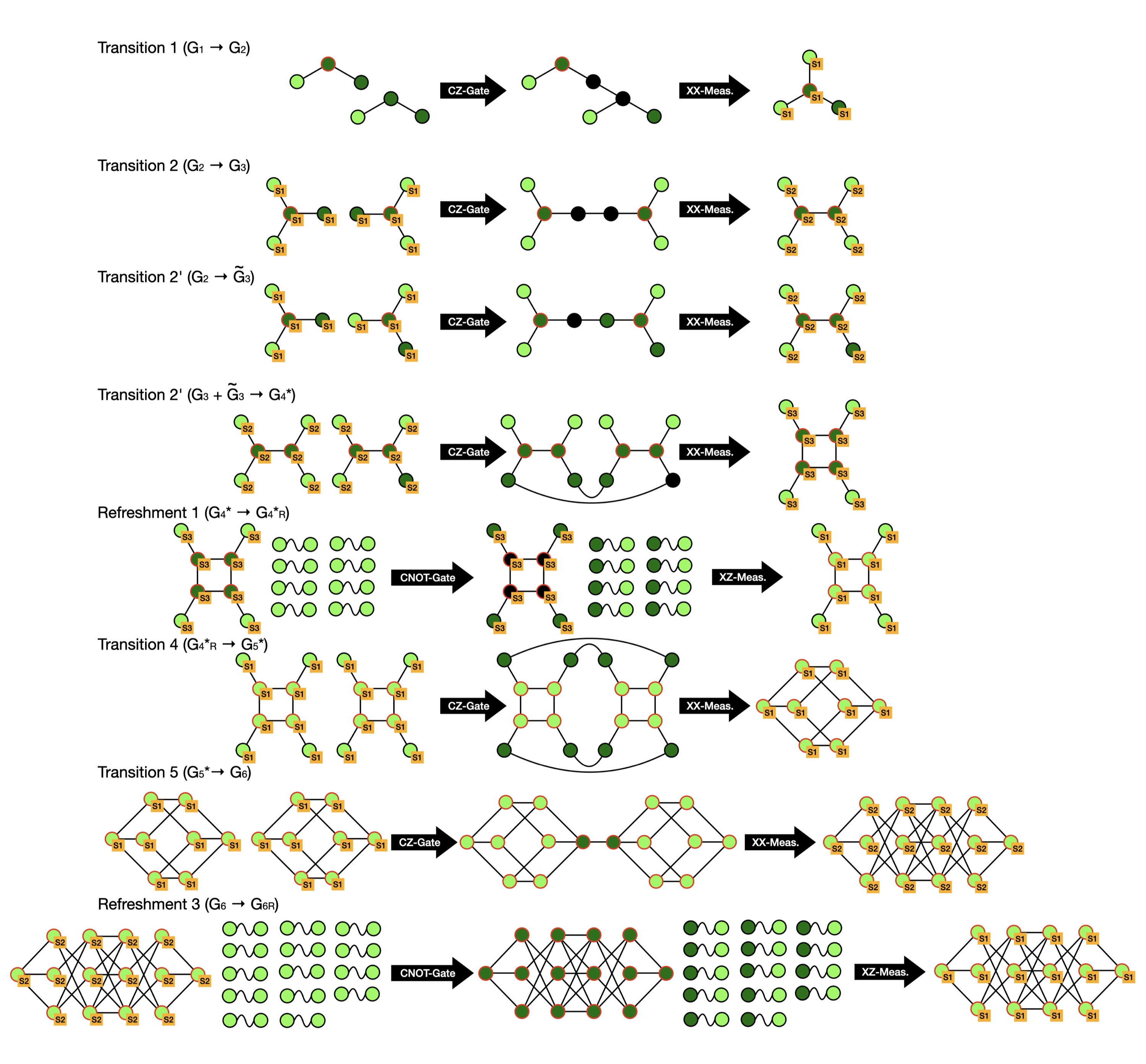}
    \caption{
    Detailed construction procedure for UW3. 
    }
    \label{fig:detailed_UW3}
\end{figure*}

We provide the detailed construction procedures for UW2 and UW3, shown in Figs.~\ref{fig:detailed_UW2} and~\ref{fig:detailed_UW3}, respectively. The transition labels are consistent with those used in Fig.~\ref{fig:3}.

The labels $\mathrm{S1}$, $\mathrm{S2}$, and $\mathrm{S3}$ in Figs.~\ref{fig:detailed_UW2} and~\ref{fig:detailed_UW3} indicate the number of optical switches that each qubit has passed through up to that point. For example, $\mathrm{S2}$ means that the corresponding qubit has passed through optical switches twice.

\end{document}